\documentclass{article}
\usepackage{fullpage}
\usepackage{hyperref}
\usepackage[backend=biber,style=numeric,doi=false,isbn=false,url=false,hyperref=true,backref=true,maxbibnames=10]{biblatex}
\usepackage{amsmath, amssymb}
\usepackage[capitalize]{cleveref}
\usepackage{mathtools}
\usepackage{amsthm}
\usepackage{enumerate}
\usepackage{url}
\usepackage{makeidx}
\usepackage{multirow}
\usepackage{caption}
\usepackage{subcaption}

\usepackage{tikz}
\usetikzlibrary{decorations.markings}
\pgfdeclarelayer{background}
\pgfdeclarelayer{nodelayer}
\pgfdeclarelayer{edgelayer}
\pgfsetlayers{background,edgelayer,nodelayer,main}

\tikzstyle{new}=[circle, minimum width=4pt,inner sep=0pt, fill=black,draw=black]
\tikzstyle{none}=[circle,fill=white,draw=black]
\tikzstyle{n}=[shape=rectangle,minimum width=1pt,inner sep=0pt, fill=none,draw=none]
\tikzstyle{emph}=[circle, minimum width=4pt,inner sep=0pt, fill=magenta,draw=magenta]

\tikzset{directed/.style={decoration={
  markings,
  mark=at position .6 with {\arrow{>}}},postaction={decorate}}}
\tikzset{directed2/.style={decoration={
  markings,
  mark=at position .6875 with {\arrow{>}},
  mark=at position .7125 with {\arrow{>}}},postaction={decorate}}}
\tikzset{directed3/.style={decoration={
  markings,
  mark=at position .7 with {\arrow{>}}},postaction={decorate}}}

\usepackage{array}
\usepackage{float}
\usepackage{wrapfig}
\usepackage{graphicx}
\usepackage{rotating}
\usepackage{nicefrac}
\usepackage{longtable}
\usepackage[table]{xcolor}
\usepackage{booktabs}

\usepackage[overload]{textcase}

\makeatletter
\def\cleardoublepage{
  \clearpage
  \if@twoside\ifodd\c@page\else
  \hbox{}
  \thispagestyle{empty}
  \newpage
  \if@twocolumn\hbox{}\newpage\fi
  \fi\fi
  }
\makeatother

\newtheoremstyle{plainsl}%
    {\topsep}
    {\topsep}
    {\slshape}
    {}
    {\normalfont\bfseries}
    {.}
    { }
    {}

\swapnumbers

{\theoremstyle{plainsl}
\newtheorem{theorem}{Theorem}[section]
\newtheorem{lemma}[theorem]{Lemma}

}
{\theoremstyle{remark}
\newtheorem{remark}[theorem]{Remark}
}

\numberwithin{equation}{section}

\newcommand\comp[1]{{\mkern2mu\overline{\mkern-2mu#1}}}

\newcommand\pmat[1]{\begin{pmatrix} #1 \end{pmatrix}}

\DeclareMathOperator{\diag}{diag}

\DeclareMathOperator\vecspan{span}

\newcommand\cx{{\mathbb C}}

\newcommand\re{{\mathbb R}}

\newcommand\cE{{\mathcal E}}

\newcommand\cP{{\mathcal P}}

\newcommand\ket[1]{| #1 \rangle}

\newcommand{\code}[1]{\texttt{#1}}

\definecolor{vcolour}{RGB}{230,97,0}
\definecolor{kcolour}{RGB}{93,58,155}
\definecolor{gcolour}{RGB}{40,155,50}
\definecolor{pcolour}{RGB}{200,40,120}

\hypersetup{
  colorlinks=true,
  linkcolor=kcolour,
  urlcolor=kcolour,
  citecolor=kcolour,
  linktoc=all,
}

\usepackage{authblk}

\title{Perfect and peak state transfer on IBM quantum hardware}
\author[2]{Priscilla Awuah}
\author[1,2]{Krystal Guo}

\affil[1]{Korteweg-de Vries Institute for Mathematics, University of Amsterdam, Amsterdam, The Netherlands}
\affil[2]{\href{https://qusoft.org/}{QuSoft} (Research center for quantum software \& technology), Amsterdam, The Netherlands}

\affil[ ]{\texttt{pawuah@aims.edu.gh}, \texttt{k.guo@uva.nl}}

\date{October 12, 2026}

\begin{document}

\maketitle

\begin{abstract}
We run continuous-time quantum walks with perfect and peak state transfer
on a present-day superconducting quantum processor, IBM Quantum's
$156$-qubit Heron device \code{ibm\_kingston}, deliberately without
error correction or error mitigation. Perfect state transfer and its
relaxations have been studied extensively in algebraic graph theory,
where one can prove exactly which graphs move a quantum state from one
vertex to another with probability one (or with the largest probability
the graph allows, in the case of peak state transfer), and at what time. Our experiments cover perfect state
transfer on the hypercubes $Q_2 = C_4$ and $Q_3$, and on the
Hamming-weighted paths $\mathrm{HP}_n$, the equitable-partition quotients
of the hypercubes $Q_n$, for $n$ up to $20$; peak state transfer on the $X_n$ family for $n$ up
to $11$;  and, for comparison, moving the state along a path of
     qubits by a sequence of SWAP gates.


\end{abstract}


\tableofcontents
\bigskip

\section{Introduction}\label{sec:intro}

Perfect state transfer (PST) and its variants have been studied intensively over the past two decades as a problem in algebraic graph theory; many connections have been found between state transfer and the spectral and combinatorial structure of the underlying graph, see \cite{ChristandlPSTQuantumSpinNet2, God2010, GodsilPeriodicGraphs11, GodsilStateTransfer12, CoutinhoGodsilGuoVanhove2, CoutinhoGodsilPSTProducts, KemLipTau2017, GodGuoKemLip2020, CouGuovBo2017, CouJulSpi2024, CGS-2025}.
This programme has produced spectral characterizations of exactly which graphs admit state transfer, closed forms for the times at which transfer occurs, and infinite families of examples. Although perfect state transfer originated as a problem in quantum computing, it has since become a standard topic in algebraic graph theory, where it features in survey treatments, see for example \cite{liu2022eigenvalues, vandam2016distanceregular}.
In this paper,  we take a selection of these graphs and execute
the corresponding walks on a present-day IBM Quantum processor. We do
so \textit{naively}, by design: no error correction, no error
mitigation, no sophisticated device-specific tricks. This is to say that each walk is transpiled as-is
and run on the noisy hardware.  Our goal is demonstration: to see what
perfect (or peak) state transfer looks like when implemented on current noisy
intermediate-scale quantum (NISQ) hardware, and thereby to connect a
mathematical literature on state transfer with experiments that anyone
with cloud access to a quantum device can reproduce.
The scripts for all the computations in this paper are available at
\url{https://doi.org/10.5281/zenodo.23300422}; see \cite{qiskitcode}.

The quantum-wire problem was posed by Bose~\cite{BoseQuantumComPaths}: move a qubit state along a chain using a fixed Hamiltonian and no control. Christandl, Datta, Ekert and Landahl~\cite{ChristandlPSTQuantumSpinNet} asked for a fixed Hamiltonian that does this perfectly, at a known time, and coined the name. In \cite{ChristandlPSTQuantumSpinNet}, the hypercubes $Q_n$ are shown to
admit PST and transfer a state across distance $n$ at time $\pi/2$,
while among unweighted paths only $P_2$ and $P_3$ do. In a subsequent
paper \cite{ChristandlPSTQuantumSpinNet2}, they observe that the quotient of $Q_n$
by its distance partition, a weighted path we will call the
\textsl{Hamming-weighted path} (\cref{sec:hamming-paths}), has the same transfer.
We run the PST computations on these graphs in \cref{sec:c4-full,sec:q3-full,sec:hamming-paths}. 

Beyond the hypercubes, PST was established on a number of special
families including circulant
graphs \cite{TamonAngeles-CanulPSTcirculant, BasicCirculant,
BasicPetkovicStevanovic}, cubelike graphs
\cite{GodsilBernasconiPSTCubelike, GodsilCheungPSTCubelike}, and
distance-regular graphs and association schemes
\cite{CoutinhoGodsilGuoVanhove2}, before a spectral characterization of PST in
 \cite{CoutinhoPhD}. PST has also been studied with respect to 
graph products, quotients and liftings \cite{CoutinhoGodsilPSTProducts,
TamonPSTQuotientandLifting} and by adding potentials
at the vertices \cite{KemLipTau2017, GodGuoKemLip2020}. 
For broader background we refer to the
surveys \cite{KayReviewPST, coutinho2018continuous, kay2018perfect},
the book of Coutinho and Godsil \cite{CouGodBook},
\cite{nikolopoulos2014quantum} for the physical side of the story, and
\cite{CoutinhoGuo2024} for a selection of open problems.

When PST does not occur, one can still ask how close $|U(t)_{v,u}|$ can
get to its largest possible value. One such relaxation is \textsl{pretty
good} state transfer, where $|U(t)_{v,u}|$ can be made arbitrarily close
to $1$, see
\cite{GodsilKirklandSeveriniSmithPGST, VinetZhedanovAlmost, CouGuovBo2017,
coutinho2023irrational}. However, for fixed $\epsilon > 0$, there is no general result which gives a time $t$ such that $|U(t)_{v,u}|$ is within $\epsilon$ of $1$. For the purposes of experiments, we look instead at \textsl{peak state transfer},
introduced in \cite{CGS-2025}, which relaxes the requirement $|U(t)_{v,u}| = 1$
to the largest value $B(A)_{v,u} = \sum_r |(E_r)_{v,u}|$, where the sum runs over
the spectral idempotents $E_r$ of the adjacency matrix $A$ (see \cref{sec:pst-peakST}); it is realized by
graphs that do not necessarily admit PST but for which $|U(t)_{v,u}|$ is nonetheless
maximized at a specific, computable time.\footnote{The two notions coincide when $B(A)_{v,u} = 1$.} For discrete-time quantum walks, peak
state transfer was defined via projected transition matrices in
\cite{GuoSch2025}.

Our experiments run on \code{ibm\_kingston}, a $156$-qubit
superconducting processor from IBM Quantum's Heron r2 family. We run
each quantum-walk circuit both in the device's default gate basis and
with its optional fractional gates enabled; we will describe the two
bases, and the error rates the device reports for them, in
\cref{sec:c4-transpile}. Entangling gates apply only to physically
adjacent pairs in a fixed coupling graph, so any continuous-time
quantum walk must first be \textsl{transpiled} into a sequence of
native gates and routed through the coupling graph before it can be run.

On \code{ibm\_kingston} we ran:
\begin{itemize}\setlength{\itemsep}{2pt}
\item PST and uniform mixing on the hypercubes $Q_2 = C_4$
  (\cref{sec:c4-full}) and $Q_3$ (\cref{sec:q3-full}) in the full vertex
  encoding, on $2^n$ qubits;
\item PST on the Hamming-weighted paths $\mathrm{HP}_n$
  (\cref{sec:hamming-paths}) for $n = 2, \ldots, 20$, the
  equitable-partition quotients of $Q_n$, on $n+1$ qubits;
\item Peak ST on the $X_n$ family of \cite[Section 6.3]{CGS-2025}
  (\cref{sec:xn-peakST}) for $n = 2, \ldots, 11$, on $2n$ qubits;
\item A SWAP-chain quantum-wire baseline (\cref{sec:swap-chain}) on
  $n = 2, \ldots, 21$ qubits.
\end{itemize}

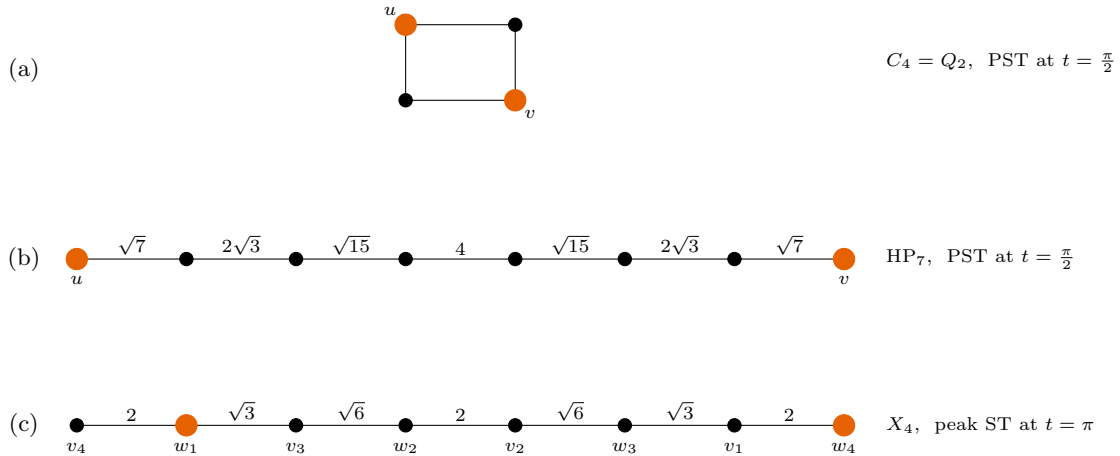
\begin{figure}[ht]
  \centering
  \begin{tikzpicture}[
    gv/.style={circle, minimum width=5pt, inner sep=0pt, fill=black, draw=black},
    pst/.style={circle, minimum width=8pt, inner sep=0pt, fill=vcolour, draw=vcolour},
    wt/.style={midway, above, font=\scriptsize, inner sep=1.5pt},
    lbl/.style={below=3pt, font=\scriptsize},
  ]
    \def\dx{1.45}
    \begin{scope}[yshift=4.3cm]
      \node[font=\small] at (-0.7, 0.4) {(a)};
      \coordinate (nw) at (3.0*\dx, 1.0);
      \coordinate (ne) at (4.0*\dx, 1.0);
      \coordinate (se) at (4.0*\dx, 0.0);
      \coordinate (sw) at (3.0*\dx, 0.0);
      \draw (nw) -- (ne) -- (se) -- (sw) -- (nw);
      \node[gv] at (ne) {};
      \node[gv] at (sw) {};
      \node[pst] at (nw) {}; \node[above left=0pt, font=\scriptsize] at (nw) {$u$};
      \node[pst] at (se) {}; \node[below right=0pt, font=\scriptsize] at (se) {$v$};
      \node[font=\scriptsize, anchor=west] at (7.3*\dx, 0.5) {$C_4 = Q_2$, \ PST at $t = \tfrac{\pi}{2}$};
    \end{scope}
    \begin{scope}[yshift=2.2cm]
      \node[font=\small] at (-0.7, 0) {(b)};
      \foreach \i in {0,...,7} {\coordinate (b\i) at (\i*\dx, 0);}
      \draw (b0)--node[wt]{$\sqrt{7}$}(b1)
            (b1)--node[wt]{$2\sqrt{3}$}(b2)
            (b2)--node[wt]{$\sqrt{15}$}(b3)
            (b3)--node[wt]{$4$}(b4)
            (b4)--node[wt]{$\sqrt{15}$}(b5)
            (b5)--node[wt]{$2\sqrt{3}$}(b6)
            (b6)--node[wt]{$\sqrt{7}$}(b7);
      \foreach \i in {1,...,6} {\node[gv] at (b\i) {};}
      \node[pst] at (b0) {}; \node[lbl] at (b0) {$u$};
      \node[pst] at (b7) {}; \node[lbl] at (b7) {$v$};
      \node[font=\scriptsize, anchor=west] at (7.3*\dx, 0) {$\mathrm{HP}_7$, \ PST at $t = \tfrac{\pi}{2}$};
    \end{scope}
    \begin{scope}[yshift=0cm]
      \node[font=\small] at (-0.7, 0) {(c)};
      \foreach \i in {0,...,7} {\coordinate (c\i) at (\i*\dx, 0);}
      \draw (c0)--node[wt]{$2$}(c1)
            (c1)--node[wt]{$\sqrt{3}$}(c2)
            (c2)--node[wt]{$\sqrt{6}$}(c3)
            (c3)--node[wt]{$2$}(c4)
            (c4)--node[wt]{$\sqrt{6}$}(c5)
            (c5)--node[wt]{$\sqrt{3}$}(c6)
            (c6)--node[wt]{$2$}(c7);
      \foreach \i/\nm in {0/{v_4},2/{v_3},3/{w_2},4/{v_2},5/{w_3},6/{v_1}}
        {\node[gv] at (c\i) {}; \node[lbl] at (c\i) {$\nm$};}
      \node[pst] at (c1) {}; \node[lbl] at (c1) {$w_1$};
      \node[pst] at (c7) {}; \node[lbl] at (c7) {$w_4$};
      \node[font=\scriptsize, anchor=west] at (7.3*\dx, 0) {$X_4$, \ peak ST at $t = \pi$};
    \end{scope}
  \end{tikzpicture}
  \caption{Three of the graphs we run, with the transfer pair
  highlighted in orange. \textbf{(a)} The $4$-cycle $C_4 = Q_2$, which
  admits PST between antipodal vertices $u, v$ at $t = \pi/2$ (all edge
  weights $1$). \textbf{(b)} The Hamming-weighted path $\mathrm{HP}_7$
  on $8$ vertices (the equitable-partition quotient of the hypercube
  $Q_7$), with edge weights $\sqrt{k(8-k)}$; PST runs between the two
  endpoints at $t = \pi/2$. \textbf{(c)} The member $X_4$ of the family
  of \cite[Section 6.3]{CGS-2025} on $8$ vertices, which admits peak ST
  between $w_1$ and $w_4$ at $t = \pi$. Panels (b) and (c) both transfer
  across $8$ qubits but are otherwise different graphs.}
  \label{fig:intro-families}
\end{figure}

\Cref{fig:intro-families} shows three of these graphs, with the pair of
vertices between which the state is transferred highlighted.

The paper is organized as follows. \Cref{sec:background} reviews
continuous-time quantum walks and the XY Hamiltonian that realizes them
in the single-excitation subspace, and \cref{sec:pst-peakST} collects
the definitions and results we need on perfect and peak state transfer.
\Cref{sec:implementation} describes the pipeline from a graph to a
circuit: the exact \code{UnitaryGate} route and the Trotterized
\code{PauliEvolutionGate} route, followed by transpilation onto the
device. The experiments occupy
\cref{sec:c4-full,sec:q3-full,sec:hamming-paths,sec:xn-peakST,sec:swap-chain}:
\cref{sec:c4-full} treats PST on $C_4$ as a worked example of the full
methodology, \cref{sec:q3-full} repeats it on $Q_3$,
\cref{sec:hamming-paths} scales the hypercube experiment to $n = 20$
via the Hamming-weighted paths, \cref{sec:xn-peakST} runs peak state
transfer on the family $X_n$, and \cref{sec:swap-chain} provides state transfer along a
     path of qubits using SWAP gates instead of a quantum walk, for comparison. Per-vertex measurement tables for the larger experiments are
collected in \cref{app:hamming-tables,app:xn-tables}.

\section{Continuous-time quantum walks and the XY Hamiltonian}\label{sec:background}

We consider $N$ qubits, one for each vertex of a graph $G$, interacting according to a time-independent Hamiltonian $H$, a Hermitian matrix on $(\cx^2)^{\otimes N}$; the edges of $G$ record which pairs interact. We allow the edges to have real weights $J_{uv}$, and write $A(G)$ for the (weighted) \textsl{adjacency
matrix} of $G$, the $N	imes N$ matrix whose $(u,v)$-entry equals $J_{uv}$ if
$uv\in E(G)$ and $0$ otherwise. If no weights are specified, we assume $J_{uv}=1$ for all edges $uv$. 

The \textsl{Pauli matrices} are
\[
X = \pmat{0 & 1 \\ 1 & 0}, \qquad
Y = \pmat{0 & -i \\ i & 0}, \qquad
Z = \pmat{1 & 0 \\ 0 & -1}.
\]
For a vertex $u\in V(G)$ we write $X_u$ for the $N$-fold Kronecker product
\[
X_u = I\otimes\cdots\otimes X \otimes\cdots\otimes I,
\]
in which $X$ occupies the $u$th factor and every other factor is the
$2\times 2$ identity $I$, and define $Y_u$ and $Z_u$ analogously; in the
notation of \cite{CGS-2025} these are $X_u = \sigma_u^x$, $Y_u = \sigma_u^y$ and
$Z_u = \sigma_u^z$. The \textsl{$XY$-Hamiltonian} of $G$ is
\[
H_{XY} = \frac{1}{2}\sum_{uv\in E(G)} J_{uv}\,(X_u X_v + Y_u Y_v).
\]

A \textsl{continuous-time quantum walk} is the time evolution generated by such a
Hamiltonian: its transition matrix is a matrix-valued function in time given by
\[
U(t) = e^{-itH}, \qquad t\in\re.
\]
A walk with initial state $\ket{\psi}$ is in the state $U(t)\ket{\psi}$ at
time $t$.

For each vertex $u$, let $\ket{e_u} = X_u\ket{0\cdots0}$ denote the computational
basis state carrying a single $\ket{1}$ on the qubit at $u$. These $N$
\textsl{single-excitation states} are in bijection with $V(G)$ and span the
\textsl{single-excitation subspace}
$\cE = \vecspan\{\ket{e_u} : u\in V(G)\}$. It is straightforward to check that
$H_{XY}$ preserves $\cE$ and that, in the basis $\{\ket{e_u}\}$, it acts exactly
as the weighted adjacency matrix $A(G)$; this is verified for $J_{uv}=1$ in
\cite[Section 2]{CGS-2025}, and the identical computation gives $A(G)$ for
arbitrary weights. Restricting the walk to $\cE$ therefore yields
\[
\left. e^{-itH_{XY}}\right|_{\cE} = e^{-itA(G)}.
\]
Thus, we may work with $e^{-itA(G)}$, which is the setting of most results on state transfer; see \cite{CouGodBook}.

\begin{remark}
We use the physics sign convention $U(t) = e^{-itH}$, matching the Qiskit
time-evolution primitives used in our implementation, whereas \cite{CouGodBook, CGS-2025}
uses $U(t) = e^{+itA}$. Since $A(G)$ is real symmetric,
$e^{-itA} = \comp{e^{+itA}}$, so the two conventions yield identical transfer
probabilities $|U(t)_{v,u}|$ and identical transfer times, differing only by an
overall complex conjugation.
\end{remark}

\section{Perfect and peak state transfer}\label{sec:pst-peakST}

Let $M$ be a real symmetric matrix, typically the adjacency matrix
$A(G)$ of a (weighted) graph $G$. For $u, v \in V(G)$, the probability
of finding the walker at $v$ at time $t$ when it starts at $u$ is
$|U(t)_{v,u}|^2$. We say that there is \textsl{perfect state transfer}
(PST) from $u$ to $v$ if
\[
|U(\tau)_{v,u}| = 1
\qquad \text{for some } \tau > 0,
\]
in which case unitarity forces $U(\tau)\ket{e_u} = \gamma \ket{e_v}$
for some $\gamma \in \cx$ with $|\gamma| = 1$. We say that $G$ admits \textsl{uniform mixing} at time $\tau$ if $|U(\tau)_{v,u}|^2 = 1/|V(G)|$ for all $u, v \in V(G)$.

When PST does not occur, we may still ask how close $|U(t)_{v,u}|$ can
get to its largest possible value. Using the spectral decomposition
$M = \sum_{r=0}^{d} \theta_r E_r$ and the triangle inequality,  with
$|e^{-it\theta_r}| = 1$ for $t\in\re$, we obtain
\[
|U(t)_{v,u}|
\;=\; \Big| \sum_{r=0}^{d} e^{-it\theta_r}\,(E_r)_{v,u} \Big|
\;\leq\; \sum_{r=0}^{d} |(E_r)_{v,u}|.
\]
The right-hand side is the $(v,u)$-entry of the \textsl{bounding matrix}
of $M$ \cite{CGS-2025},
\begin{equation}\label{eq:bounding-matrix}
B(M) \;\coloneqq\; \sum_{r=0}^{d} |E_r|,
\end{equation}
where $|E_r|$ is the entrywise absolute value. Following
\cite[Definition 3.1]{CGS-2025}, we say that there is \textsl{peak state
transfer} (peak ST) from $u$ to $v$ if $|U(\tau)_{v,u}| = B(M)_{v,u}$
for some $\tau > 0$. We call $B(M)_{v,u}$ the \textsl{peak transfer rate}
of $(u, v)$ and $B(M)_{v,u}^2$ the \textsl{peak transfer probability}.
PST is the special case $B(M)_{v,u} = 1$.

We implement PST on the hypercubes $Q_n$ (taking $J_{uv} = 1$), which
admit PST between every pair of antipodal vertices at $t = \pi/2$
\cite{ChristandlPSTQuantumSpinNet}: the $Q_2 = C_4$ case in
\cref{sec:c4-full} and $Q_3$ in \cref{sec:q3-full}.
In \cref{sec:hamming-paths}, we will use that the equitable partition of
$Q_n$ by Hamming distance from a fixed source has the weighted-path
quotient $\mathrm{HP}_n$ on $n + 1$ vertices, to which PST descends
between the two endpoints at the same time by Lemma~\ref{lem:quotient-transfer};
this drops the qubit count from $2^n$ to $n + 1$.

Let $\cP$ be an equitable partition of $G$ with cells $C_0, \ldots, C_m$.
Its quotient matrix $A_{\cP}$ is the $(m+1)\times(m+1)$ matrix whose
$(i,j)$-entry is the number of neighbours in $C_j$ of a vertex of $C_i$.
Let $D = \diag(|C_0|, \ldots, |C_m|)$. The \textsl{symmetrized quotient}
$\tilde A_{\cP} = D^{1/2} A_{\cP} D^{-1/2}$ is symmetric.

\begin{lemma}\label{lem:quotient-transfer}
Let $\cP$ be an equitable partition of $G$, and let $u, v \in V(G)$ be
such that $\{u\}$ and $\{v\}$ are cells of $\cP$. Then
$U_{A(G)}(t)_{v,u} = U_{\tilde A_{\cP}}(t)_{v,u}$ for all $t$. In
particular, perfect and peak state transfer between $u$ and $v$ pass to
$\tilde A_{\cP}$ with the same time and the same value.
\end{lemma}

\begin{proof}
The normalized characteristic matrix of $\cP$ intertwines $A(G)$ and
$\tilde A_{\cP}$, and the result follows.
\end{proof}

We implement peak ST on the family $X_n$ of \cite[Section 6.3]{CGS-2025}:
$X_n$ is the weighted path on $2n$ vertices visiting
$v_n, w_1, v_{n-1}, w_2, \ldots, v_1, w_n$ with the edge between $v_j$
and $w_k$ of weight $\sqrt{j\,k}$. By \cite[Theorem 6.5]{CGS-2025}, $X_n$
admits peak state transfer between $w_1$ and $w_n$ at time $t = \pi$,
with peak transfer rate
\[
B(A(X_n))_{w_1, w_n}
\;=\; \frac{2 \cdot 4^{n-1}}{\sqrt{n}\,(n+1)\,C_n}
\;\xrightarrow{n \to \infty}\; \frac{\sqrt{\pi}}{2},
\]
where $C_n$ is the $n$-th Catalan number, so the peak transfer
probability converges to $\pi/4 \approx 0.785$ from above and is
strictly less than $1$, making $X_n$ a strict instance of peak ST that
is not PST. We treat it in \cref{sec:xn-peakST}.

\section{Implementing continuous-time quantum walks on a quantum computer}\label{sec:implementation}

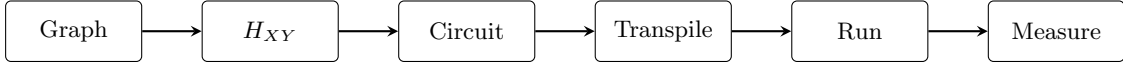
\begin{figure}[ht]
  \centering
  \begin{tikzpicture}[
    box/.style={draw, rounded corners=2pt, minimum height=8mm,
                minimum width=18mm, inner sep=4pt, font=\small},
    arr/.style={->, >=stealth, thick}
  ]
    \node[box] (G) at ( 0.0,0) {Graph};
    \node[box] (H) at ( 2.6,0) {$H_{XY}$};
    \node[box] (C) at ( 5.2,0) {Circuit};
    \node[box] (T) at ( 7.8,0) {Transpile};
    \node[box] (R) at (10.4,0) {Run};
    \node[box] (M) at (13.0,0) {Measure};
    \draw[arr] (G) -- (H);
    \draw[arr] (H) -- (C);
    \draw[arr] (C) -- (T);
    \draw[arr] (T) -- (R);
    \draw[arr] (R) -- (M);

  \end{tikzpicture}
  \caption{The implementation pipeline, from a graph $G$ to measurement on
  IBM Quantum hardware.}
  \label{fig:method-pipeline}
\end{figure}

To run a continuous-time quantum walk on the IBM Quantum hardware, we use one qubit for
each vertex of the graph, so $N$ qubits for a graph on $N$ vertices. The CTQW we wish to
implement is obtained by preparing $\ket{e_{u_0}}$ with a single $X$ gate,
applying the unitary $U(t) = e^{-itH_{XY}}$, and measuring in the
computational basis; see \cref{fig:method-pipeline}. Any measured
bit-string outside the single-excitation set is an artifact of hardware
noise.

To do this, we feed the time-evolution unitary to Qiskit by one of two routes. With
\code{UnitaryGate}, we build the $2^N \times 2^N$ matrix $U(t)$ exactly and
append it to the circuit; the transpiler later decomposes this block into
native gates. With \code{PauliEvolutionGate}, we describe
$H_{XY}$ symbolically as a sum of Pauli strings and ask for the evolution,
which Qiskit approximates by a second-order Suzuki--Trotter formula. \code{UnitaryGate}
encodes the transition matrix of the quantum walk exactly, but becomes
infeasible to execute on hardware, as we will see for our small examples.
\code{PauliEvolutionGate} has a free parameter, the Trotter rep
count~$r$, which we will need to optimize: the choice balances the error
coming from the Trotterization and the error resulting from too many
two-qubit gates.

Whichever route is used, the Qiskit transpiler then rewrites the
circuit in \code{ibm\_kingston}'s native gate basis $\{\code{cz},
\code{sx}, \code{rz}, \code{x}\}$
\cite{ibmquantum_qpu_info}, and routes it onto the IBM device's coupling
map. If a logical two-qubit gate uses two non-adjacent physical qubits,
the transpiler uses \code{swap}s to implement the gate.

Heron-family devices additionally offer optional native
\textsl{fractional gates} \cite{ibmquantum_native_gates}. We enable them
for all of our quantum-walk circuits and report both transpilation
regimes throughout; we will describe the fractional gates, and the
caveats that come with them, in \cref{sec:c4-transpile}. The one
exception is the SWAP-chain baseline of \cref{sec:swap-chain}, which is
built from full-angle gates: transpiling the chain with fractional gates
enabled produces an identical circuit, so the baseline is compiled
once, to the default basis.

We will explain our methodology in more detail in \cref{sec:c4-full}, using $C_4$ as a worked example.

\section{Perfect state transfer on the 4-cycle}\label{sec:c4-full}

In this section, we treat the 4-cycle $C_4$ in full detail, using it as the running example 
through which we lay out our methodology. The $4$-cycle admits PST at 
time $t = \frac{\pi}{2}$ and uniform mixing at time $t = \frac{\pi}{4}$. After the walk-through 
of our methodology with PST at time $t = \frac{\pi}{2}$ on $C_4$, we also implement uniform 
mixing in \Cref{sec:c4-mixing} and we implement PST a second way, using uniform mixing
as an intermediate state (\Cref{sec:c4-pst-via-mixing}).  
We will then compare the outputs of these different implementations of PST.

\begin{figure}[ht]
  \centering
  \begin{tikzpicture}[scale=1.4, every label/.style={font=\small, label distance=2pt}]
    \node[new, label={above left:{$00\;\;(q_0)$}}] (v00) at (0, 2) {};
    \node[new, label={above right:{$01\;\;(q_1)$}}] (v01) at (2, 2) {};
    \node[new, label={below right:{$11\;\;(q_2)$}}] (v11) at (2, 0) {};
    \node[new, label={below left:{$10\;\;(q_3)$}}] (v10) at (0, 0) {};
    \draw (v00) -- (v01) -- (v11) -- (v10) -- (v00);
  \end{tikzpicture}
  \caption{The 4-cycle $C_4 = Q_2$: vertices are the binary strings
  $\{00, 01, 11, 10\}$, with edges connecting strings that differ in one
  position. We assign one qubit per vertex ($q_0, \ldots, q_3$); the
  single-excitation state at vertex $u$ is
  $\ket{e_u} = X_u\ket{0000}$, so e.g.\
  $\ket{e_{00}} = \ket{0001}$ in Qiskit's little-endian display.}
  \label{fig:c4-graph}
\end{figure}
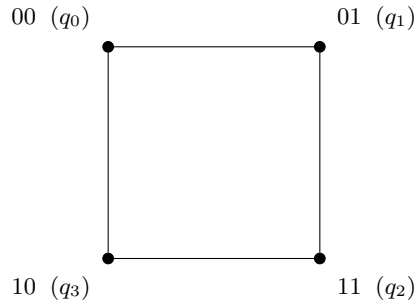

\subsection{Mathematical setup}\label{sec:c4-math}

We view the 4-cycle $C_4$ as the binary hypercube $Q_2$: the vertices are
the four binary strings $\{00,\,01,\,11,\,10\}$ and the edges are between pairs of vertices
that differ in exactly one position.

We model a single qubit at each vertex of $C_4$, indexed by
$u \in V(C_4)$, with the Pauli operators $X_u, Y_u, Z_u$ of
\cref{sec:background}. The XY-Hamiltonian on $C_4$ is
\begin{equation}\label{eq:c4-H}
\begin{aligned}
  H \;=\; \tfrac12 \bigl( & X_{00} X_{01} + Y_{00} Y_{01} \,+\, X_{01} X_{11} + Y_{01} Y_{11} \\
                      \,+\, & X_{11} X_{10} + Y_{11} Y_{10} \,+\, X_{10} X_{00} + Y_{10} Y_{00} \bigr).
\end{aligned}
\end{equation}
The unitary transition matrix is $U(t) = e^{-itH}$, acting on $(\cx^2)^{\otimes 4}$.

 For each vertex $u \in V(C_4)$, the single-excitation state is given by
  $\ket{e_u} \coloneqq X_u \ket{0000}$. For instance, $\ket{e_{00}} = \ket{0001}$; in Qiskit's 
  convention, the rightmost bit corresponds to qubit~$0$. The four states
  \[
    \ket{e_{00}} = \ket{0001},\quad
    \ket{e_{01}} = \ket{0010},\quad
    \ket{e_{11}} = \ket{0100},\quad
    \ket{e_{10}} = \ket{1000}
  \]
  span an invariant subspace of $H$ (the single-excitation subspace), and the restriction of $H$ to this
  subspace, written in the basis above, is exactly the adjacency matrix
  $A(C_4)$,  which is the setting of the algebraic-graph-theory literature on state transfer, see \cite{CouGodBook, GodsilStateTransfer12}.
Since the transition matrix $U(t)$ preserves the single-excitation subspace, 
any measurement bit-string corresponding to states outside of this subspace 
is a sign of hardware noise. Thus, for our computations on IBM hardware, we report the 
in-subspace fraction, which is the proportion of shots returning a single-excitation state.

Recall from \cref{sec:pst-peakST} that any hypercube has perfect state transfer between 
every pair of antipodal vertices at time $\pi/2$. In particular, perfect state 
transfer occurs in the $4$-cycle from vertex $00$ to vertex $11$ at time
$t = \pi/2$.

\begin{figure}[ht]
  \centering
  \includegraphics[width=0.6\linewidth]{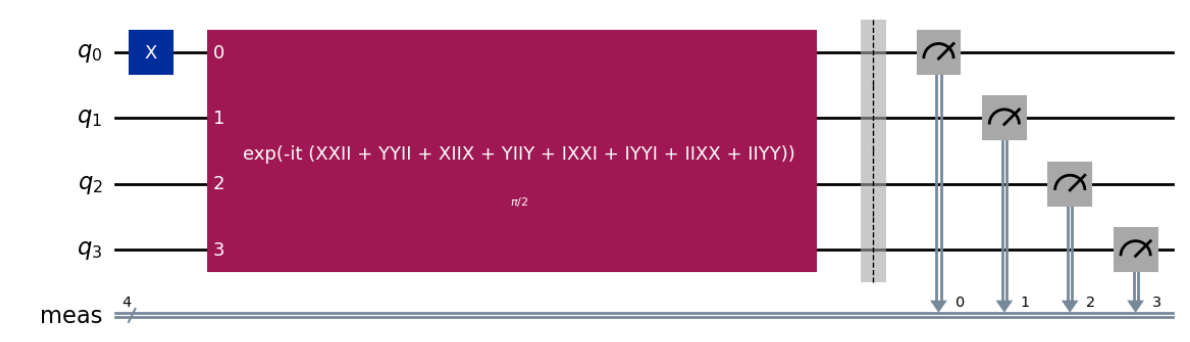}
  \caption{Logical circuit for $V_{\mathrm{exp}}(\pi/2) = e^{-i\pi H/2}\cdot X_{00}$ on
  four qubits.
  The barrier separates the evolution from the measurements on the four
  qubits.}
  \label{fig:c4-logical}
\end{figure}

To see this in action on quantum hardware, we will implement the unitary
\[
  V_{\mathrm{exp}}(\pi/2) \;=\; U(\pi/2) \cdot X_{00} \;=\; e^{-i\pi H/2} \cdot X_{00}
\]
on four qubits; that is, we start in $\ket{0000}$, apply $X$ on the qubit
associated with vertex $00$ (to bring us to the single-excitation state
$\ket{e_{00}}$), evolve under $H$ for time $\pi/2$, and measure in the
computational basis. See \Cref{fig:c4-logical}.  On the IBM hardware, we will execute 
each circuit for 4096 \textsl{shots}; that is, 4096 independent runs of the program, which 
produce  4096 bit-strings.
In a noiseless setting, we would obtain 4096 copies of the bit-string \texttt{0100}, 
corresponding to
$\ket{e_{11}} = \ket{0100}$. 

\subsection{Inputting into Qiskit}\label{sec:c4-qiskit}

We need to express the unitary $V_{\mathrm{exp}}(\pi/2) = e^{-i\pi H/2}\cdot X_{00}$, shown as a circuit in 
\Cref{fig:c4-logical}, 
as a Qiskit circuit that can be fed into the transpilation step. For the time-evolution 
block, Qiskit offers two natural constructions, \code{UnitaryGate} and
\code{PauliEvolutionGate}, which we describe below.

\paragraph{\code{UnitaryGate}.}
We compute the $16 \times 16$ matrix $U(\pi/2) = e^{-i\pi H/2}$ numerically
in \code{scipy} (\code{scipy.linalg.expm}), wrap it as a Qiskit
\code{UnitaryGate}, and append it to the circuit. The resulting logical
circuit consists of two operations: $X_{00}$ followed by the opaque
\texttt{UnitaryGate} block. 
The transpiler then decomposes this quantum circuit into the device's native gates; 
this synthesis makes no use of the fact that $U(\pi/2)$ came from a Hamiltonian on an underlying graph. 

\paragraph{\code{PauliEvolutionGate}.}
Alternatively, we describe $H$ to Qiskit symbolically as a sum of Pauli
operators (\code{SparsePauliOp}) and ask for the time evolution $e^{-itH}$
via the \code{PauliEvolutionGate}, which approximates the evolution by a
second-order Suzuki--Trotter product formula. Concretely, we will write the
Hamiltonian \eqref{eq:c4-H} as $H = \sum_{k=1}^{8} h_k$, where $h_1, \ldots, h_8$
are the eight terms
\[
  \tfrac12 X_{00}X_{01},\ \tfrac12 Y_{00}Y_{01},\ \tfrac12 X_{01}X_{11},\ \tfrac12 Y_{01}Y_{11},\
  \tfrac12 X_{11}X_{10},\ \tfrac12 Y_{11}Y_{10},\ \tfrac12 X_{10}X_{00},\ \tfrac12 Y_{10}Y_{00}
\]
in the order in which they appear in \eqref{eq:c4-H}; this is also the order in
which our code passes them to Qiskit. Each $e^{-ish_k}$ is a single two-qubit
Pauli rotation. Terms on two edges that share a vertex need not commute (for
example, $X_{00}X_{01}$ and $Y_{01}Y_{11}$ anticommute), so $e^{-itH}$ is not the
product of the $e^{-ith_k}$. The idea of Trotterization is
that $e^{-itH}$ can nevertheless be approximated to arbitrary accuracy by
\[
  e^{-itH} \;\approx\; \Bigl( e^{-i\frac{t}{2r}h_1}\, e^{-i\frac{t}{2r}h_2} \cdots e^{-i\frac{t}{2r}h_8}\;
  e^{-i\frac{t}{2r}h_8}\, e^{-i\frac{t}{2r}h_7} \cdots e^{-i\frac{t}{2r}h_1} \Bigr)^{r},
\]
by increasing the number of repetitions $r$; for more background, see
\cite[\S4.7]{nielsen2010quantum}. This is the second-order formula that
\code{PauliEvolutionGate} with \code{SuzukiTrotter(order=2, reps=r)} implements:
each repetition applies the terms in the listed order and then in reverse, each
for time $t/(2r)$. We will refer to $r$ as the \textsl{Trotter rep count}
throughout. At $t = \pi/2$, the formula gives the target probability
$P(\ket{0100}) = 0.78125$ at $r = 1$, and $0.999948$ at $r = 8$, as in
\cref{tab:c4-aer-pi2}.

\begin{remark}
The order of the terms matters. Write $H = H_A + H_B$, where
\[
  H_A = \tfrac12 (X_{00}X_{01} + Y_{00}Y_{01}) + \tfrac12(X_{11}X_{10} + Y_{11}Y_{10}),
  \quad
  H_B = \tfrac12 (X_{01}X_{11} + Y_{01}Y_{11}) + \tfrac12(X_{10}X_{00} + Y_{10}Y_{00}),
\]
so that $H_A$ and $H_B$ correspond to the two perfect matchings of $C_4$. All the terms within
$H_A$ commute, as do all the terms within $H_B$. The operators $H_A$ and $H_B$ do not commute on
$(\cx^2)^{\otimes 4}$, but both preserve the single-excitation subspace, and their
restrictions to it are the permutation matrices of the two matchings, which
commute. Thus on the single-excitation subspace
$e^{-itH} = e^{-itH_A} e^{-itH_B}$ exactly, and a product formula that groups the
terms by matching has no Trotter error for any $r$, including $r = 1$. The same
holds for every hypercube $Q_n$: grouping the terms by the $n$ directions gives $n$
operators whose restrictions are the commuting coordinate-flip permutation
matrices whose sum is $A(Q_n)$. The circuits in this report use the
order of \eqref{eq:c4-H}, and the gate counts we report are for that order.
\end{remark}

\subsection{Transpilation}\label{sec:c4-transpile}

The Qiskit \textsl{transpiler} converts a circuit written in arbitrary gates
into one written in the basis natively executable on a given backend. For
\code{ibm\_kingston} (a $156$-qubit Heron r2 processor) the default basis is
\[
  \{\code{cz},\ \code{sx},\ \code{rz},\ \code{x}\},
\]
where \code{cz} (controlled-$Z$) is the only two-qubit operation and the
rest act on single qubits \cite{ibmquantum_qpu_info}. Heron-family
devices additionally offer two optional native \textsl{fractional
gates} \cite{ibmquantum_native_gates}: the single-qubit rotation
$\code{rx}(\theta)$, and the two-qubit rotation
\[
  \code{rzz}(\theta) = e^{-i \frac{\theta}{2} Z \otimes Z},
  \qquad 0 \le \theta \le \pi/2,
\]
realized as a partial entangling interaction rather than a sequence of
full \code{cz} gates. As discussed in \cref{sec:implementation}, we
enable the fractional gates for the experiments of this section; since
each Trotter factor in our circuits is a two-qubit Pauli
rotation by a small angle, this is exactly the workload that fractional
gates are designed for.

Three caveats explain why these gates are off by default. IBM Quantum
classifies them as experimental. They are incompatible with much of the
error-suppression and error-mitigation stack of the runtime (dynamic
circuits, Pauli twirling, probabilistic error cancellation, zero-noise
extrapolation); this restriction costs us nothing by design, since we
run every circuit without error mitigation. Finally, while the device
calibration reports a lower median error for \code{rzz} than for
\code{cz} (see below), the angle at which \code{rzz} is benchmarked is
not specified, so the reported figure should be read as representative
rather than as the error of the small-angle rotations that dominate our
Trotter circuits.
 The hardware only accepts \code{rzz} angles in
$[0, \pi/2]$, and the transpiler can produce angles outside this range,
so we apply the \code{FoldRzzAngle} pass of Qiskit IBM Runtime as a
final step; it wraps every \code{rzz} angle into range at the cost of
some additional single-qubit gates.

The
device has a fixed \textsl{coupling map}: a graph on the $156$ physical
qubits, with each physical qubit being adjacent to at
most three others. A two-qubit gate can only be applied to a pair
of adjacent qubits in the coupling map. To execute a
logical two-qubit gate between two qubits that are not physically
adjacent, the transpiler inserts \code{swap} operations to route them
through the coupling map. Here, we use \code{optimization\_level=3}, the
highest-effort preset.

  The two-qubit gate count after transpilation is a key indicator of
  hardware-noise sensitivity: present-day single-qubit gates on
  \code{ibm\_kingston} have median error rates around $2.6 \times 10^{-4}$
  for \code{sx} and \code{x}, while the two-qubit \code{cz} sits at roughly
  $2.1 \times 10^{-3}$ (median) and the fractional \code{rzz} at
  $1.8 \times 10^{-3}$ (median), see
  \cite{ibmquantum_qpu_info}.\footnote{All numbers are read from the live device
  calibration via \code{backend.target}; the values quoted here are from
  the calibration of 2026-06-12.}

\Cref{tab:c4-transpile} reports the depth, total gate count, and two-qubit
gate counts for our two $t = \pi/2$ circuits, transpiled to the live
\code{ibm\_kingston} backend both with and without fractional gates.
The two regimes are instructive. For \code{PauliEvolutionGate}, the
two-qubit \textsl{count} is identical ($133$): each Trotter factor
$e^{-i\theta(XX+YY)/2}$ needs two entangling gates either way. What
changes is what those gates are: in the default basis all $133$ are
full-angle \code{cz} gates, while in the fractional basis $124$ of them
become \code{rzz} gates at the small angles dictated by the Trotter
step, each cheaper than a full \code{cz}. (The extra depth in the
fractional row is the single-qubit overhead of \code{FoldRzzAngle}.)
For \code{UnitaryGate} the picture is reversed: the exact synthesis of
a generic four-qubit unitary has no small-angle structure to exploit,
and the fractional-basis count is
actually \textsl{larger} ($211$ against $187$). 

\begin{table}[ht]
\centering
\small
\begin{tabular}{llrrrrr}
\toprule
Configuration & basis & depth & total & \code{cz} & \code{rzz} & 2q \\
\midrule
$t = \pi/2$, \code{UnitaryGate}                 & \code{cz}  & 687 & 1160 & 187 & n/a & 187 \\
$t = \pi/2$, \code{UnitaryGate}                 & fractional & 559 &  904 & 168 &  43 & 211 \\
$t = \pi/2$, \code{PauliEvolutionGate} ($r=8$)  & \code{cz}  & 500 &  762 & 133 & n/a & 133 \\
$t = \pi/2$, \code{PauliEvolutionGate} ($r=8$)  & fractional & 562 &  846 &   9 & 124 & 133 \\
\bottomrule
\end{tabular}
\caption{Transpiled gate counts for the $t = \pi/2$ PST circuits on $C_4$,
compiled against the live \code{ibm\_kingston} backend
(\code{optimization\_level=3}, \code{seed\_transpiler=0}), in the
default \code{cz} basis and with fractional gates enabled (including
the \code{FoldRzzAngle} post-pass). Numbers
exclude the four final measurements and the barrier inserted by
\code{measure\_all}.}
\label{tab:c4-transpile}
\end{table}

\begin{figure}[p]
  \centering
  \includegraphics[width=0.85\linewidth]{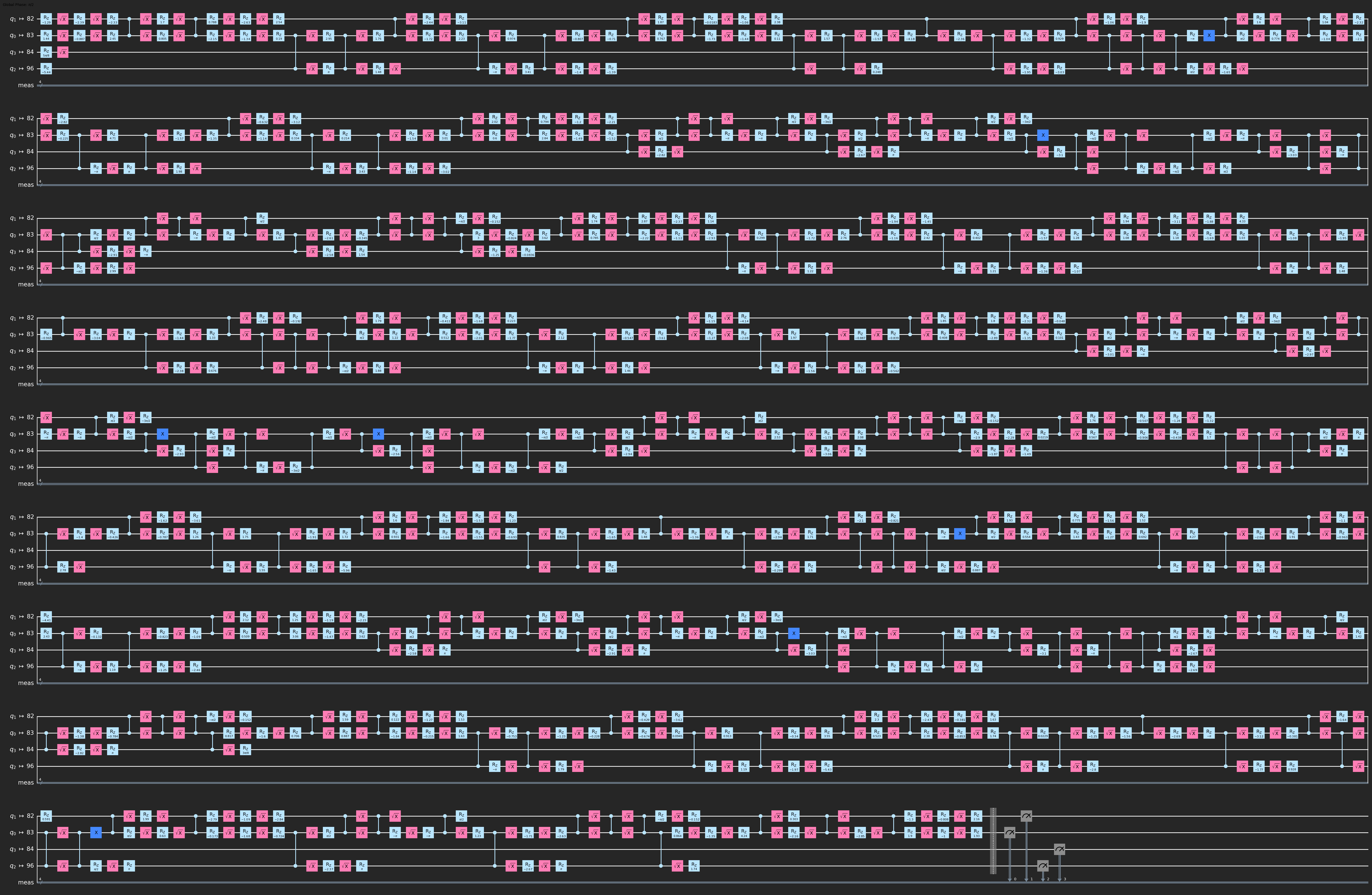}

  \vspace{14pt}

  \includegraphics[width=0.85\linewidth]{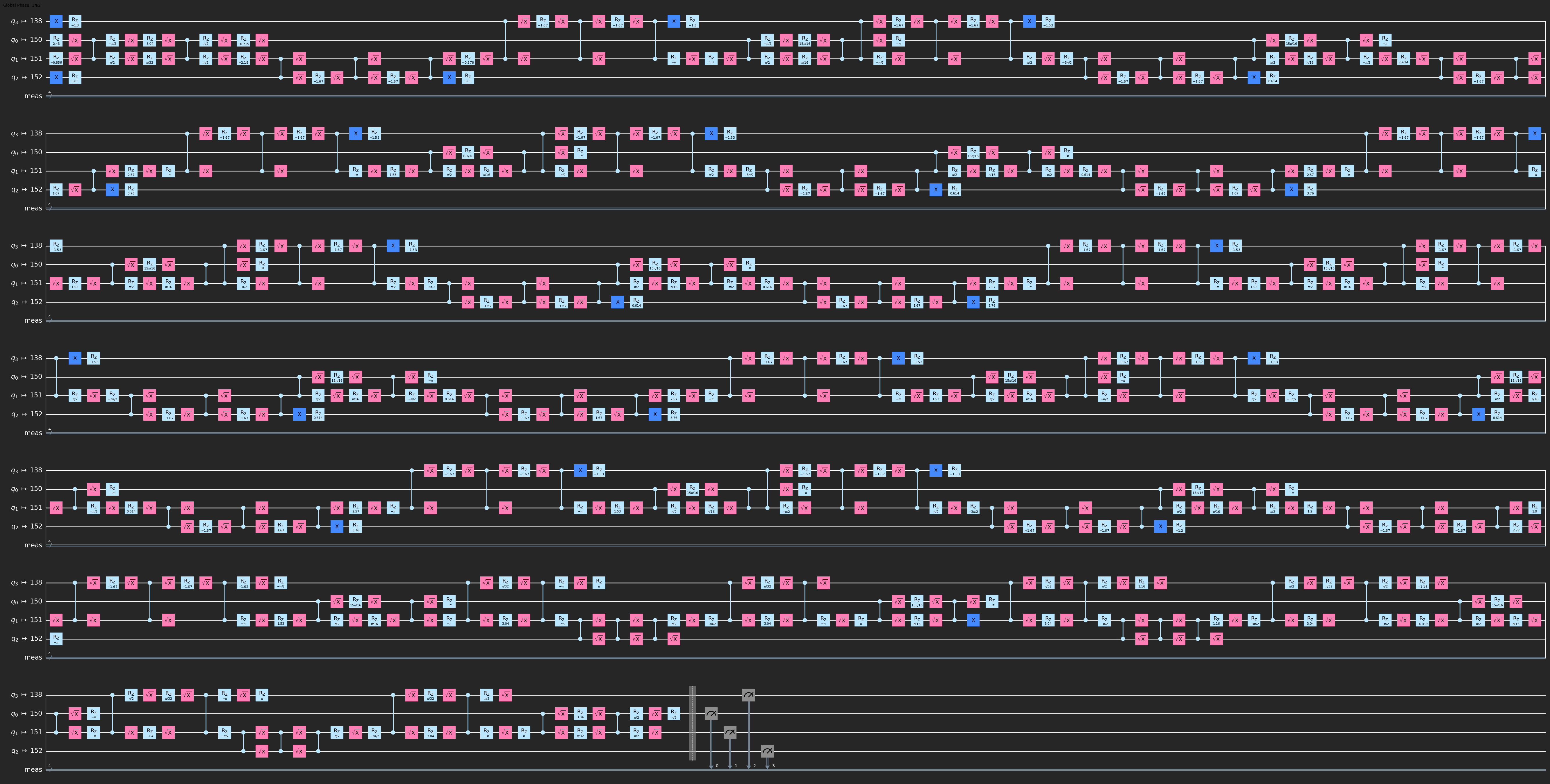}
\caption{The full transpiled $C_4$ PST circuits at $t = \pi/2$ on
\code{ibm\_kingston}, in the default \code{cz} basis. \textbf{(Top)}
\code{UnitaryGate}. \textbf{(Bottom)}
\code{PauliEvolutionGate} at $r = 8$.}  \label{fig:c4-transpiled}
\end{figure}

\Cref{fig:c4-transpiled} shows the full transpiled circuits for
\code{UnitaryGate} and \code{PauliEvolutionGate} on \code{ibm\_kingston}'s coupling map, 
 so the reader can see the visual scale of what it takes in native gates to realize our 
 circuit.
We note that the four logical qubits of $C_4$ are mapped by
the transpiler to a connected line of physical qubits, and the
wraparound edge $(10, 00)$ is realized by inserted \code{swap}s that
route it through the coupling map. 

\subsection{Aer simulator results}\label{sec:c4-aer}

Qiskit Aer, a classical statevector simulator, runs the device-transpiled circuit (the same gate
sequence that runs on \code{ibm\_kingston}) and computes the exact output amplitudes to
 machine precision, with no sampling and no error model.
We report the output probabilities in the computational basis with six
decimal places of precision in \cref{tab:c4-aer-pi2}. The noiseless
distribution is the same whether the circuit is transpiled in the
default \code{cz} basis or with fractional gates enabled: replacing
full \code{cz} gates by \code{rzz} rotations is an exact rewriting of
the same unitary, so the only residual is the synthesis-level Trotter
residual of \code{PauliEvolutionGate}, which is introduced before the
choice of basis.

\begin{table}[ht]
\centering
\small
\begin{tabular}{lcrr}
\toprule
Bit-string & Vertex & \code{UnitaryGate}, $t = \pi/2$ & \code{PauliEvolutionGate} ($r=8$), $t = \pi/2$ \\
\midrule
\texttt{0000} &       & $0.000000$ & $0.000000$ \\
\rowcolor{vcolour!8} \texttt{0001} & $00$  & $0.000000$ & $0.000023$ \\
\rowcolor{vcolour!8}\texttt{0010} & $01$  & $0.000000$ & $0.000014$ \\
\texttt{0011} &       & $0.000000$ & $0.000000$ \\
\rowcolor{kcolour!8}\texttt{0100} & $11$  & $1.000000$ & $0.999948$ \\
\texttt{0101} &       & $0.000000$ & $0.000000$ \\
\texttt{0110} &       & $0.000000$ & $0.000000$ \\
\texttt{0111} &       & $0.000000$ & $0.000000$ \\
\rowcolor{vcolour!8}\texttt{1000} & $10$  & $0.000000$ & $0.000014$ \\
\texttt{1001} &       & $0.000000$ & $0.000000$ \\
\texttt{1010} &       & $0.000000$ & $0.000000$ \\
\texttt{1011} &       & $0.000000$ & $0.000000$ \\
\texttt{1100} &       & $0.000000$ & $0.000000$ \\
\texttt{1101} &       & $0.000000$ & $0.000000$ \\
\texttt{1110} &       & $0.000000$ & $0.000000$ \\
\texttt{1111} &       & $0.000000$ & $0.000000$ \\
\bottomrule
\end{tabular}
\caption{Aer noiseless output probabilities for the $t = \pi/2$ PST
circuits on $C_4$, simulated from the device-transpiled gate sequence.
The distributions are identical, to all six decimal places, for the
\code{cz}-basis and fractional-basis transpilations of
\cref{sec:c4-transpile}. The states of the single-excitation subspaces
have been highlighted, with the target state in its own colour.}
\label{tab:c4-aer-pi2}
\end{table}

\subsection{IBM hardware results}\label{sec:c4-hw}

All hardware runs use the IBM Quantum backend \code{ibm\_kingston}
(Heron r2, $156$ qubits), at $4096$ shots per
circuit.\footnote{The \code{cz}-basis $C_4$ runs were taken on
2026-04-29; the fractional runs on 2026-06-12 (the $t = \pi/2$ PST
circuits) and 2026-06-13 (uniform mixing and PST-via-mixing).} We ran each
$t = \pi/2$ circuit twice: once transpiled to the default basis
$\{\code{cz}, \code{sx}, \code{rz}, \code{x}\}$, and once with the
fractional gates \code{rx} and \code{rzz} enabled, as described in
\cref{sec:c4-transpile}. We
emphasize two particular outputs:
\begin{itemize}\setlength{\itemsep}{2pt}
\item the \textsl{in-subspace fraction}, the total proportion of shots that
  measured a single-excitation bit-string (one of
  $\ket{0001},\ket{0010},\ket{0100},\ket{1000}$); and 
\item the \textsl{target probability} $P(\ket{e_{11}}) = P(\ket{0100})$, the
  fraction of shots that gave the antipodal-vertex PST outcome.
\end{itemize}

We report the output distribution in \Cref{tab:c4-hw}; recall that the  noiseless prediction 
is $1.000000$ on $\ket{0100}$
and $0.000000$ on every other string. 

\begin{table}[ht]
\centering
\small
\setlength{\tabcolsep}{4pt}
\begin{tabular}{lcrrrr}
\toprule
& & \multicolumn{2}{c}{\code{cz} basis} & \multicolumn{2}{c}{fractional basis} \\
\cmidrule(lr){3-4}\cmidrule(lr){5-6}
Bit-string & Vertex & {\footnotesize\code{UnitaryGate}} & {\footnotesize\code{PauliEvolutionGate} ($r{=}8$)} & {\footnotesize\code{UnitaryGate}} & {\footnotesize\code{PauliEvolutionGate} ($r{=}8$)} \\
\midrule
\texttt{0000} &       & $0.028076$ & $0.052002$ & $0.038330$ & $0.073730$ \\
\rowcolor{vcolour!8}\texttt{0001} & $00$  & $0.046631$ & $0.036377$ & $0.041992$ & $0.025879$ \\
\rowcolor{vcolour!8}\texttt{0010} & $01$  & $0.024414$ & $0.038818$ & $0.032227$ & $0.022461$ \\
\texttt{0011} &       & $0.056396$ & $0.028809$ & $0.043457$ & $0.022949$ \\
\rowcolor{kcolour!8}\texttt{0100} & $11$  & $0.320068$ & $0.532959$ & $0.382568$ & $0.420166$ \\
\texttt{0101} &       & $0.032959$ & $0.040771$ & $0.041504$ & $0.068604$ \\
\texttt{0110} &       & $0.031738$ & $0.050049$ & $0.033447$ & $0.052979$ \\
\texttt{0111} &       & $0.105225$ & $0.027100$ & $0.067139$ & $0.032959$ \\
\rowcolor{vcolour!8}\texttt{1000} & $10$  & $0.028809$ & $0.031982$ & $0.038818$ & $0.048584$ \\
\texttt{1001} &       & $0.029297$ & $0.030762$ & $0.034424$ & $0.051514$ \\
\texttt{1010} &       & $0.023438$ & $0.023193$ & $0.037354$ & $0.026855$ \\
\texttt{1011} &       & $0.025879$ & $0.010254$ & $0.026611$ & $0.015625$ \\
\texttt{1100} &       & $0.121826$ & $0.038086$ & $0.067871$ & $0.082275$ \\
\texttt{1101} &       & $0.060303$ & $0.018799$ & $0.042480$ & $0.017090$ \\
\texttt{1110} &       & $0.028809$ & $0.027588$ & $0.031738$ & $0.028320$ \\
\texttt{1111} &       & $0.036133$ & $0.012451$ & $0.040039$ & $0.010010$ \\
\midrule
\multicolumn{2}{l}{In-subspace sum} & $0.419922$ & $0.640625$ & $0.495605$ & $0.517090$ \\
\multicolumn{2}{l}{QPU usage (s)} & $2$ & $2$ & $3$ & $3$ \\
\bottomrule
\end{tabular}
\caption{$t = \pi/2$ hardware runs on \code{ibm\_kingston}, $4096$ shots
each, \code{UnitaryGate} and \code{PauliEvolutionGate} with $r = 8$, in
the default \code{cz} basis and in the fractional basis of
\cref{sec:c4-transpile}. The two bases were run on different days (the
fractional pair on 2026-06-12, immediately after a device maintenance
window). We also report the QPU usage: the quantum execution time, in seconds, billed by IBM Quantum's runtime usage API and rounded to the nearest integer. The states of the single-excitation subspaces have been highlighted, with the target state in its own colour.}
\label{tab:c4-hw}
\end{table}

We make two observations about \cref{tab:c4-hw}. Within each basis,
\code{PauliEvolutionGate} beats \code{UnitaryGate}, as the gate counts
of \cref{tab:c4-transpile} predict. Between the bases, the fractional
runs did \textsl{not} improve the Trotter circuit ($P(\ket{0100})$
dropped from $0.533$ to $0.420$), while \code{UnitaryGate} moved the
other way ($0.320$ to $0.383$) despite its larger fractional-basis gate
count. The two bases were however measured in different calibration
epochs, and run-to-run drift on these devices is of the same order as
these differences, so we read \cref{tab:c4-hw} as evidence that the
fractional basis is at best comparable to the default basis on this
circuit, not as a controlled per-gate comparison; the family-wide
sweeps of \cref{sec:hamming-paths,sec:xn-peakST} probe the question at
scale.

\subsection{Uniform mixing at \texorpdfstring{$t = \pi/4$}{t = pi/4}}\label{sec:c4-mixing}

\begin{figure}[ht]
  \centering
  \includegraphics[width=0.6\linewidth]{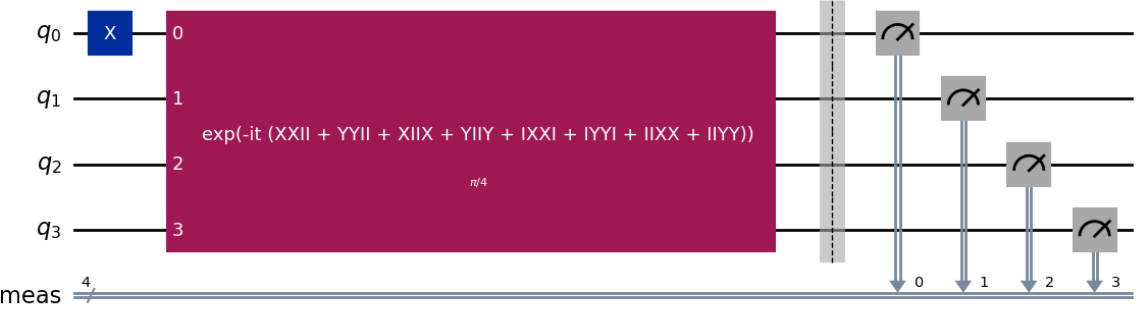}
  \caption{Logical circuit for $V_{\mathrm{exp}}(\pi/4) = e^{-i\pi H/4}\cdot X_{00}$,
  the uniform-mixing circuit on $C_4$. Identical to \cref{fig:c4-logical}
  except for the evolution time, which is $\pi/4$ rather than $\pi/2$.}
  \label{fig:c4-logical-pi4}
\end{figure}

Now that we have explained our workflow, we can run the same Hamiltonian $H$ from \eqref{eq:c4-H} but at $t = \pi/4$, when the quantum walk on $C_4$ admits uniform mixing.
The unitary we apply is $V_{\mathrm{exp}}(\pi/4) =
e^{-i\pi H/4} \cdot X_{00}$, see \cref{fig:c4-logical-pi4}. 

As with the PST computation, we used both options to encode the transition matrix.
For \code{PauliEvolutionGate} we use $r = 4$ Trotter reps.
The expected output is therefore $P(\ket{0001}) = P(\ket{0010}) =
P(\ket{0100}) = P(\ket{1000}) = 0.250000$ with all other twelve bit-strings
at $0.000000$. We report the Aer simulation of the transpiled circuits and the results from the IBM hardware in \cref{tab:c4-mixing}.

\begin{table}[p]
\centering
\footnotesize
\setlength{\tabcolsep}{3.5pt}
\begin{tabular}{lcrrrrrr}
\toprule
& & \multicolumn{2}{c}{Aer} & \multicolumn{2}{c}{Hardware (\code{cz})} & \multicolumn{2}{c}{Hardware (frac.)} \\
\cmidrule(lr){3-4}\cmidrule(lr){5-6}\cmidrule(lr){7-8}
Bit-string & Vertex & exact & Trotter & exact & Trotter & exact & Trotter \\
\midrule
\texttt{0000} &  & $0.000000$ & $0.000000$ & $0.039795$ & $0.042725$ & $0.041748$ & $0.036865$ \\
\rowcolor{vcolour!8} \texttt{0001} & $00$ & $0.250000$ & $0.254318$ & $0.078369$ & $0.165283$ & $0.126709$ & $0.215332$ \\
\rowcolor{vcolour!8} \texttt{0010} & $01$ & $0.250000$ & $0.246388$ & $0.070801$ & $0.203125$ & $0.215820$ & $0.181641$ \\
\texttt{0011} &  & $0.000000$ & $0.000000$ & $0.038574$ & $0.023682$ & $0.028076$ & $0.019043$ \\
\rowcolor{vcolour!8} \texttt{0100} & $11$ & $0.250000$ & $0.248103$ & $0.174072$ & $0.207764$ & $0.110596$ & $0.210449$ \\
\texttt{0101} &  & $0.000000$ & $0.000000$ & $0.023926$ & $0.020020$ & $0.029541$ & $0.012451$ \\
\texttt{0110} &  & $0.000000$ & $0.000000$ & $0.048340$ & $0.024658$ & $0.035156$ & $0.025391$ \\
\texttt{0111} &  & $0.000000$ & $0.000000$ & $0.045410$ & $0.015869$ & $0.031494$ & $0.006348$ \\
\rowcolor{vcolour!8} \texttt{1000} & $10$ & $0.250000$ & $0.251191$ & $0.170898$ & $0.179688$ & $0.071533$ & $0.194092$ \\
\texttt{1001} &  & $0.000000$ & $0.000000$ & $0.067139$ & $0.026367$ & $0.064697$ & $0.020508$ \\
\texttt{1010} &  & $0.000000$ & $0.000000$ & $0.026855$ & $0.023682$ & $0.045410$ & $0.031006$ \\
\texttt{1011} &  & $0.000000$ & $0.000000$ & $0.034424$ & $0.011475$ & $0.033203$ & $0.008545$ \\
\texttt{1100} &  & $0.000000$ & $0.000000$ & $0.048828$ & $0.020752$ & $0.070801$ & $0.019775$ \\
\texttt{1101} &  & $0.000000$ & $0.000000$ & $0.030029$ & $0.013428$ & $0.027344$ & $0.007812$ \\
\texttt{1110} &  & $0.000000$ & $0.000000$ & $0.056152$ & $0.012451$ & $0.033691$ & $0.006836$ \\
\texttt{1111} &  & $0.000000$ & $0.000000$ & $0.046387$ & $0.009033$ & $0.034180$ & $0.003906$ \\
\midrule
\multicolumn{2}{l}{In-subspace sum} & $1.000000$ & $1.000000$ & $0.494141$ & $0.755859$ & $0.524658$ & $0.801514$ \\
\multicolumn{2}{l}{QPU usage (s)} & n/a & n/a & $2$ & $2$ & $3$ & $3$ \\
\bottomrule
\end{tabular}
\caption{Output probabilities for the $t = \pi/4$ uniform-mixing circuits
on $C_4$, in the default \code{cz} basis and in the fractional basis
(\cref{sec:c4-transpile}). Columns labelled \emph{exact} use
\code{UnitaryGate}; \emph{Trotter} uses \code{PauliEvolutionGate} at
$r = 4$. Shaded rows are the single-excitation subspace. The Aer
distributions are identical, to all six decimal places, for the two
bases, so one Aer pair serves both. QPU usage is the IBM-billed
\code{job.usage()} time.}
\label{tab:c4-mixing}
\end{table}

\subsection{Perfect state transfer via uniform mixing}\label{sec:c4-pst-via-mixing}

\begin{figure}[htb]
  \centering
  \includegraphics[width=0.6\linewidth]{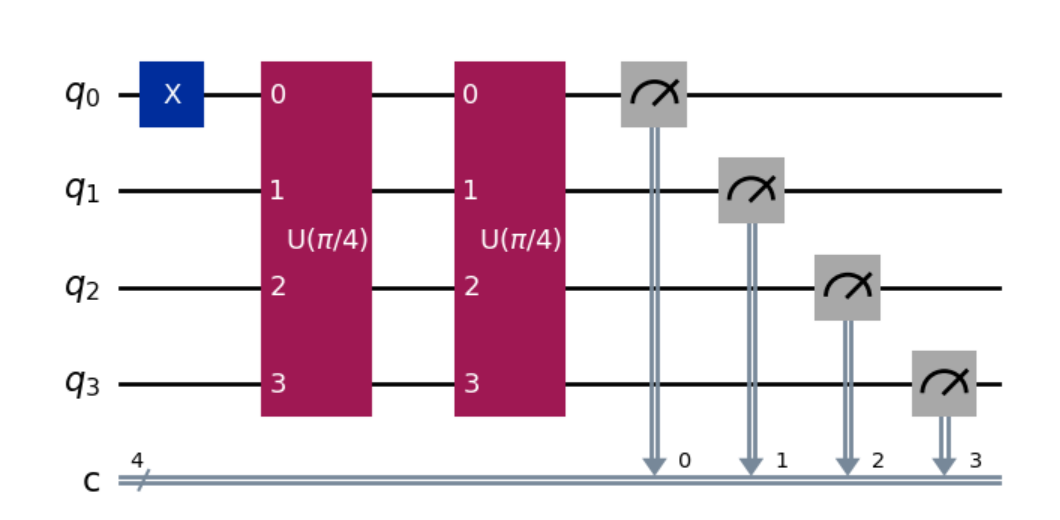}
  \caption{Logical circuit for $V_{\mathrm{exp}}^{(2)} = U(\pi/4) \cdot
  U(\pi/4) \cdot X_{00}$: PST on $C_4$ reached as the composition of two
  uniform-mixing blocks. Each $U(\pi/4)$ block is the same as in
  \cref{fig:c4-logical-pi4}.}
  \label{fig:c4-logical-pst-via-um}
\end{figure}

Since $U(\pi/4)^2 = U(\pi/2)$, applying two consecutive blocks of
$e^{-i\pi H/4}$ recovers the PST evolution from \Cref{sec:c4-math}. We
therefore consider the circuit
\[
  V_{\mathrm{exp}}^{(2)} \;=\; U(\pi/4) \cdot U(\pi/4) \cdot X_{00},
\]
which, noiselessly, produces $\ket{0100}$ with probability $1$, see \Cref{fig:c4-logical-pst-via-um}. 
The point of running this is to
compare against the direct $\pi/2$ block at fixed Hamiltonian. We report the output distributions of the simulations and hardware in \cref{tab:c4-twice}. 

\begin{table}[p]
\centering
\footnotesize
\setlength{\tabcolsep}{3.5pt}
\begin{tabular}{lcrrrrrr}
\toprule
& & \multicolumn{2}{c}{Aer} & \multicolumn{2}{c}{Hardware (\code{cz})} & \multicolumn{2}{c}{Hardware (frac.)} \\
\cmidrule(lr){3-4}\cmidrule(lr){5-6}\cmidrule(lr){7-8}
Bit-string & Vertex & exact & Trotter & exact & Trotter & exact & Trotter \\
\midrule
\texttt{0000} &  & $0.000000$ & $0.000000$ & $0.055664$ & $0.047363$ & $0.039062$ & $0.129639$ \\
\rowcolor{vcolour!8} \texttt{0001} & $00$ & $0.000000$ & $0.000023$ & $0.035645$ & $0.029297$ & $0.050537$ & $0.022705$ \\
\rowcolor{vcolour!8} \texttt{0010} & $01$ & $0.000000$ & $0.000014$ & $0.093506$ & $0.037109$ & $0.043945$ & $0.034424$ \\
\texttt{0011} &  & $0.000000$ & $0.000000$ & $0.059570$ & $0.028076$ & $0.059326$ & $0.042236$ \\
\rowcolor{kcolour!8} \texttt{0100} & $11$ & $1.000000$ & $0.999948$ & $0.129395$ & $0.571289$ & $0.153076$ & $0.360596$ \\
\texttt{0101} &  & $0.000000$ & $0.000000$ & $0.038330$ & $0.032959$ & $0.055176$ & $0.026123$ \\
\texttt{0110} &  & $0.000000$ & $0.000000$ & $0.045654$ & $0.043457$ & $0.053223$ & $0.067627$ \\
\texttt{0111} &  & $0.000000$ & $0.000000$ & $0.053711$ & $0.022949$ & $0.040527$ & $0.018311$ \\
\rowcolor{vcolour!8} \texttt{1000} & $10$ & $0.000000$ & $0.000014$ & $0.052979$ & $0.031982$ & $0.059326$ & $0.053955$ \\
\texttt{1001} &  & $0.000000$ & $0.000000$ & $0.036865$ & $0.031494$ & $0.055420$ & $0.032227$ \\
\texttt{1010} &  & $0.000000$ & $0.000000$ & $0.042236$ & $0.020264$ & $0.056396$ & $0.034424$ \\
\texttt{1011} &  & $0.000000$ & $0.000000$ & $0.079346$ & $0.012207$ & $0.043457$ & $0.011475$ \\
\texttt{1100} &  & $0.000000$ & $0.000000$ & $0.116699$ & $0.040039$ & $0.133301$ & $0.102783$ \\
\texttt{1101} &  & $0.000000$ & $0.000000$ & $0.046631$ & $0.018311$ & $0.043701$ & $0.020996$ \\
\texttt{1110} &  & $0.000000$ & $0.000000$ & $0.041016$ & $0.019775$ & $0.041748$ & $0.031738$ \\
\texttt{1111} &  & $0.000000$ & $0.000000$ & $0.072754$ & $0.013428$ & $0.071777$ & $0.010742$ \\
\midrule
\multicolumn{2}{l}{In-subspace sum} & $1.000000$ & $1.000000$ & $0.311523$ & $0.669678$ & $0.306885$ & $0.471680$ \\
\multicolumn{2}{l}{QPU usage (s)} & n/a & n/a & $3$ & $2$ & $3$ & $3$ \\
\bottomrule
\end{tabular}
\caption{Output probabilities for the $(\pi/4)\times 2$ PST-via-mixing
circuits on $C_4$, in the default \code{cz} basis and in the fractional
basis (\cref{sec:c4-transpile}). Columns labelled \emph{exact} use
\code{UnitaryGate}; \emph{Trotter} uses \code{PauliEvolutionGate} at
$r = 4$ per block. The Aer distributions are identical, to all six
decimal places, for the two bases. Shaded rows are the
single-excitation subspace; the row of PST target
$\ket{e_{11}} = \ket{0100}$ has its own colour.}
\label{tab:c4-twice}
\end{table}

\Cref{tab:c4-summary} summarizes the $t = \pi/2$ hardware target
probabilities across the four ways of reaching it (direct or via
uniform mixing; encoded with \code{UnitaryGate} or
\code{PauliEvolutionGate}), together with the fractional-basis runs of
the two single-block circuits.

\begin{table}[H]
\centering
\small
\begin{tabular}{llrr}
\toprule
Method                                    & basis & 2q gates & $P(\ket{0100})$ \\
\midrule
\code{PauliEvolutionGate} (Trotter, $r{=}4$ each, twice)   & \code{cz}  & 133 & $0.571289$ \\
\code{PauliEvolutionGate} (Trotter, $r{=}8$, one block)    & \code{cz}  & 133 & $0.532959$ \\
\code{PauliEvolutionGate} (Trotter, $r{=}8$, one block)    & fractional & 133 & $0.420166$ \\
\code{UnitaryGate} (exact, one block)               & fractional & 211 & $0.382568$ \\
\code{UnitaryGate} (exact, one block)               & \code{cz}  & 187 & $0.320068$ \\
\code{UnitaryGate} (exact, twice)                   & \code{cz}  & 368 & $0.129395$ \\
\bottomrule
\end{tabular}
\caption{Summary for PST on $C_4$ on \code{ibm\_kingston}, sorted by
$P(\ket{0100})$. The \code{cz}-basis and fractional-basis runs were
taken on different days.}
\label{tab:c4-summary}
\end{table}

\section{Perfect state transfer on the 3-cube}\label{sec:q3-full}

We now run the same family of experiments on the 3-cube $Q_3$, on 8 qubits. Here, 
\code{UnitaryGate} results in $29{,}655$ two-qubit (\code{cz}) gates, and thus we present only
\code{PauliEvolutionGate} hardware results.
We follow the methodology established in \cref{sec:c4-full}; explanations of
encoding, \code{UnitaryGate} and \code{PauliEvolutionGate}, and transpilation are not repeated here.

Here the Hamiltonian is given by 
\[
H = \frac{1}{2} \sum_{uv \in E(Q_3)} (X_u X_v + Y_u Y_v).
\]
We assign qubit $q_k$ to vertex $k \in \{0,\ldots,7\}$ and define the
single-excitation states $\ket{e_k} = X_{q_k}\ket{0\cdots 0}$. The cube also admits
uniform mixing at time $\pi/4$ and PST between antipodal vertices at time $\pi/2$. Again,
in Qiskit's little-endian convention, the rightmost bit corresponds to $q_0$, so the
single-excitation state $\ket{e_k}$ is the bit-string $\ket{2^k}$ in binary,
e.g.\ $\ket{e_0} = \ket{00000001}$, $\ket{e_7} = \ket{10000000}$. The vertices have been
ordered so that $0$ and $7$ are antipodal vertices.

\begin{figure}[ht]
  \centering
  \begin{tikzpicture}[scale=0.85, every label/.style={font=\footnotesize, label distance=4pt}]
    \node[new, label={above left:{$000\;(q_0)$}}] (v0) at (-3, 3) {};
    \node[new, label={above right:{$001\;(q_1)$}}] (v1) at (3, 3) {};
    \node[new, label={below right:{$011\;(q_3)$}}] (v3) at (3, -3) {};
    \node[new, label={below left:{$010\;(q_2)$}}] (v2) at (-3, -3) {};
    \node[new, label={above:{$100\;(q_4)$}}] (v4) at (-1, 1) {};
    \node[new, label={above:{$101\;(q_5)$}}] (v5) at (1, 1) {};
    \node[new, label={below:{$111\;(q_7)$}}] (v7) at (1, -1) {};
    \node[new, label={below:{$110\;(q_6)$}}] (v6) at (-1, -1) {};
    \draw (v0) -- (v1) -- (v3) -- (v2) -- (v0);
    \draw (v4) -- (v5) -- (v7) -- (v6) -- (v4);
    \draw (v0) -- (v4);
    \draw (v1) -- (v5);
    \draw (v3) -- (v7);
    \draw (v2) -- (v6);
  \end{tikzpicture}
  \caption{The 3-cube $Q_3$, drawn as nested squares: outer vertices
  have leading bit $0$, inner vertices have leading bit $1$, and the
  four radial edges flip that bit. Vertices are labelled by their 3-bit
  binary string, with the integer index $k \in \{0,\ldots,7\}$ equal
  to that binary value. We assign qubit $q_k$ to vertex $k$, so the
  single-excitation state at vertex $k$ is
  $\ket{e_k} = X_{q_k}\ket{0\cdots 0}$, e.g.\
  $\ket{e_0} = \ket{00000001}$ and $\ket{e_7} = \ket{10000000}$ in
  Qiskit's little-endian display. Vertices $0$ and $7$ are antipodal,
  differing in all three bits.}
  \label{fig:q3-graph}
\end{figure}
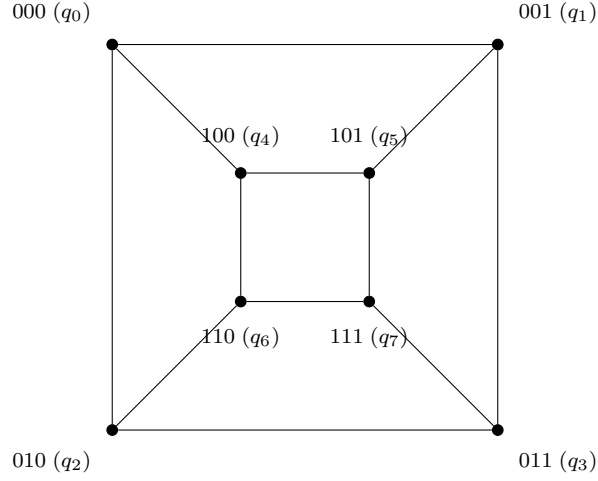

For the transpilation, \Cref{tab:q3-transpile} reports gate counts for both options at the three
times we consider. The ``logical'' columns transpile to
$\{\code{cz}, \code{sx}, \code{rz}, \code{x}\}$ with no coupling map; the
``device'' columns transpile against the live \code{ibm\_kingston}
backend, in the default \code{cz} basis and with fractional gates
enabled (\cref{sec:c4-transpile}). Unlike the weighted paths of
\cref{sec:hamming-paths,sec:xn-peakST}, the cube does not embed in the
heavy-hex coupling map, and the routing overhead stays in full-angle
gates: at $r = 8$, of the $822$ two-qubit gates in the fractional
basis, $363$ become \code{rzz} rotations and $459$ remain \code{cz}.

\begin{table}[ht]
\centering
\small
\setlength{\tabcolsep}{3.5pt}
\begin{tabular}{lrrrrrrrrr}
\toprule
& \multicolumn{3}{c}{Logical (no routing)} & \multicolumn{3}{c}{Device (\code{cz})} & \multicolumn{3}{c}{Device (fractional)} \\
\cmidrule(lr){2-4}\cmidrule(lr){5-7}\cmidrule(lr){8-10}
Configuration                                   & depth & total & 2q & depth & total & 2q & depth & total & 2q \\
\midrule
$t = \pi/4$, \code{UnitaryGate}                           & 139{,}757 & 235{,}440 & $\mathbf{29{,}655}$ & n/a & n/a & n/a & n/a & n/a & n/a \\
$t = \pi/4$, \code{PauliEvolutionGate} ($r=4$)                   & 364 & 1{,}390 & 178 & 780 & 1{,}827 & $\mathbf{356}$ & 697 & 1{,}706 & $\mathbf{356}$ \\
\addlinespace
$t = \pi/2$, \code{UnitaryGate}                           & 139{,}770 & 235{,}458 & $\mathbf{29{,}655}$ & n/a & n/a & n/a & n/a & n/a & n/a \\
$t = \pi/2$, \code{PauliEvolutionGate} ($r=8$)                   & 708 & 2{,}726 & 354 & 1{,}681 & 4{,}200 & $\mathbf{822}$ & 1{,}311 & 3{,}247 & $\mathbf{822}$ \\
\addlinespace
$(\pi/4)$ twice, \code{UnitaryGate}                       & 279{,}497 & 470{,}835 & $\mathbf{59{,}308}$ & n/a & n/a & n/a & n/a & n/a & n/a \\
$(\pi/4)$ twice, \code{PauliEvolutionGate} ($r=4$ each)          & 708 & 2{,}726 & 354 & 1{,}681 & 4{,}200 & $\mathbf{822}$ & 1{,}311 & 3{,}247 & $\mathbf{822}$ \\
\bottomrule
\end{tabular}
\caption{Transpiled gate counts for the $Q_3$ PST circuits. Device
columns are compiled against the live \code{ibm\_kingston} backend
(\code{optimization\_level=3}, \code{seed\_transpiler=0}, calibration of
2026-06-12), in the default \code{cz} basis and with fractional gates
enabled (including the \code{FoldRzzAngle} post-pass); numbers exclude
the final measurements and the \code{measure\_all} barrier.
\code{UnitaryGate} device columns are empty: at $\sim 30{,}000$
two-qubit gates the routed circuit is impractical to run.}
\label{tab:q3-transpile}
\end{table}

As we can see in \cref{tab:q3-transpile}, encoding with \code{UnitaryGate} produces too many gates for a feasible run. Thus, we will continue only with 
\code{PauliEvolutionGate}, for PST and for uniform mixing on the cube. We skip the third computation, PST via uniform mixing, since we cannot implement it with \code{UnitaryGate} and \code{PauliEvolutionGate} will return the same number of gates when applying $e^{-i\frac{\pi}{4} H}$ twice as opposed to $e^{-i\frac{\pi}{2} H}$ once. 

We run two circuits. For perfect state transfer at $t = \pi/2$, the
circuit is $V_{\mathrm{exp}}(\pi/2) = e^{-i\frac{\pi}{2} H} X_{0}$, and the noiseless
target is $\ket{e_7} = \ket{10000000}$ with probability $1$. For uniform
mixing at $t = \pi/4$, the circuit is
$V_{\mathrm{exp}}(\pi/4) = e^{-i\frac{\pi}{4} H} X_{0}$; without noise,
and up to Trotter error, we expect a uniform distribution over all eight
single-excitation states.

The merged \cref{tab:q3-hw} reports the hardware results for both $Q_3$
experiments side-by-side, each run in the default \code{cz} basis and in
the fractional basis (\cref{sec:c4-transpile}).\footnote{The
\code{cz}-basis $Q_3$ runs were taken on 2026-05-20 and the fractional
runs on 2026-06-13.} As in
\cref{sec:c4-aer}, the Aer column is the exact noiseless distribution of
the device-transpiled circuit; for $Q_3$ the Trotter residual is visible
at six decimal places ($P(\ket{e_7}) = 0.998874$ at $r = 8$ rather than
$1$, and the mixing distribution is only approximately uniform), and it
is identical for the two bases, so a single Aer column serves both.  For
the PST run at $t = \pi/2$, the single-excitation in-subspace fraction is
$0.134$ in the \code{cz} basis and $\ket{e_7}$ is the largest peak ($0.0251$ vs.\ the
next-largest at $0.0229$). The fractional basis raises the target peak to
$0.0432$ and the in-subspace fraction to $0.139$, the same direction of
improvement we will see for the weighted-path families in
\cref{sec:hamming-paths,sec:xn-peakST}, although $Q_3$ in the
full vertex encoding does not embed in the coupling map, so only $363$ of
its $822$ two-qubit gates become \code{rzz} rotations
(\cref{tab:q3-transpile}) and the gain is correspondingly smaller. The
uniform-mixing run is essentially unchanged between the bases (in-subspace
$0.339$ vs.\ $0.320$), as expected when there is no single target peak to
sharpen.

\begin{table}[htb]
\centering
\footnotesize
\setlength{\tabcolsep}{3.5pt}
\begin{tabular}{lcrrrrrr}
\toprule
& & \multicolumn{3}{c}{PST at $t=\pi/2$ {\tiny ($r=8$)}} & \multicolumn{3}{c}{Mixing at $t=\pi/4$ {\tiny ($r=4$)}} \\
\cmidrule(lr){3-5}\cmidrule(lr){6-8}
Vertex & Bit-string & Aer & HW (\code{cz}) & HW (frac.) & Aer & HW (\code{cz}) & HW (frac.) \\
\midrule
$0$ & \texttt{00000001} & $0.000501$ & $0.014893$ & $0.008545$ & $0.132398$ & $0.052734$ & $0.037598$ \\
$1$ & \texttt{00000010} & $0.000087$ & $0.010254$ & $0.016846$ & $0.127879$ & $0.047119$ & $0.027100$ \\
$2$ & \texttt{00000100} & $0.000094$ & $0.015625$ & $0.010254$ & $0.127797$ & $0.048096$ & $0.038818$ \\
$3$ & \texttt{00001000} & $0.000229$ & $0.015869$ & $0.014160$ & $0.119384$ & $0.036621$ & $0.036377$ \\
$4$ & \texttt{00010000} & $0.000099$ & $0.022949$ & $0.011230$ & $0.128377$ & $0.048096$ & $0.060059$ \\
$5$ & \texttt{00100000} & $0.000058$ & $0.014160$ & $0.017334$ & $0.119813$ & $0.036621$ & $0.041504$ \\
$6$ & \texttt{01000000} & $0.000058$ & $0.014648$ & $0.017822$ & $0.117210$ & $0.033691$ & $0.037109$ \\
\cellcolor{kcolour!8} $7$ & \cellcolor{kcolour!8} \texttt{10000000} & \cellcolor{kcolour!8} $\mathbf{0.998874}$ & \cellcolor{kcolour!8} $\mathbf{0.025146}$ & \cellcolor{kcolour!8} $\mathbf{0.043213}$ & $0.127143$ & $0.035889$ & $0.041260$ \\
\midrule
\multicolumn{2}{l}{In-subspace sum} & $1.000000$ & $0.133545$ & $0.139404$ & $1.000000$ & $0.338867$ & $0.319824$ \\
\multicolumn{2}{l}{QPU usage (s)} & n/a & $3$ & $3$ & n/a & $3$ & $3$ \\
\bottomrule
\end{tabular}
\caption{$Q_3$ hardware results for the two time settings, side by side.
The \emph{Aer} column in each group is the exact noiseless distribution
of the device-transpiled circuit, identical to all six decimal places
for the \code{cz}-basis and fractional-basis transpilations of
\cref{sec:c4-transpile}. \emph{HW (\code{cz})} and \emph{HW (frac.)} are
$4096$-shot runs on \code{ibm\_kingston} with \code{PauliEvolutionGate}
in the default \code{cz} basis and with fractional gates enabled. The
shaded cells are vertex $7$, the antipodal PST target at $t = \pi/2$. For
the uniform-mixing columns, vertex $7$ has no special status; its
hardware values are among eight nearly-equal entries, all attenuated to
roughly $1/3$ of the noiseless $\approx 0.125$.}
\label{tab:q3-hw}
\end{table}

\section{Perfect state transfer on hypercubes via Hamming-weighted paths}\label{sec:hamming-paths}

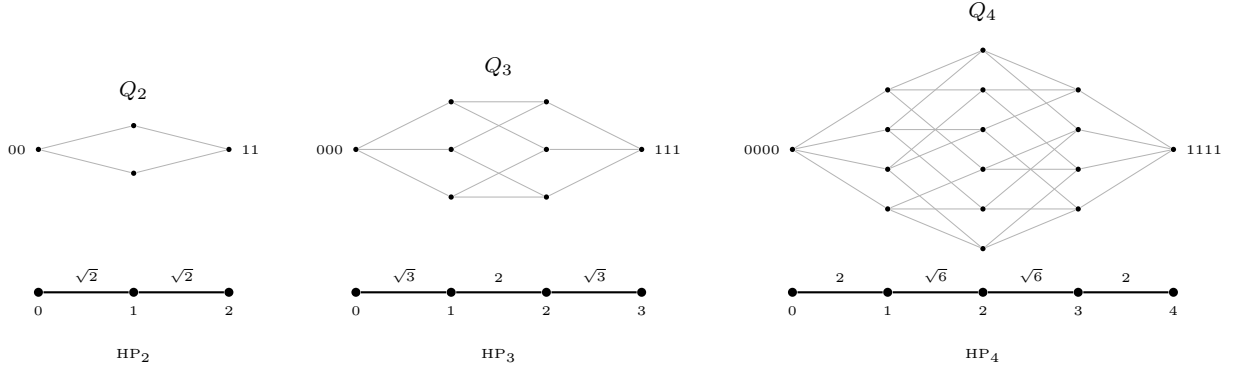
\begin{figure}[ht]
\centering
\begin{tikzpicture}[
  cv/.style={circle, fill=black, inner sep=0.7pt},
  hv/.style={circle, fill=black, inner sep=1.2pt},
  ce/.style={very thin, gray!60},
  he/.style={thick},
  scale=1.05
]

\begin{scope}[xshift=0cm]
  \node[font=\footnotesize, anchor=south] at (1.2, 0.5) {$Q_2$};
  \node[cv, label={left:{\tiny $00$}}] (a00) at (0, 0) {};
  \node[cv] (a01) at (1.2, 0.3) {};
  \node[cv] (a10) at (1.2, -0.3) {};
  \node[cv, label={right:{\tiny $11$}}] (a11) at (2.4, 0) {};
  \draw[ce] (a00) -- (a01); \draw[ce] (a00) -- (a10);
  \draw[ce] (a01) -- (a11); \draw[ce] (a10) -- (a11);
  \node[hv, label={below:{\tiny $0$}}] (b0) at (0, -1.8) {};
  \node[hv, label={below:{\tiny $1$}}] (b1) at (1.2, -1.8) {};
  \node[hv, label={below:{\tiny $2$}}] (b2) at (2.4, -1.8) {};
  \draw[he] (b0) -- node[above, font=\tiny] {$\sqrt{2}$} (b1);
  \draw[he] (b1) -- node[above, font=\tiny] {$\sqrt{2}$} (b2);
  \node[font=\tiny, anchor=north] at (1.2, -2.4) {$\mathrm{HP}_2$};
\end{scope}

\begin{scope}[xshift=4cm]
  \node[font=\footnotesize, anchor=south] at (1.8, 0.8) {$Q_3$};
  \node[cv, label={left:{\tiny $000$}}] (b000) at (0, 0) {};
  \node[cv] (b001) at (1.2, 0.6) {};
  \node[cv] (b010) at (1.2, 0) {};
  \node[cv] (b100) at (1.2, -0.6) {};
  \node[cv] (b011) at (2.4, 0.6) {};
  \node[cv] (b101) at (2.4, 0) {};
  \node[cv] (b110) at (2.4, -0.6) {};
  \node[cv, label={right:{\tiny $111$}}] (b111) at (3.6, 0) {};
  \draw[ce] (b000) -- (b001); \draw[ce] (b000) -- (b010); \draw[ce] (b000) -- (b100);
  \draw[ce] (b001) -- (b011); \draw[ce] (b001) -- (b101);
  \draw[ce] (b010) -- (b011); \draw[ce] (b010) -- (b110);
  \draw[ce] (b100) -- (b101); \draw[ce] (b100) -- (b110);
  \draw[ce] (b011) -- (b111); \draw[ce] (b101) -- (b111); \draw[ce] (b110) -- (b111);
  \node[hv, label={below:{\tiny $0$}}] (c0) at (0, -1.8) {};
  \node[hv, label={below:{\tiny $1$}}] (c1) at (1.2, -1.8) {};
  \node[hv, label={below:{\tiny $2$}}] (c2) at (2.4, -1.8) {};
  \node[hv, label={below:{\tiny $3$}}] (c3) at (3.6, -1.8) {};
  \draw[he] (c0) -- node[above, font=\tiny] {$\sqrt{3}$} (c1);
  \draw[he] (c1) -- node[above, font=\tiny] {$2$} (c2);
  \draw[he] (c2) -- node[above, font=\tiny] {$\sqrt{3}$} (c3);
  \node[font=\tiny, anchor=north] at (1.8, -2.4) {$\mathrm{HP}_3$};
\end{scope}

\begin{scope}[xshift=9.5cm]
  \node[font=\footnotesize, anchor=south] at (2.4, 1.5) {$Q_4$};
  \node[cv, label={left:{\tiny $0000$}}] (c0000) at (0, 0) {};
  \node[cv] (c0001) at (1.2, 0.75) {};
  \node[cv] (c0010) at (1.2, 0.25) {};
  \node[cv] (c0100) at (1.2, -0.25) {};
  \node[cv] (c1000) at (1.2, -0.75) {};
  \node[cv] (c0011) at (2.4, 1.25) {};
  \node[cv] (c0101) at (2.4, 0.75) {};
  \node[cv] (c0110) at (2.4, 0.25) {};
  \node[cv] (c1001) at (2.4, -0.25) {};
  \node[cv] (c1010) at (2.4, -0.75) {};
  \node[cv] (c1100) at (2.4, -1.25) {};
  \node[cv] (c0111) at (3.6, 0.75) {};
  \node[cv] (c1011) at (3.6, 0.25) {};
  \node[cv] (c1101) at (3.6, -0.25) {};
  \node[cv] (c1110) at (3.6, -0.75) {};
  \node[cv, label={right:{\tiny $1111$}}] (c1111) at (4.8, 0) {};
  \draw[ce] (c0000) -- (c0001); \draw[ce] (c0000) -- (c0010);
  \draw[ce] (c0000) -- (c0100); \draw[ce] (c0000) -- (c1000);
  \draw[ce] (c0001) -- (c0011); \draw[ce] (c0001) -- (c0101); \draw[ce] (c0001) -- (c1001);
  \draw[ce] (c0010) -- (c0011); \draw[ce] (c0010) -- (c0110); \draw[ce] (c0010) -- (c1010);
  \draw[ce] (c0100) -- (c0101); \draw[ce] (c0100) -- (c0110); \draw[ce] (c0100) -- (c1100);
  \draw[ce] (c1000) -- (c1001); \draw[ce] (c1000) -- (c1010); \draw[ce] (c1000) -- (c1100);
  \draw[ce] (c0011) -- (c0111); \draw[ce] (c0011) -- (c1011);
  \draw[ce] (c0101) -- (c0111); \draw[ce] (c0101) -- (c1101);
  \draw[ce] (c0110) -- (c0111); \draw[ce] (c0110) -- (c1110);
  \draw[ce] (c1001) -- (c1011); \draw[ce] (c1001) -- (c1101);
  \draw[ce] (c1010) -- (c1011); \draw[ce] (c1010) -- (c1110);
  \draw[ce] (c1100) -- (c1101); \draw[ce] (c1100) -- (c1110);
  \draw[ce] (c0111) -- (c1111); \draw[ce] (c1011) -- (c1111);
  \draw[ce] (c1101) -- (c1111); \draw[ce] (c1110) -- (c1111);
  \node[hv, label={below:{\tiny $0$}}] (d0) at (0, -1.8) {};
  \node[hv, label={below:{\tiny $1$}}] (d1) at (1.2, -1.8) {};
  \node[hv, label={below:{\tiny $2$}}] (d2) at (2.4, -1.8) {};
  \node[hv, label={below:{\tiny $3$}}] (d3) at (3.6, -1.8) {};
  \node[hv, label={below:{\tiny $4$}}] (d4) at (4.8, -1.8) {};
  \draw[he] (d0) -- node[above, font=\tiny] {$2$} (d1);
  \draw[he] (d1) -- node[above, font=\tiny] {$\sqrt{6}$} (d2);
  \draw[he] (d2) -- node[above, font=\tiny] {$\sqrt{6}$} (d3);
  \draw[he] (d3) -- node[above, font=\tiny] {$2$} (d4);
  \node[font=\tiny, anchor=north] at (2.4, -2.4) {$\mathrm{HP}_4$};
\end{scope}

\end{tikzpicture}
\caption{The hypercubes $Q_2, Q_3, Q_4$ (upper, grey), arranged by
distance from the leftmost vertex, and their symmetrized quotients
$\mathrm{HP}_2, \mathrm{HP}_3, \mathrm{HP}_4$ (lower).}
\label{fig:hp-quotient}
\end{figure}

The hypercube $Q_n$ has $2^n$ vertices and diameter $n$. Starting from a vertex $u$, we can partition the vertices by their distance from $u$. This partition is an \textsl{equitable partition}; that is, every vertex at distance $i$ from $u$ has some constant $c_{ij}$ neighbours at distance $j$ from $u$. By Lemma~\ref{lem:quotient-transfer}, the symmetrized quotient of this partition is a weighted path with PST between its end vertices at time $\pi/2$.
\Cref{fig:hp-quotient} shows the construction at $n = 2, 3, 4$.

More precisely, for each integer $n \geq 1$, the \textsl{Hamming-weighted path}, denoted $\mathrm{HP}_n$, is the weighted path
on $n+1$ vertices $\{0, 1, \ldots, n\}$ with the edge $(k-1, k)$ having weight
\begin{equation}\label{eq:hamming-weights}
  w_k \;=\; \sqrt{k\,(n-k+1)}, \qquad k = 1, \ldots, n.
\end{equation}
The adjacency matrix $A(\mathrm{HP}_n)$ is the symmetrized quotient of
$A(Q_n)$ with respect to the distance partition from $u$
(\cref{sec:pst-peakST}). Its eigenvalues are $n - 2k$ for
$k = 0, 1, \ldots, n$, each simple; these are the distinct eigenvalues of
$A(Q_n)$. Since $u$ and its antipodal vertex are singleton cells, the PST
between them in $Q_n$ at time $\pi/2$ gives PST from $0$ to $n$ in
$\mathrm{HP}_n$ at the same time, by Lemma~\ref{lem:quotient-transfer}.

\paragraph{Implementation.}
We use one qubit for each vertex of $\mathrm{HP}_n$, so $n+1$ qubits. The
Hamiltonian is
\[
  H_{\mathrm{HP}_n} \;=\; \tfrac12\sum_{k=1}^n w_k\,(X_{k-1}X_k + Y_{k-1}Y_k),
\]
which acts on the single-excitation subspace as $A(\mathrm{HP}_n)$
(\cref{sec:background}). \code{PauliEvolutionGate} accepts arbitrary real
coefficients, so the edge weights only change the rotation angles in the
circuit.

\subsection{Gate-count survey at \texorpdfstring{$t = \pi/2$}{t = pi/2}}\label{sec:hamming-survey}

\Cref{tab:hamming-survey} gives the gate counts of the PST circuit on
$\mathrm{HP}_n$ for $n = 2, \ldots, 30$, transpiled without a coupling map.

\begin{table}[ht]
\centering
\footnotesize
\begin{tabular}{rrrrrrrr}
\toprule
& & \multicolumn{3}{c}{\code{UnitaryGate}} & \multicolumn{3}{c}{\code{PauliEvolutionGate} ($r{=}8$)} \\
\cmidrule(lr){3-5}\cmidrule(lr){6-8}
$n$ & qubits & depth & total & $n_{2q}$ & depth & total & $n_{2q}$ \\
\midrule
2  & 3  & 29         & 46         & \textbf{7}       & 169   & 258   & \textbf{34} \\
3  & 4  & 433        & 723        & \textbf{91}      & 289   & 507   & \textbf{66} \\
4  & 5  & 1{,}079    & 1{,}812    & \textbf{230}     & 411   & 752   & \textbf{98} \\
5  & 6  & 8{,}389    & 14{,}186   & \textbf{1{,}767} & 562   & 991   & \textbf{130} \\
6  & 7  & 19{,}087   & 31{,}352   & \textbf{4{,}110} & 689   & 1{,}262 & \textbf{162} \\
7  & 8  & 139{,}444  & 234{,}938  & \textbf{29{,}573} & 863  & 1{,}549 & \textbf{194} \\
8  & 9  & 313{,}194  & 513{,}168  & \textbf{67{,}724} & 969  & 1{,}818 & \textbf{226} \\
9  & 10 & 2{,}254{,}563 & 3{,}794{,}371 & \textbf{479{,}063} & 1{,}137 & 2{,}039 & \textbf{258} \\
\addlinespace
10 & 11 & \multicolumn{3}{c}{\textit{QSD \texttt{NoConvergence}}}      & 1{,}224 & 2{,}248 & \textbf{290} \\
11 & 12 & \multicolumn{3}{c}{n/a} & 1{,}393 & 2{,}615 & \textbf{322} \\
12 & 13 & \multicolumn{3}{c}{n/a} & 1{,}490 & 2{,}748 & \textbf{354} \\
13 & 14 & \multicolumn{3}{c}{n/a} & 1{,}669 & 3{,}089 & \textbf{386} \\
14 & 15 & \multicolumn{3}{c}{n/a} & 1{,}748 & 3{,}214 & \textbf{418} \\
15 & 16 & \multicolumn{3}{c}{n/a} & 1{,}936 & 3{,}573 & \textbf{450} \\
16 & 17 & \multicolumn{3}{c}{n/a} & 2{,}032 & 3{,}722 & \textbf{482} \\
17 & 18 & \multicolumn{3}{c}{n/a} & 2{,}222 & 4{,}095 & \textbf{514} \\
18 & 19 & \multicolumn{3}{c}{n/a} & 2{,}285 & 4{,}302 & \textbf{546} \\
19 & 20 & \multicolumn{3}{c}{n/a} & 2{,}444 & 4{,}537 & \textbf{578} \\
20 & 21 & \multicolumn{3}{c}{n/a} & 2{,}586 & 4{,}760 & \textbf{610} \\
21 & 22 & \multicolumn{3}{c}{n/a} & 2{,}715 & 5{,}089 & \textbf{642} \\
22 & 23 & \multicolumn{3}{c}{n/a} & 2{,}814 & 5{,}276 & \textbf{674} \\
23 & 24 & \multicolumn{3}{c}{n/a} & 3{,}074 & 5{,}626 & \textbf{706} \\
24 & 25 & \multicolumn{3}{c}{n/a} & 3{,}079 & 5{,}808 & \textbf{738} \\
25 & 26 & \multicolumn{3}{c}{n/a} & 3{,}272 & 6{,}107 & \textbf{770} \\
26 & 27 & \multicolumn{3}{c}{n/a} & 3{,}357 & 6{,}226 & \textbf{802} \\
27 & 28 & \multicolumn{3}{c}{n/a} & 3{,}532 & 6{,}621 & \textbf{834} \\
28 & 29 & \multicolumn{3}{c}{n/a} & 3{,}641 & 6{,}844 & \textbf{866} \\
29 & 30 & \multicolumn{3}{c}{n/a} & 3{,}799 & 7{,}099 & \textbf{898} \\
30 & 31 & \multicolumn{3}{c}{n/a} & 3{,}890 & 7{,}312 & \textbf{930} \\
\bottomrule
\end{tabular}
\caption{Gate-count survey of PST circuits on $\mathrm{HP}_n$ at
$t = \pi/2$. Transpile to $\{\code{cz}, \code{sx}, \code{rz}, \code{x}\}$
at \code{optimization\_level=3}, no coupling map. \code{UnitaryGate}
synthesizes $e^{-i\pi H/2}$ via Quantum Shannon Decomposition (QSD).
\code{PauliEvolutionGate} applies second-order Suzuki--Trotter at $r = 8$
reps. At $n = 10$ Qiskit's QSD synthesis pass raises
\texttt{PanicException: NoConvergence} in the underlying Cosine-Sine
Decomposition step; this is a numerical failure of the algorithm at large
$n$, not a \code{UnitaryGate} circuit that costs ``too many gates'' to write down.
Beyond that, \code{UnitaryGate}'s gate counts are also growing as
$(23/48)\cdot 4^n$ (the QSD count), so the synthesis was infeasible 
in practice well before the numerical failure: by $n = 9$ a single PST
circuit costs nearly half a million two-qubit gates.}
\label{tab:hamming-survey}
\end{table}

\subsection{Choosing the Trotter rep count \texorpdfstring{$r$}{r}}\label{sec:hamming-rsweep}

So far we have used $r = 8$ at $t = \pi/2$. We now choose $r$ for each
$n$, balancing the Trotter error against the error from the two-qubit
gates (\cref{sec:implementation}).

Even without hardware noise, the Suzuki--Trotter circuit only
approximates $e^{-itH}$; for $r$ reps of the second-order formula, the
error is roughly
\[
  \varepsilon_{\mathrm{Trotter}} \;\sim\; \frac{t^3\,\lVert H\rVert^3}{r^2}.
\]
On the single-excitation subspace, $\lVert H\rVert = n$, so at fixed $r$
the approximation gets worse as $n$ grows.

For the hardware, we take the median \code{cz} error
$p_{2q} = 2.0 \times 10^{-3}$ of \code{ibm\_kingston} and assume that each
two-qubit gate fails independently with probability $p_{2q}$. The
probability that a circuit with $n_{2q}$ two-qubit gates has no
two-qubit gate error is then
\[
  S \;=\; (1 - p_{2q})^{n_{2q}}.
\]
We call $S$ the \textsl{survival fraction}. We ignore single-qubit gate
errors, which are an order of magnitude smaller
(\cref{sec:c4-transpile}). If the circuit runs without error with
probability $S$ and otherwise returns a uniformly random bit-string,
then, neglecting the contribution $(1 - S)\,2^{-(n+1)}$ of the random
bit-strings, the target probability is
\[
  \hat{P}_{\mathrm{HW}} \;=\; P_{\mathrm{noiseless}} \cdot S,
\]
where $P_{\mathrm{noiseless}}$ is the target probability of the Trotter
circuit without noise, computed with Aer. We call $\hat{P}_{\mathrm{HW}}$
the \textsl{predicted hardware target probability}. Increasing $r$
decreases the Trotter error and, since $n_{2q}$ grows linearly in $r$,
also decreases $S$. For each $n$, we will choose the $r$ that maximizes
$\hat{P}_{\mathrm{HW}}$. The model also ignores readout error and
decoherence, so $\hat{P}_{\mathrm{HW}}$ overestimates the measured target
probability; we use it only to choose $r$.

\Cref{tab:hamming-rsweep} gives $n_{2q}$, $P_{\mathrm{noiseless}}$, $S$
and $\hat{P}_{\mathrm{HW}}$ for $n \in \{4, 8, 12, 16, 20\}$ and
$r \in \{2, 4, 8, 16, 32\}$.

\begin{table}[ht]
\centering
\small
\begin{tabular}{rrrrrrr}
\toprule
$n$ & $r$ & qubits & $n_{2q}$ & $P_{\mathrm{noiseless}}$ & $S$ & $\hat{P}_{\mathrm{HW}}$ \\
\midrule
$ 4$ &  $2$ &  $5$ &    $26$ & $0.702241$ & $0.949279$ & $0.666623$ \\
$ 4$ &  $4$ &  $5$ &    $50$ & $0.967533$ & $0.904747$ & $\mathbf{0.875372}$ \\
$ 4$ &  $8$ &  $5$ &    $98$ & $0.997757$ & $0.821851$ & $0.820008$ \\
$ 4$ & $16$ &  $5$ &   $194$ & $0.999857$ & $0.678149$ & $0.678052$ \\
$ 4$ & $32$ &  $5$ &   $386$ & $0.999991$ & $0.461731$ & $0.461727$ \\
\addlinespace
$ 8$ &  $2$ &  $9$ &    $58$ & $0.022924$ & $0.890372$ & $0.020411$ \\
$ 8$ &  $4$ &  $9$ &   $114$ & $0.549182$ & $0.795943$ & $0.437117$ \\
$ 8$ &  $8$ &  $9$ &   $226$ & $0.958462$ & $0.636066$ & $\mathbf{0.609645}$ \\
$ 8$ & $16$ &  $9$ &   $450$ & $0.997203$ & $0.406203$ & $0.405067$ \\
$ 8$ & $32$ &  $9$ &   $898$ & $0.999822$ & $0.165663$ & $0.165634$ \\
\addlinespace
$12$ &  $2$ & $13$ &    $90$ & $0.006569$ & $0.835120$ & $0.005486$ \\
$12$ &  $4$ & $13$ &   $178$ & $0.006154$ & $0.700223$ & $0.004309$ \\
$12$ &  $8$ & $13$ &   $354$ & $0.717693$ & $0.492279$ & $\mathbf{0.353305}$ \\
$12$ & $16$ & $13$ &   $706$ & $0.978545$ & $0.243311$ & $0.238091$ \\
$12$ & $32$ & $13$ &  $1410$ & $0.998619$ & $0.059438$ & $0.059356$ \\
\addlinespace
$16$ &  $2$ & $17$ &    $98$ & $0.000000$ & $0.821851$ & $0.000000$ \\
$16$ &  $4$ & $17$ &   $242$ & $0.022496$ & $0.616015$ & $0.013858$ \\
$16$ &  $8$ & $17$ &   $482$ & $0.191375$ & $0.380996$ & $0.072913$ \\
$16$ & $16$ & $17$ &   $962$ & $0.906199$ & $0.145741$ & $\mathbf{0.132070}$ \\
$16$ & $32$ & $17$ &  $1922$ & $0.993717$ & $0.021326$ & $0.021192$ \\
\addlinespace
$20$ &  $2$ & $21$ &   $154$ & $0.000000$ & $0.734689$ & $0.000000$ \\
$20$ &  $4$ & $21$ &   $306$ & $0.047412$ & $0.541933$ & $0.025694$ \\
$20$ &  $8$ & $21$ &   $610$ & $0.011032$ & $0.294870$ & $0.003253$ \\
$20$ & $16$ & $21$ &  $1218$ & $0.724704$ & $0.087297$ & $\mathbf{0.063264}$ \\
$20$ & $32$ & $21$ &  $2434$ & $0.979660$ & $0.007651$ & $0.007496$ \\
\bottomrule
\end{tabular}
\caption{Trotter-rep sweep for the Hamming-weighted-path PST circuit at
$t = \pi/2$. For each $(n, r)$: $n_{2q}$ is the two-qubit gate count
after transpilation to \code{ibm\_kingston}'s native \code{cz} basis;
$P_{\mathrm{noiseless}}$ is the exact Aer prediction (no hardware noise,
only Trotter approximation); $S = (1 - 2.0\times 10^{-3})^{n_{2q}}$ is
the device survival fraction; $\hat{P}_{\mathrm{HW}} = P_{\mathrm{noiseless}}
\cdot S$ is the predicted hardware target probability. The
boldfaced entry in each $n$-block is the maximizer of $\hat{P}_{\mathrm{HW}}$
over $r$.}
\label{tab:hamming-rsweep}
\end{table}

We then computed $\hat{P}_{\mathrm{HW}}$ for
$r \in \{2, 3, 4, 5, 6, 7, 8, 10, 12, 16\}$ and $n = 2, \ldots, 20$.
\Cref{tab:hp-r-optimal} gives, for each $n$, the value $r^{*}$ of $r$
that maximizes $\hat{P}_{\mathrm{HW}}$. For $n = 13, 15, 17, 19$, $r^{*}$
is $3$ or $4$; at these values of $r$, $P_{\mathrm{noiseless}}$ is between
$0.36$ and $0.54$ and $n_{2q}$ is at most $290$. We will use $r^{*}$ for
the hardware runs in \cref{sec:hamming-hw}.

\begin{table}[ht]
\centering
\small
\begin{tabular}{rrrrrr}
\toprule
$n$ & qubits & $r^{*}$ & $n_{2q}$ & $P_{\mathrm{noiseless}}$ & $\hat{P}_{\mathrm{HW}}$ \\
\midrule
$ 2$ & $ 3$ & $ 3$ & $  14$ & $0.994421$ & $0.966936$ \\
$ 3$ & $ 4$ & $ 4$ & $  34$ & $0.988575$ & $0.923523$ \\
$ 4$ & $ 5$ & $ 4$ & $  50$ & $0.967533$ & $0.875372$ \\
$ 5$ & $ 6$ & $ 5$ & $  82$ & $0.967695$ & $0.821189$ \\
$ 6$ & $ 7$ & $ 5$ & $ 102$ & $0.933560$ & $0.761127$ \\
$ 7$ & $ 8$ & $ 6$ & $ 146$ & $0.933957$ & $0.697246$ \\
$ 8$ & $ 9$ & $ 7$ & $ 198$ & $0.931470$ & $0.626637$ \\
$ 9$ & $10$ & $ 7$ & $ 226$ & $0.881124$ & $0.560453$ \\
$10$ & $11$ & $ 8$ & $ 290$ & $0.879425$ & $0.492103$ \\
$11$ & $12$ & $ 8$ & $ 322$ & $0.810534$ & $0.425408$ \\
$12$ & $13$ & $10$ & $ 442$ & $0.871119$ & $0.359564$ \\
$13$ & $14$ & $ 3$ & $ 146$ & $0.535430$ & $0.399726$ \\
$14$ & $15$ & $10$ & $ 522$ & $0.732475$ & $0.257594$ \\
$15$ & $16$ & $ 4$ & $ 226$ & $0.362132$ & $0.230340$ \\
$16$ & $17$ & $12$ & $ 722$ & $0.735073$ & $0.173213$ \\
$17$ & $18$ & $ 4$ & $ 258$ & $0.424774$ & $0.253418$ \\
$18$ & $19$ & $12$ & $ 818$ & $0.559489$ & $0.108786$ \\
$19$ & $20$ & $ 4$ & $ 290$ & $0.402472$ & $0.225213$ \\
$20$ & $21$ & $ 5$ & $ 382$ & $0.162311$ & $0.075547$ \\
\bottomrule
\end{tabular}
\caption{The Trotter rep count $r^{*}$ maximizing $\hat{P}_{\mathrm{HW}}$, for each $n$, for
Hamming-weighted-path PST at $t = \pi/2$, over the discrete sweep
$r \in \{2, 3, 4, 5, 6, 7, 8, 10, 12, 16\}$. Survival is computed at
$p_{2q} = 2.0\times 10^{-3}$ (the \code{ibm\_kingston} median \code{cz} error), as in
\cref{tab:hamming-rsweep}. Several rows ($n = 13, 15, 17, 19$) win at
small $r$ via a Trotter-pocket coincidence; the constructive-interference
pocket is at lower $n_{2q}$ but the noiseless $P(\mathrm{target})$ is
correspondingly more sensitive to gate noise.}
\label{tab:hp-r-optimal}
\end{table}

\subsection{Hardware results}\label{sec:hamming-hw}

We ran the PST circuit on $\mathrm{HP}_n$ for $n = 2, \ldots, 20$ on
\code{ibm\_kingston}, using \code{PauliEvolutionGate} with $r = r^{*}$,
source vertex $0$, target vertex $n$ and $4096$ shots, in both the
default \code{cz} basis and the fractional basis
(\cref{sec:c4-transpile}).\footnote{The \code{cz}-basis sweep was taken on
2026-05-22 and the fractional sweep on 2026-06-13.} The transpiler placed
the $n+1$ qubits on a line of physical qubits, so no \code{swap} gates
were needed, and in the fractional basis every two-qubit gate is an
\code{rzz} rotation. The two-qubit gate count is the same in both bases.
We did not choose $r$ separately for the fractional basis, since the
reported \code{rzz} error is not given for a specific angle
(\cref{sec:implementation}).

\begin{figure}[ht]
  \centering
  \includegraphics[width=\linewidth]{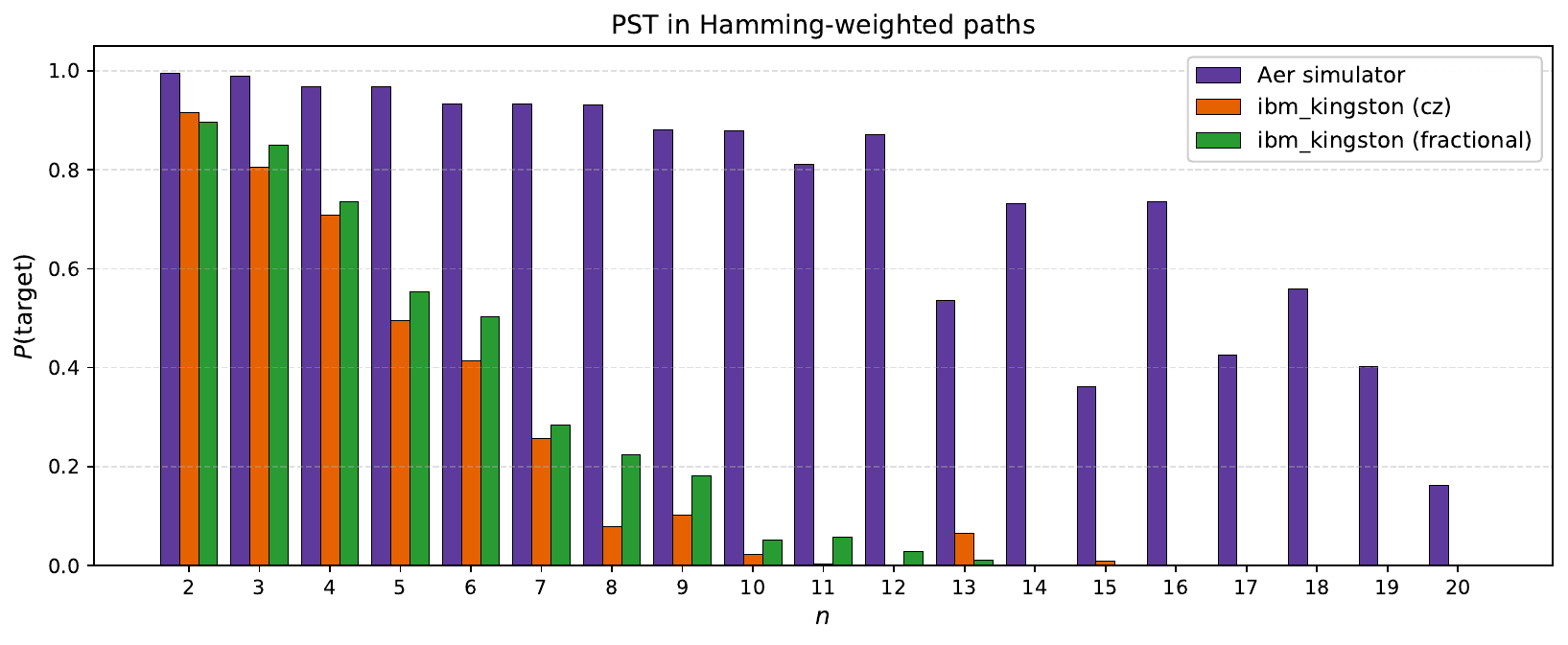}
  \caption{Target-vertex probability $P(\ket{e_n})$ for the
  Hamming-weighted-path PST circuit at $t = \pi/2$ and the per-$n$
  $r^{*}$ of \cref{tab:hp-r-optimal}, as a function of $n$. Purple:
  noiseless Aer prediction (statevector). Orange: \code{ibm\_kingston}
  hardware in the default \code{cz} basis. Green: \code{ibm\_kingston}
  hardware in the fractional basis. Both hardware curves sit below the
  Aer curve by a factor that grows with the two-qubit gate count, and
  the fractional curve dominates the \code{cz} curve on the
  large-$r^{*}$ circuits (e.g. $n = 8$--$12$); the exceptions are the
  Trotter-pocket sizes $n = 13, 15, 17, 19$. Per-vertex breakdowns for
  each $n$ are tabulated in \cref{app:hamming-tables}.}
  \label{fig:hp-pst-target-vs-n}
\end{figure}

\Cref{tab:hamming-hw} and \cref{fig:hp-pst-target-vs-n} give the
target probability for each $n$; the probabilities of all
single-excitation states are in \cref{app:hamming-tables}.

\begin{table}[ht]
\centering
\small
\setlength{\tabcolsep}{4.5pt}
\begin{tabular}{rrrrrrrrr}
\toprule
& & & & & \multicolumn{2}{c}{\code{cz} basis} & \multicolumn{2}{c}{fractional basis} \\
\cmidrule(lr){6-7}\cmidrule(lr){8-9}
$n$ & qubits & $r^{*}$ & $n_{2q}$ & Aer & $P(\text{target})$ & QPU & $P(\text{target})$ & QPU \\
\midrule
$2$ & $3$ & $3$ & $14$ & $0.994421$ & $\mathbf{0.915527}$ & $3$ & $0.896729$ & $3$ \\
$3$ & $4$ & $4$ & $34$ & $0.988575$ & $0.805420$ & $3$ & $\mathbf{0.849365}$ & $3$ \\
$4$ & $5$ & $4$ & $50$ & $0.967533$ & $0.708496$ & $3$ & $\mathbf{0.735107}$ & $3$ \\
$5$ & $6$ & $5$ & $82$ & $0.967695$ & $0.495361$ & $3$ & $\mathbf{0.553955}$ & $3$ \\
$6$ & $7$ & $5$ & $102$ & $0.933560$ & $0.414307$ & $3$ & $\mathbf{0.502441}$ & $3$ \\
$7$ & $8$ & $6$ & $146$ & $0.933957$ & $0.256836$ & $3$ & $\mathbf{0.284912}$ & $3$ \\
$8$ & $9$ & $7$ & $198$ & $0.931470$ & $0.078857$ & $3$ & $\mathbf{0.223389}$ & $3$ \\
$9$ & $10$ & $7$ & $226$ & $0.881124$ & $0.101562$ & $3$ & $\mathbf{0.182373}$ & $3$ \\
$10$ & $11$ & $8$ & $290$ & $0.879425$ & $0.022217$ & $3$ & $\mathbf{0.051758}$ & $3$ \\
$11$ & $12$ & $8$ & $322$ & $0.810534$ & $0.003418$ & $3$ & $\mathbf{0.057617}$ & $3$ \\
$12$ & $13$ & $10$ & $442$ & $0.871119$ & $0.002197$ & $3$ & $\mathbf{0.029297}$ & $3$ \\
$13$ & $14$ & $3$ & $146$ & $0.535430$ & $\mathbf{0.064453}$ & $3$ & $0.010986$ & $3$ \\
$14$ & $15$ & $10$ & $522$ & $0.732475$ & $0.000488$ & $4$ & $\mathbf{0.000732}$ & $3$ \\
$15$ & $16$ & $4$ & $226$ & $0.362132$ & $\mathbf{0.008789}$ & $3$ & $0.001953$ & $3$ \\
$16$ & $17$ & $12$ & $722$ & $0.735073$ & $0.000000$ & $4$ & $\mathbf{0.000488}$ & $4$ \\
$17$ & $18$ & $4$ & $258$ & $0.424774$ & $\mathbf{0.002197}$ & $3$ & $0.001221$ & $3$ \\
$18$ & $19$ & $12$ & $818$ & $0.559489$ & $0.000000$ & $4$ & $0.000000$ & $4$ \\
$19$ & $20$ & $4$ & $290$ & $0.402472$ & $\mathbf{0.001953}$ & $3$ & $0.000244$ & $3$ \\
$20$ & $21$ & $5$ & $382$ & $0.162311$ & $\mathbf{0.000488}$ & $3$ & $0.000000$ & $3$ \\
\bottomrule
\end{tabular}
\caption{Per-$n$ hardware target probability $P(\ket{e_n})$ for the
Hamming-weighted-path PST circuit at $t = \pi/2$ and the
$\hat{P}_{\mathrm{HW}}$-maximizing $r^{*}$ of \cref{tab:hp-r-optimal}, run on
\code{ibm\_kingston} ($4096$ shots) in the default \code{cz} basis and
in the fractional basis (\cref{sec:c4-transpile}). \emph{Aer} is the
exact noiseless target probability, identical for the two bases; the
bold entry in each row is the larger of the two hardware values.
\emph{QPU} is the IBM-billed \code{job.usage()} quantum time in seconds.
The \code{cz}-basis runs are from an earlier calibration epoch; the
fractional runs were taken in a single session on 2026-06-13.}
\label{tab:hamming-hw}
\end{table}

At $n = 2$, the \code{cz} basis gives a slightly larger target
probability than the fractional basis. For $3 \leq n \leq 12$, the
fractional basis gives the larger value; the ratio is at most $1.21$ for
$n \leq 7$, between $1.8$ and $2.8$ for $8 \leq n \leq 10$, and more than
$13$ for $n = 11, 12$, where the \code{cz}-basis values are $0.003$ and
$0.002$. For $n = 13, 15, 17, 19$, where $r^{*}$ is small, the \code{cz}
basis gives the larger value, by a factor between $1.8$ and $8$. For
$n = 14, 16, 18, 20$, both values are at most $0.001$. Since the two bases
were run on different days, some of the difference between them may be
due to changes in the device calibration. In the \code{cz} basis, the
measured target probability is below $\hat{P}_{\mathrm{HW}}$
(\cref{tab:hp-r-optimal}) for every $n$, by a factor of more than $5$ for
$n \geq 8$.

\paragraph{Comparison with $Q_n$.}
\Cref{tab:hamming-vs-full} compares PST on $Q_n$
(\cref{sec:c4-full,sec:q3-full}) with PST on $\mathrm{HP}_n$ for
$n = 2, 3, 4$, with $r = 8$. The target probability on $\mathrm{HP}_n$ is
larger by a factor of about $1.5$ at $n = 2$ and $28$ at $n = 3$. We did
not run PST on $Q_4$: with
$r = 4$, the \code{PauliEvolutionGate} circuit transpiles to $1{,}386$
two-qubit gates on a Heron coupling map, and the \code{UnitaryGate}
synthesis already fails at $11$ qubits (\cref{tab:hamming-survey}).

\begin{table}[ht]
\centering
\small
\begin{tabular}{lcccc}
\toprule
Encoding of $Q_n$ PST at $t = \pi/2$ & $n_{\text{qubits}}$ & $n_{2q}$ & $P(\text{target})$ & QPU usage (s) \\
\midrule
\multicolumn{5}{l}{$n = 2$ (the $4$-cycle):}\\
\quad full vertex encoding   & $4$ & $133$ & $0.532959$ (\cref{tab:c4-hw}) & $2$ \\
\quad Hamming-weighted path  & $3$ & $34$  & $\mathbf{0.782471}$ & $3$ \\
\addlinespace
\multicolumn{5}{l}{$n = 3$ (the $3$-cube):}\\
\quad full vertex encoding   & $8$ & $822$ & $0.025146$ (\cref{tab:q3-hw}) & $3$ \\
\quad Hamming-weighted path  & $4$ & $66$  & $\mathbf{0.697754}$ & $3$ \\
\addlinespace
\multicolumn{5}{l}{$n = 4$ (the $4$-cube):}\\
\quad full vertex encoding   & $16$ & n/a & \textsl{infeasible} & n/a \\
\quad Hamming-weighted path  & $5$ & $98$  & $\mathbf{0.523926}$ & $3$ \\
\bottomrule
\end{tabular}
\caption{Side-by-side hardware comparison of $Q_n$ PST under the two
encodings, both at \code{PauliEvolutionGate} Suzuki--Trotter with $r = 8$ on
\code{ibm\_kingston}. The Hamming-weighted-path runs at $r = 8$ were made separately from
the sweep of \cref{tab:hamming-hw}, between 2026-05-20 and 2026-05-22. The Hamming-weighted-path encoding uses roughly
$2^{n}/(n+1)$ times fewer qubits and at $n = 3$ saves a factor of $\sim 12$
in two-qubit gate count. The resulting hardware target probability improves by a
factor of $\sim 28$ on the same physical experiment. IBM-reported QPU
usage (integer-billed quantum seconds) is essentially constant across all
six runs, with the C$_4$ direct $\pi/2$ block at $2$~s and everything
else at $3$~s.}
\label{tab:hamming-vs-full}
\end{table}

\section{Peak state transfer on the $X_n$ family}\label{sec:xn-peakST}

In this section, we run the walk on the family $X_n$ of
\cite[Section 6.3]{CGS-2025}, which has peak state transfer
(\cref{sec:pst-peakST}).

\subsection{The family $X_n$}\label{sec:xn-family}

Recall that, for $n \geq 2$, $X_n$ is the weighted path on $2n$ vertices
that visits, in order,
\[
  v_n,\; w_1,\; v_{n-1},\; w_2,\; \ldots,\; v_2,\; w_{n-1},\; v_1,\; w_n,
\]
with the edge between $v_j$ and $w_k$ having weight $\sqrt{j\,k}$. We will
write $A_n = A(X_n)$. The eigenvalues of $A_n$ are
$\pm 1, \pm 2, \ldots, \pm n$, each simple. By
\cite[Theorem 6.5]{CGS-2025}, $X_n$ has peak state transfer from $w_1$ to
$w_n$ at time $\pi$. The peak transfer probability
$B(A_n)_{w_1, w_n}^2$ (\cref{sec:pst-peakST}) is less than $1$ and
decreases to $\pi/4$ as $n \to \infty$.

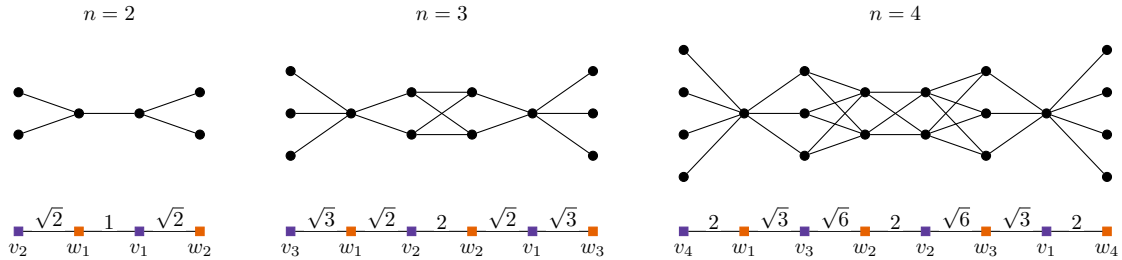
\begin{figure}[htbp]
    \centering
\begin{tikzpicture}[scale=0.8, every node/.style={scale=0.8}]
\tikzstyle{graph node}=[circle, fill=black, draw=black, inner sep=1.5pt]
\tikzstyle{path node v}=[rectangle, fill=kcolour, draw=kcolour, inner sep=2pt]
\tikzstyle{path node w}=[rectangle, fill=vcolour, draw=vcolour, inner sep=2pt]
\tikzstyle{edge label}=[midway, fill=white, inner sep=1pt]

\node at (1.5,0.6) {$n=2$};
\node[graph node] (a1) at (0,-0.70) {};
\node[graph node] (a2) at (0,-1.40) {};
\node[graph node] (b1) at (1,-1.05) {};
\node[graph node] (c1) at (2,-1.05) {};
\node[graph node] (d1) at (3,-0.70) {};
\node[graph node] (d2) at (3,-1.40) {};
\draw (a1) -- (b1); \draw (a2) -- (b1);
\draw (b1) -- (c1);
\draw (c1) -- (d1); \draw (c1) -- (d2);
\node[path node v] (p1) at (0,-3) {};
\node[path node w] (p2) at (1,-3) {};
\node[path node v] (p3) at (2,-3) {};
\node[path node w] (p4) at (3,-3) {};
\draw (p1) -- node[edge label, above] {$\sqrt{2}$} (p2);
\draw (p2) -- node[edge label, above] {$1$} (p3);
\draw (p3) -- node[edge label, above] {$\sqrt{2}$} (p4);
\node[below] at (p1.south) {$v_2$};
\node[below] at (p2.south) {$w_1$};
\node[below] at (p3.south) {$v_1$};
\node[below] at (p4.south) {$w_2$};

\node at (7.0,0.6) {$n=3$};
\node[graph node] (e1) at (4.5,-0.35) {};
\node[graph node] (e2) at (4.5,-1.05) {};
\node[graph node] (e3) at (4.5,-1.75) {};
\node[graph node] (f1) at (5.5,-1.05) {};
\node[graph node] (g1) at (6.5,-0.70) {};
\node[graph node] (g2) at (6.5,-1.40) {};
\node[graph node] (h1) at (7.5,-0.70) {};
\node[graph node] (h2) at (7.5,-1.40) {};
\node[graph node] (i1) at (8.5,-1.05) {};
\node[graph node] (j1) at (9.5,-0.35) {};
\node[graph node] (j2) at (9.5,-1.05) {};
\node[graph node] (j3) at (9.5,-1.75) {};
\draw (e1) -- (f1); \draw (e2) -- (f1); \draw (e3) -- (f1);
\draw (f1) -- (g1); \draw (f1) -- (g2);
\draw (g1) -- (h1); \draw (g1) -- (h2); \draw (g2) -- (h1); \draw (g2) -- (h2);
\draw (h1) -- (i1); \draw (h2) -- (i1);
\draw (i1) -- (j1); \draw (i1) -- (j2); \draw (i1) -- (j3);
\node[path node v] (q1) at (4.5,-3) {};
\node[path node w] (q2) at (5.5,-3) {};
\node[path node v] (q3) at (6.5,-3) {};
\node[path node w] (q4) at (7.5,-3) {};
\node[path node v] (q5) at (8.5,-3) {};
\node[path node w] (q6) at (9.5,-3) {};
\draw (q1) -- node[edge label, above] {$\sqrt{3}$} (q2);
\draw (q2) -- node[edge label, above] {$\sqrt{2}$} (q3);
\draw (q3) -- node[edge label, above] {$2$} (q4);
\draw (q4) -- node[edge label, above] {$\sqrt{2}$} (q5);
\draw (q5) -- node[edge label, above] {$\sqrt{3}$} (q6);
\node[below] at (q1.south) {$v_3$};
\node[below] at (q2.south) {$w_1$};
\node[below] at (q3.south) {$v_2$};
\node[below] at (q4.south) {$w_2$};
\node[below] at (q5.south) {$v_1$};
\node[below] at (q6.south) {$w_3$};

\node at (14.5,0.6) {$n=4$};
\node[graph node] (x11) at (11,0) {};
\node[graph node] (x12) at (11,-0.7) {};
\node[graph node] (x13) at (11,-1.4) {};
\node[graph node] (x14) at (11,-2.1) {};
\node[graph node] (x21) at (12,-1.05) {};
\node[graph node] (x31) at (13,-0.35) {};
\node[graph node] (x32) at (13,-1.05) {};
\node[graph node] (x33) at (13,-1.75) {};
\node[graph node] (x41) at (14,-0.7) {};
\node[graph node] (x42) at (14,-1.4) {};
\node[graph node] (x51) at (15,-0.7) {};
\node[graph node] (x52) at (15,-1.4) {};
\node[graph node] (x61) at (16,-0.35) {};
\node[graph node] (x62) at (16,-1.05) {};
\node[graph node] (x63) at (16,-1.75) {};
\node[graph node] (x71) at (17,-1.05) {};
\node[graph node] (x81) at (18,0) {};
\node[graph node] (x82) at (18,-0.7) {};
\node[graph node] (x83) at (18,-1.4) {};
\node[graph node] (x84) at (18,-2.1) {};
\draw (x11) -- (x21);
\draw (x12) -- (x21);
\draw (x13) -- (x21);
\draw (x14) -- (x21);
\draw (x21) -- (x31);
\draw (x21) -- (x32);
\draw (x21) -- (x33);
\draw (x41) -- (x31);
\draw (x41) -- (x32);
\draw (x41) -- (x33);
\draw (x42) -- (x31);
\draw (x42) -- (x32);
\draw (x42) -- (x33);
\draw (x41) -- (x51);
\draw (x42) -- (x51);
\draw (x41) -- (x52);
\draw (x42) -- (x52);
\draw (x51) -- (x61);
\draw (x51) -- (x62);
\draw (x51) -- (x63);
\draw (x52) -- (x61);
\draw (x52) -- (x62);
\draw (x52) -- (x63);
\draw (x71) -- (x61);
\draw (x71) -- (x62);
\draw (x71) -- (x63);
\draw (x71) -- (x81);
\draw (x71) -- (x82);
\draw (x71) -- (x83);
\draw (x71) -- (x84);
\node[path node v] (y1) at (11,-3) {};
\node[path node w] (y2) at (12,-3) {};
\node[path node v] (y3) at (13,-3) {};
\node[path node w] (y4) at (14,-3) {};
\node[path node v] (y5) at (15,-3) {};
\node[path node w] (y6) at (16,-3) {};
\node[path node v] (y7) at (17,-3) {};
\node[path node w] (y8) at (18,-3) {};
\draw (y1) -- node[edge label, above] {$2$} (y2);
\draw (y2) -- node[edge label, above] {$\sqrt{3}$} (y3);
\draw (y3) -- node[edge label, above] {$\sqrt{6}$} (y4);
\draw (y4) -- node[edge label, above] {$2$} (y5);
\draw (y5) -- node[edge label, above] {$\sqrt{6}$} (y6);
\draw (y6) -- node[edge label, above] {$\sqrt{3}$} (y7);
\draw (y7) -- node[edge label, above] {$2$} (y8);
\node[below] at (y1.south) {$v_4$};
\node[below] at (y2.south) {$w_1$};
\node[below] at (y3.south) {$v_3$};
\node[below] at (y4.south) {$w_2$};
\node[below] at (y5.south) {$v_2$};
\node[below] at (y6.south) {$w_3$};
\node[below] at (y7.south) {$v_1$};
\node[below] at (y8.south) {$w_4$};
\end{tikzpicture}
    \caption{The unweighted graphs $Y_n$ (upper) with their quotients
    $X_n$ (lower) for $n = 2, 3, 4$. Each weighted vertex $v_j, w_j$ of
    $X_n$ lifts to a cell $V_j, W_j$ of $j$ vertices in $Y_n$; the edge
    weight $\sqrt{jk}$ is the symmetrized quotient value of the complete
    bipartite join between cells $V_j$ and $W_k$. Source $w_1$ and target
    $w_n$ are the $X_n$ peak-ST endpoints.}
    \label{fig:xn-quotient}
\end{figure}

\paragraph{Quotient construction.}
Let $Y_n$ be the unweighted graph obtained from $X_n$ by replacing each
vertex $v_j$ and $w_j$ by a set $V_j$ or $W_j$ of $j$ vertices, and each
edge $v_jw_k$ by all edges between $V_j$ and $W_k$. The sets $V_j$ and
$W_j$ form an equitable partition of $Y_n$, and $A_n$ is the symmetrized
quotient of $A(Y_n)$ with respect to this partition.
\Cref{fig:xn-quotient} shows $Y_n$ and $X_n$ for $n = 2, 3, 4$.

\paragraph{Implementation.}
We use one qubit for each vertex of $X_n$, numbered $0, 1, \ldots, 2n-1$
in the order of the path, so the source $w_1$ is qubit $1$ and the target
$w_n$ is qubit $2n-1$. The Hamiltonian is
\[
  H_{X_n} \;=\; \tfrac12 \sum_{v_jw_k \in E(X_n)} \sqrt{jk}\,
  \bigl( X_{v_j} X_{w_k} + Y_{v_j} Y_{w_k} \bigr).
\]
Here $X_n$ is the graph and $X_{v_j}$, $X_{w_k}$ are the Pauli operators
of \cref{sec:background} on the qubits at $v_j$ and $w_k$.
As for $\mathrm{HP}_n$, we use \code{PauliEvolutionGate} with the
second-order Suzuki--Trotter formula; we will choose $r$ in
\cref{sec:xn-rsweep}.

\subsection{Gate-count survey at \texorpdfstring{$t = \pi$}{t = pi}}\label{sec:xn-survey}

\Cref{tab:xn-survey} gives the gate counts of the \code{PauliEvolutionGate}
circuit for $n = 2, \ldots, 20$ at $r = 8$, transpiled without a coupling
map. Since $X_n$ has $2n - 1$ edges, the two-qubit gate count
$n_{2q} = 64n - 62$ grows twice as fast in $n$ as for $\mathrm{HP}_n$.

\begin{table}[ht]
\centering
\small
\begin{tabular}{rrrrrr|rrrrrr}
\toprule
$n$ & $q$ & edges & depth & total & $n_{2q}$ & $n$ & $q$ & edges & depth & total & $n_{2q}$ \\
\midrule
 2 &  4 &  3 &   302 &    519 &  \textbf{66} & 12 & 24 & 23 & 3{,}041 & 5{,}652 & \textbf{706} \\
 3 &  6 &  5 &   589 &  1{,}017 & \textbf{130} & 13 & 26 & 25 & 3{,}271 & 6{,}135 & \textbf{770} \\
 4 &  8 &  7 &   851 &  1{,}533 & \textbf{194} & 14 & 28 & 27 & 3{,}532 & 6{,}615 & \textbf{834} \\
 5 & 10 &  9 & 1{,}115 & 2{,}029 & \textbf{258} & 15 & 30 & 29 & 3{,}817 & 7{,}135 & \textbf{898} \\
 6 & 12 & 11 & 1{,}392 & 2{,}585 & \textbf{322} & 16 & 32 & 31 & 3{,}959 & 7{,}537 & \textbf{962} \\
 7 & 14 & 13 & 1{,}650 & 3{,}023 & \textbf{386} & 17 & 34 & 33 & 4{,}303 & 8{,}133 & \textbf{1{,}026} \\
 8 & 16 & 15 & 1{,}973 & 3{,}555 & \textbf{450} & 18 & 36 & 35 & 4{,}609 & 8{,}650 & \textbf{1{,}090} \\
 9 & 18 & 17 & 2{,}221 & 4{,}071 & \textbf{514} & 19 & 38 & 37 & 4{,}927 & 9{,}090 & \textbf{1{,}154} \\
10 & 20 & 19 & 2{,}419 & 4{,}461 & \textbf{578} & 20 & 40 & 39 & 5{,}170 & 9{,}632 & \textbf{1{,}218} \\
11 & 22 & 21 & 2{,}781 & 5{,}109 & \textbf{642} & & & & & & \\
\bottomrule
\end{tabular}
\caption{\code{PauliEvolutionGate} gate counts with second-order
Suzuki--Trotter at $r = 8$ for the $X_n$ peak-ST circuit at $t = \pi$. Transpile to
$\{\code{cz}, \code{sx}, \code{rz}, \code{x}\}$ at \code{optimization\_level=3},
no coupling map. $n_{2q}(n) = 64n - 62$ on the nose: roughly twice the
slope of the Hamming-weighted-path family because the path has
$2n - 1$ edges (vs.\ $n$).}
\label{tab:xn-survey}
\end{table}

\paragraph{\code{UnitaryGate}.}
The \code{UnitaryGate} circuit has $92$, $1{,}752$, $29{,}651$ and
$479{,}063$ two-qubit gates for $n = 2, 3, 4, 5$. We did not transpile it
for $n \geq 6$, since the synthesis already failed at $11$ qubits for
$\mathrm{HP}_n$ (\cref{tab:hamming-survey}).

\subsection{Choosing the Trotter rep count \texorpdfstring{$r$}{r}}\label{sec:xn-rsweep}

On the single-excitation subspace, $\lVert H_{X_n} \rVert = n$, as for
$\mathrm{HP}_n$, but now $t = \pi$ rather than $\pi/2$, so the estimate of
the Trotter error in \cref{sec:hamming-rsweep} is $8$ times larger. We
repeated the computation of \cref{sec:hamming-rsweep} for
$n \in \{2, 4, 6, 8, 10\}$ and $r \in \{2, 4, 8, 16, 32\}$, with the same
$p_{2q} = 2.0 \times 10^{-3}$; see \cref{tab:xn-rsweep}.

\begin{table}[ht]
\centering
\small
\begin{tabular}{rrrrrrr}
\toprule
$n$ & $r$ & qubits & $n_{2q}$ & $P_{\mathrm{noiseless}}$ & $S$ & $\hat{P}_{\mathrm{HW}}$ \\
\midrule
$ 2$ &  $2$ &  $4$ &    $18$ & $0.390233$ & $0.964606$ & $0.376421$ \\
$ 2$ &  $4$ &  $4$ &    $34$ & $0.882862$ & $0.934197$ & $\mathbf{0.824767}$ \\
$ 2$ &  $8$ &  $4$ &    $66$ & $0.898952$ & $0.876225$ & $0.787685$ \\
$ 2$ & $16$ &  $4$ &   $130$ & $0.892137$ & $0.770851$ & $0.687704$ \\
$ 2$ & $32$ &  $4$ &   $258$ & $0.889746$ & $0.596595$ & $0.530818$ \\
\addlinespace
$ 4$ &  $2$ &  $8$ &    $26$ & $0.000000$ & $0.949279$ & $0.000000$ \\
$ 4$ &  $4$ &  $8$ &    $98$ & $0.373242$ & $0.821851$ & $0.306749$ \\
$ 4$ &  $8$ &  $8$ &   $194$ & $0.809383$ & $0.678149$ & $\mathbf{0.548882}$ \\
$ 4$ & $16$ &  $8$ &   $386$ & $0.841214$ & $0.461731$ & $0.388415$ \\
$ 4$ & $32$ &  $8$ &   $770$ & $0.838096$ & $0.214051$ & $0.179395$ \\
\addlinespace
$ 6$ &  $2$ & $12$ &    $82$ & $0.000474$ & $0.848603$ & $0.000402$ \\
$ 6$ &  $4$ & $12$ &   $162$ & $0.435207$ & $0.723016$ & $\mathbf{0.314662}$ \\
$ 6$ &  $8$ & $12$ &   $322$ & $0.504559$ & $0.524849$ & $0.264817$ \\
$ 6$ & $16$ & $12$ &   $642$ & $0.804754$ & $0.276572$ & $0.222572$ \\
$ 6$ & $32$ & $12$ &  $1282$ & $0.820863$ & $0.076799$ & $0.063041$ \\
\addlinespace
$ 8$ &  $2$ & $16$ &   $100$ & $0.000000$ & $0.818567$ & $0.000000$ \\
$ 8$ &  $4$ & $16$ &   $226$ & $0.228183$ & $0.636066$ & $\mathbf{0.145139}$ \\
$ 8$ &  $8$ & $16$ &   $450$ & $0.002013$ & $0.406203$ & $0.000818$ \\
$ 8$ & $16$ & $16$ &   $898$ & $0.696971$ & $0.165663$ & $0.115462$ \\
$ 8$ & $32$ & $16$ &  $1794$ & $0.806760$ & $0.027554$ & $0.022230$ \\
\addlinespace
$10$ &  $2$ & $20$ &   $118$ & $0.000000$ & $0.789594$ & $0.000000$ \\
$10$ &  $4$ & $20$ &   $290$ & $0.001932$ & $0.559573$ & $0.001081$ \\
$10$ &  $8$ & $20$ &   $578$ & $0.602116$ & $0.314379$ & $\mathbf{0.189292}$ \\
$10$ & $16$ & $20$ &  $1154$ & $0.429116$ & $0.099230$ & $0.042581$ \\
$10$ & $32$ & $20$ &  $2306$ & $0.780868$ & $0.009886$ & $0.007720$ \\
\bottomrule
\end{tabular}
\caption{Trotter-rep sweep for the $X_n$ peak-ST circuit at $t = \pi$.
Same conventions as \cref{tab:hamming-rsweep}: $n_{2q}$ is the
two-qubit (CZ) gate count after transpilation, $P_{\mathrm{noiseless}}$ is
the Aer prediction for the transpiled Trotter circuit, $S$ is the device
survival fraction at $p_{2q} = 2.0\times 10^{-3}$ (the \code{ibm\_kingston} median \code{cz} error), and
$\hat{P}_{\mathrm{HW}} = P_{\mathrm{noiseless}} \cdot S$ is the predicted
hardware target probability. The bolded entry per $n$ is the $r$ that
maximizes $\hat{P}_{\mathrm{HW}}$.}
\label{tab:xn-rsweep}
\end{table}

In \cref{tab:xn-rsweep}, the smallest $r$ with
$P_{\mathrm{noiseless}} \geq \pi/4$ is $4$, $8$, $16$ and $32$ for
$n = 2, 4, 6, 8$; at $n = 10$, no $r \leq 32$ reaches $\pi/4$. The value
of $r$ that maximizes $\hat{P}_{\mathrm{HW}}$ is $4$ or $8$ for every $n$,
and it is not monotone in $n$; at $n = 8$, for example,
$P_{\mathrm{noiseless}}$ is $0.228$ at $r = 4$ and $0.002$ at $r = 8$.

We then computed $\hat{P}_{\mathrm{HW}}$ for
$r \in \{2, 3, 4, 5, 6, 7, 8, 10, 12, 16\}$ and $n = 2, \ldots, 11$;
\cref{tab:xn-r-optimal} gives the maximizing value $r^{*}$ for each $n$.
We will use $r^{*}$ for the hardware runs in \cref{sec:xn-hw}; at $n = 8$,
for example, $\hat{P}_{\mathrm{HW}}$ is $0.404$ at $r^{*} = 5$ and $0.0008$
at $r = 8$.

\begin{table}[ht]
\centering
\small
\begin{tabular}{rrrrrr}
\toprule
$n$ & qubits & $r^{*}$ & $n_{2q}$ & $P_{\mathrm{noiseless}}$ & $\hat{P}_{\mathrm{HW}}$ \\
\midrule
$ 2$ &  $4$ &  $5$ &  $42$ & $0.899077$ & $0.826570$ \\
$ 3$ &  $6$ &  $6$ &  $98$ & $0.836018$ & $0.687082$ \\
$ 4$ &  $8$ &  $7$ & $170$ & $0.780275$ & $0.555188$ \\
$ 5$ & $10$ &  $8$ & $258$ & $0.708782$ & $0.422856$ \\
$ 6$ & $12$ &  $4$ & $162$ & $0.435207$ & $0.314662$ \\
$ 7$ & $14$ &  $4$ & $194$ & $0.698554$ & $0.473724$ \\
$ 8$ & $16$ &  $5$ & $282$ & $0.710666$ & $0.404090$ \\
$ 9$ & $18$ &  $6$ & $386$ & $0.655381$ & $0.302610$ \\
$10$ & $20$ &  $7$ & $506$ & $0.581118$ & $0.211017$ \\
$11$ & $22$ &  $8$ & $642$ & $0.504908$ & $0.139643$ \\
\bottomrule
\end{tabular}
\caption{The Trotter rep count $r^{*}$ maximizing $\hat{P}_{\mathrm{HW}}$, for each $n$, for
$X_n$ peak-ST at $t = \pi$, over the discrete sweep
$r \in \{2, 3, 4, 5, 6, 7, 8, 10, 12, 16\}$. Survival is computed at
$p_{2q} = 2.0\times 10^{-3}$ (the \code{ibm\_kingston} median \code{cz} error) as in \cref{tab:xn-rsweep}.
At $p_{2q} = 2.0\times 10^{-3}$ the optima sit in $r \in \{4, \ldots, 8\}$
for every $n$.}
\label{tab:xn-r-optimal}
\end{table}

\subsection{Hardware results}\label{sec:xn-hw}

\begin{figure}[H]
  \centering
  \includegraphics[width=\linewidth]{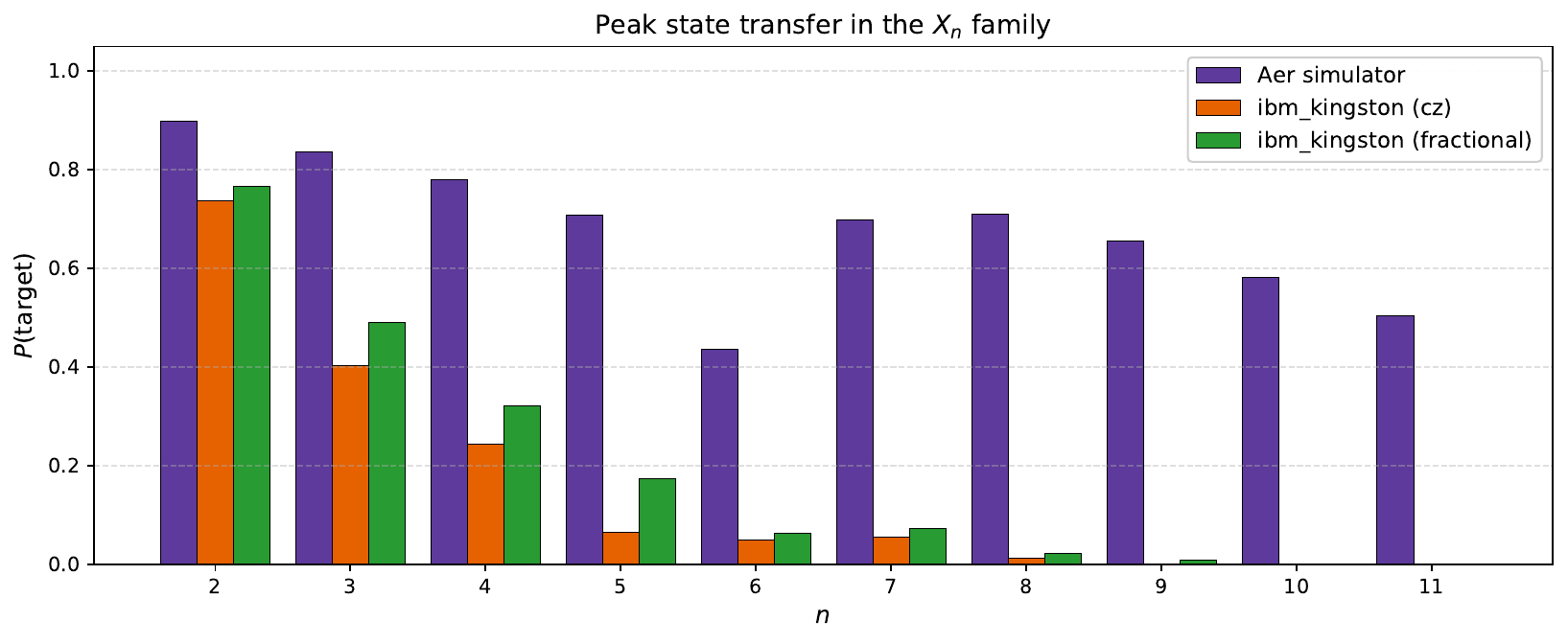}
  \caption{Target-vertex probability $P(\ket{e_{w_n}})$ for the $X_n$
  peak-ST circuit at $t = \pi$ and the per-$n$ $r^{*}$ of
  \cref{tab:xn-r-optimal}. Purple: noiseless Aer-simulator output (the
  noiseless reference for hardware, which differs from the analytical
  peak transfer probability $B(A_n)_{w_1, w_n}^2$ where Suzuki--Trotter at $r^{*}$
  is imperfect). Orange: \code{ibm\_kingston} hardware in the default
  \code{cz} basis. Green: \code{ibm\_kingston} hardware in the
  fractional basis. The hardware signal decays with the two-qubit gate
  count, faster than the Hamming-weighted-path family of
  \cref{fig:hp-pst-target-vs-n} because $X_n$ has roughly twice as
  many edges, and the fractional basis dominates the \code{cz} basis
  at every $n$ where either resolves the target.}
  \label{fig:xn-pst-target-vs-n}
\end{figure}

We ran the circuit for $n = 2, \ldots, 11$ on \code{ibm\_kingston}, with
$r = r^{*}$ and $4096$ shots, in both the default \code{cz} basis and the
fractional basis.\footnote{The \code{cz}-basis sweep was taken on
2026-05-22 and the fractional sweep on 2026-06-13.} As for
$\mathrm{HP}_n$, we used the same $r^{*}$ in both bases, and the
transpiler placed the qubits on a line of physical qubits, so every
two-qubit gate in the fractional basis is an \code{rzz} rotation.
\Cref{tab:xn-hw}
and \cref{fig:xn-pst-target-vs-n} give the target probability for each
$n$; the probabilities of all single-excitation states are in
\cref{app:xn-tables}.

\begin{table}[ht]
\centering
\small
\setlength{\tabcolsep}{4.5pt}
\begin{tabular}{rrrrrrrrrr}
\toprule
& & & & & & \multicolumn{2}{c}{\code{cz} basis} & \multicolumn{2}{c}{fractional basis} \\
\cmidrule(lr){7-8}\cmidrule(lr){9-10}
$n$ & qubits & $r^{*}$ & $n_{2q}$ & peak & Aer & $P(\text{target})$ & QPU & $P(\text{target})$ & QPU \\
\midrule
$2$ & $4$ & $5$ & $42$ & $0.888889$ & $0.899077$ & $0.738037$ & $3$ & $\mathbf{0.766113}$ & $3$ \\
$3$ & $6$ & $6$ & $98$ & $0.853333$ & $0.836018$ & $0.402588$ & $3$ & $\mathbf{0.490234}$ & $3$ \\
$4$ & $8$ & $7$ & $170$ & $0.835918$ & $0.780275$ & $0.243408$ & $3$ & $\mathbf{0.321533}$ & $3$ \\
$5$ & $10$ & $8$ & $258$ & $0.825598$ & $0.708782$ & $0.065186$ & $3$ & $\mathbf{0.173340}$ & $3$ \\
$6$ & $12$ & $4$ & $162$ & $0.818775$ & $0.435207$ & $0.050049$ & $3$ & $\mathbf{0.062744}$ & $3$ \\
$7$ & $14$ & $4$ & $194$ & $0.813930$ & $0.698554$ & $0.055420$ & $3$ & $\mathbf{0.072021}$ & $3$ \\
$8$ & $16$ & $5$ & $282$ & $0.810313$ & $0.710666$ & $0.012451$ & $3$ & $\mathbf{0.022949}$ & $3$ \\
$9$ & $18$ & $6$ & $386$ & $0.807509$ & $0.655381$ & $0.000977$ & $3$ & $\mathbf{0.009277}$ & $3$ \\
$10$ & $20$ & $7$ & $506$ & $0.805272$ & $0.581118$ & $0.000000$ & $4$ & $\mathbf{0.000732}$ & $3$ \\
$11$ & $22$ & $8$ & $642$ & $0.803446$ & $0.504908$ & $\mathbf{0.000000}$ & $4$ & $0.000000$ & $4$ \\
\bottomrule
\end{tabular}
\caption{Per-$n$ hardware target probability $P(\ket{e_{w_n}})$ for the
$X_n$ peak-ST circuit at $t = \pi$ and the $\hat{P}_{\mathrm{HW}}$-maximizing
$r^{*}$ of \cref{tab:xn-r-optimal}, run on \code{ibm\_kingston}
($4096$ shots) in the default \code{cz} basis and in the fractional
basis (\cref{sec:c4-transpile}). \emph{Peak} is the
peak transfer probability $B(A_n)_{w_1, w_n}^2$ of \cite{CGS-2025}, which
is the target probability of the exact $U(\pi)$. \emph{Aer} is the exact
noiseless target probability of the Trotter circuit, identical for the two
bases; the bold entry in each row is the larger of the two hardware values.
\emph{QPU} is the IBM-billed \code{job.usage()} quantum time in seconds. The
\code{cz}-basis runs are from an earlier calibration epoch; the
fractional runs were taken in a single session on 2026-06-13.}
\label{tab:xn-hw}
\end{table}

At $r^{*}$, the noiseless target probability is between $0.44$ and $0.90$
(\cref{tab:xn-hw}). At $n = 2$, it is larger than the peak transfer
probability $B(A_2)_{w_1, w_2}^2 = 8/9$; the Trotter circuit is not
$U(\pi)$, so its target probability is not bounded by $B(A_n)_{w_1, w_n}^2$.
At $n = 6$, where $r^{*} = 4$, the noiseless target probability $0.435$ is
about half the peak transfer probability $0.819$. Thus for $n = 6$ the
Trotter error alone halves the target probability, before any hardware
noise. In the \code{cz} basis, the target probability is at
least $0.24$ for $n \leq 4$, between $0.01$ and $0.07$ for
$5 \leq n \leq 8$, and at most $0.001$ for $n \geq 9$. The fractional
basis gives the larger target probability for every $n \leq 10$; the
ratio is between $1.04$ and $1.32$ for $n \leq 4$, $2.7$ at $n = 5$ and
$9.5$ at $n = 9$. Unlike for $\mathrm{HP}_n$, this includes the values of
$n$ with small $r^{*}$ ($n = 6, 7$).

\section{A chain of SWAP gates as a quantum wire baseline}\label{sec:swap-chain}

The quantum-wire problem of \cite{BoseQuantumComPaths} asks for state
transfer along a chain with a fixed Hamiltonian, without the
step-by-step control that a sequence of \code{swap} gates requires. As a
baseline for the experiments of
\cref{sec:hamming-paths,sec:xn-peakST}, we implement the controlled
alternative: we move a single excitation along a line of $n$ qubits
using $n - 1$ \code{swap} gates.

\subsection{Circuit}

\begin{figure}[H]
  \centering
  \includegraphics[width=\linewidth]{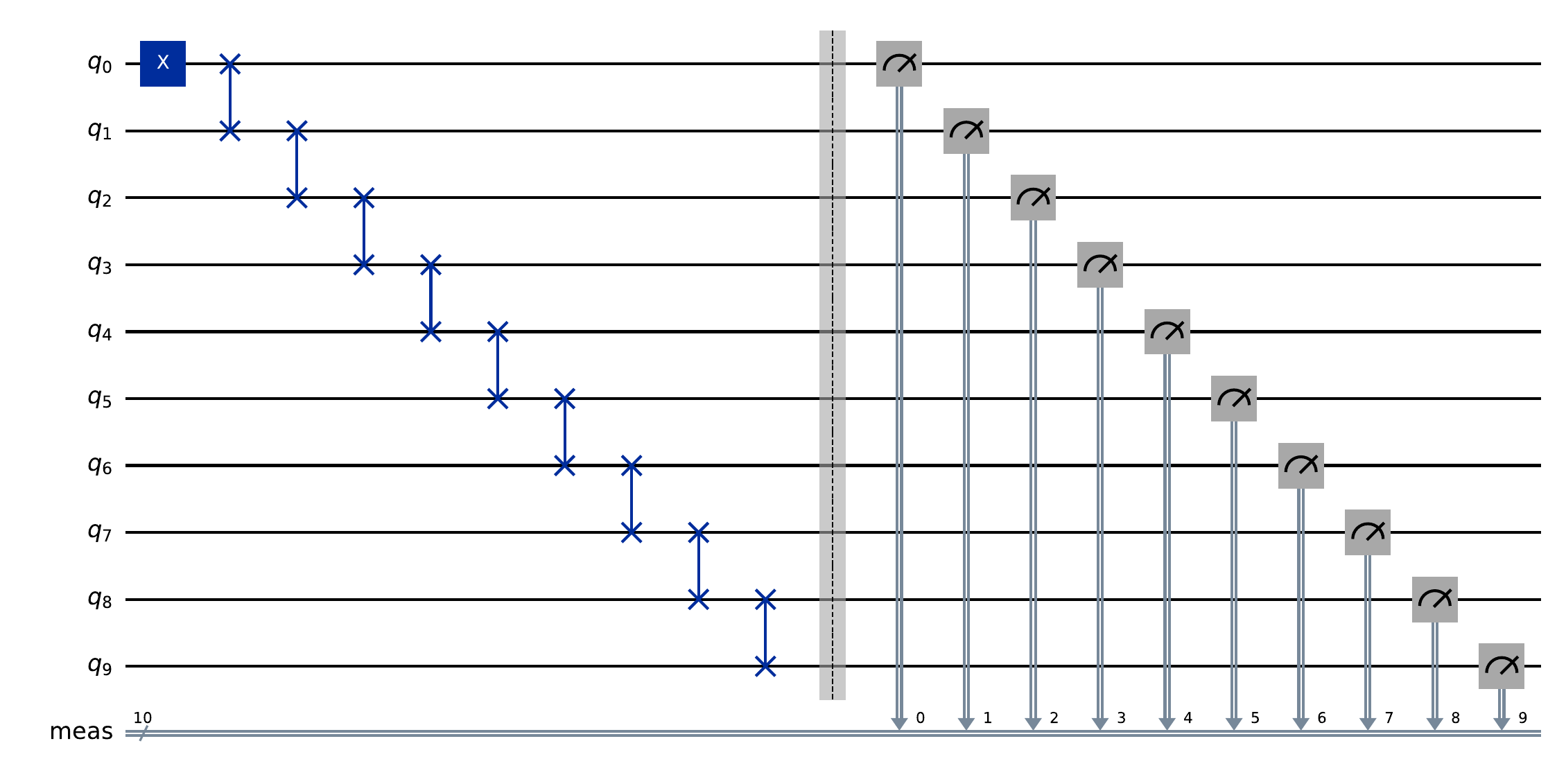}
  \caption{The $n = 10$ SWAP-chain circuit. After an initial $X$ on $q_0$
  the excitation walks one qubit at a time through nine \code{swap}
  gates, ending on $q_9$. Each \code{swap} compiles to three \code{cz}
  gates on \code{ibm\_kingston}.}
  \label{fig:swap-chain-n10}
\end{figure}

For each $n \in \{2, \ldots, 21\}$, we prepare the excitation on $q_0$,
apply $n - 1$ nearest-neighbour \code{swap} gates,
\[
  \ket{0\cdots 0} \;\xrightarrow{X_{q_0}}\;
  \ket{0\cdots 0 1} \;\xrightarrow{\code{swap}(q_0,q_1)}\;
  \ket{0\cdots 010} \;\xrightarrow{\code{swap}(q_1,q_2)}\;
  \cdots \;\xrightarrow{\code{swap}(q_{n-2},q_{n-1})}\;
  \ket{10\cdots 0},
\]
and measure all qubits; \cref{fig:swap-chain-n10} shows the circuit for
$n = 10$. The noiseless output is $\ket{e_{n-1}} = \ket{10\cdots 0}$ with
probability $1$; the excitation is then on $q_{n-1}$, which is the
leftmost bit in Qiskit's little-endian convention.

\subsection{Hardware setup}

We ran the circuits on \code{ibm\_kingston} with $4096$
shots.\footnote{The SWAP-chain runs were taken on 2026-05-27.} At
\code{optimization\_level=3}, the transpiler replaces the \code{swap}
gates by a relabelling of the physical qubits, which leaves a circuit
with no two-qubit gates. We therefore used \code{optimization\_level=0},
and placed the $n$ qubits on the first $n$ qubits of a fixed line of
physical qubits. Each \code{swap} then becomes three \code{cz} gates, so
$n_{2q} = 3(n - 1)$.

\subsection{Results}

\Cref{tab:swap-chain-hw} gives the in-subspace fraction and the target
probability $P(\ket{e_{n-1}})$ for each $n$.

\begin{table}[ht]
\centering
\small
\begin{tabular}{rrrrr}
\toprule
$n$ & $n_{2q}$ & in-subspace & $P(\ket{e_{n-1}})$ & QPU usage (s) \\
\midrule
 2 &  3 & $0.7759$ & $\mathbf{0.754395}$ & $3$ \\
 3 &  6 & $0.8464$ & $\mathbf{0.836182}$ & $3$ \\
 4 &  9 & $0.8328$ & $\mathbf{0.814941}$ & $3$ \\
 5 & 12 & $0.7939$ & $\mathbf{0.773438}$ & $3$ \\
 6 & 15 & $0.7922$ & $\mathbf{0.767822}$ & $3$ \\
 7 & 18 & $0.7878$ & $\mathbf{0.760742}$ & $3$ \\
 8 & 21 & $0.7715$ & $\mathbf{0.738281}$ & $3$ \\
 9 & 24 & $0.7537$ & $\mathbf{0.718506}$ & $3$ \\
10 & 27 & $0.7322$ & $\mathbf{0.690918}$ & $3$ \\
11 & 30 & $0.7471$ & $\mathbf{0.704834}$ & $3$ \\
12 & 33 & $0.7043$ & $\mathbf{0.655029}$ & $3$ \\
13 & 36 & $0.7034$ & $\mathbf{0.645752}$ & $3$ \\
14 & 39 & $0.6670$ & $\mathbf{0.603760}$ & $3$ \\
15 & 42 & $0.6648$ & $\mathbf{0.600830}$ & $3$ \\
16 & 45 & $0.6560$ & $\mathbf{0.588623}$ & $3$ \\
17 & 48 & $0.6287$ & $\mathbf{0.552002}$ & $3$ \\
18 & 51 & $0.5522$ & $\mathbf{0.469482}$ & $3$ \\
19 & 54 & $0.5125$ & $\mathbf{0.421143}$ & $3$ \\
20 & 57 & $0.4883$ & $\mathbf{0.394531}$ & $3$ \\
21 & 60 & $0.4785$ & $\mathbf{0.389404}$ & $3$ \\
\bottomrule
\end{tabular}
\caption{SWAP-chain target probabilities on \code{ibm\_kingston},
$4096$ shots per job, \code{optimization\_level=0}, pinned to a
connected line of $n$ physical qubits in the coupling map. $n_{2q} = 3(n-1)$ is
the two-qubit (\code{cz}) gate count after decomposition. $P(\ket{e_{n-1}})$ is
the fraction of shots whose entire measured bit-string is exactly the
target $\ket{10\cdots 0}$; any single bit-flip on any qubit (including
the $n - 1$ ``stationary'' \texttt{0}s) makes the shot miss the
target.}
\label{tab:swap-chain-hw}
\end{table}

The model $S = (1 - p_{2q})^{n_{2q}}$ of \cref{sec:hamming-rsweep}, with
$p_{2q} = 2.0\times 10^{-3}$, gives at least $0.88$ for every $n \leq 21$,
while the measured target probability falls below $0.5$ at $n = 18$ and is
$0.39$ at $n = 21$. The model ignores readout error, which affects all $n$
measured bits; with a median readout error of about $10^{-2}$ per qubit,
readout alone limits the target probability to about $(1 - 10^{-2})^{n}$,
which is $0.81$ at $n = 21$. Together, these two factors give $0.72$ at
$n = 21$, still well above the measured value; the model also ignores
decoherence and the differences in error rates between physical qubits.

The target probability at $n = 2$ ($0.754$) is smaller than at $n = 3$ and
$n = 4$. At $n = 2$, the target is the second qubit of the line, and
$14.4\%$ of the shots return $\ket{00}$; three \code{cz} gates and a
readout error of about $10^{-2}$ do not account for this. We attribute it
to that physical qubit; for $n \geq 3$, the target is a different qubit.

\subsection{Comparison at fixed transfer distance}
\label{sec:swap-vs-pst}

\begin{figure}[H]
  \centering
  \includegraphics[width=\linewidth]{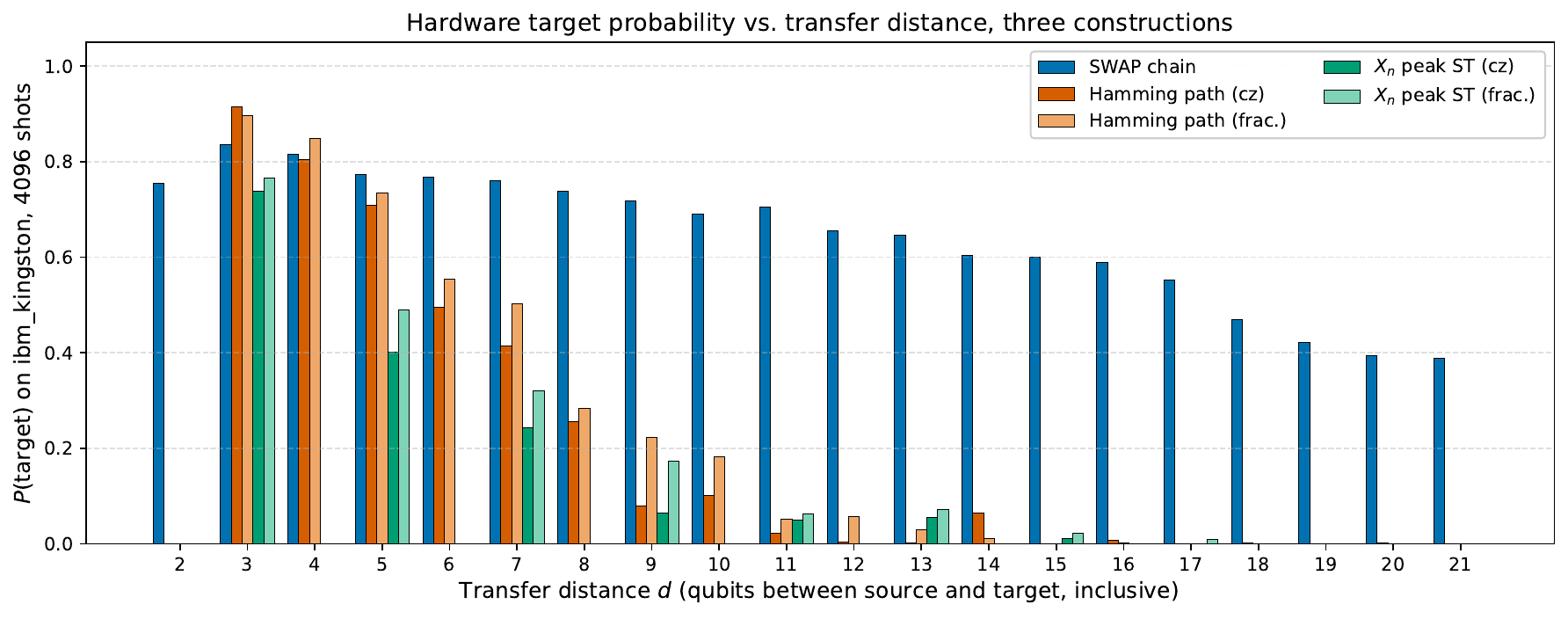}
  \caption{Measured hardware $P(\text{target})$ on \code{ibm\_kingston}
  at transfer distance $d$ for the three constructions. The two
  quantum-walk constructions are shown in both the default \code{cz}
  basis (darker bars) and the fractional basis (lighter bars); the
  SWAP chain transpiles identically in the two bases and so has a single
  bar. At every even $d$ there is no $X_n$ counterpart (the $X_n$ family
  only reaches odd distances). The \code{cz} bars are the data of
  \cref{tab:transfer-distance}; the fractional bars are from
  \cref{tab:hamming-hw,tab:xn-hw}.}
  \label{fig:transfer-distance-comparison}
\end{figure}

To compare the SWAP chain with the walks on $\mathrm{HP}_n$ and $X_n$, we
use the \textsl{transfer distance} $d$, the number of qubits on the path
from the source to the target, including both; this is one more than the
distance between the source and the target in the path. We have $d = n$
for the SWAP chain on $n$ qubits, $d = n + 1$ for $\mathrm{HP}_n$, and
$d = 2n - 1$ for $X_n$, since the source $w_1$ is the second vertex of the
path; thus $X_n$ only gives odd values of $d$.
\Cref{tab:transfer-distance} and \cref{fig:transfer-distance-comparison}
give the target probabilities for $d = 2, \ldots, 21$.

\begin{table}[ht]
\centering
\small
\begin{tabular}{rrrrccc}
\toprule
& \multicolumn{3}{c}{family (size)} & \multicolumn{3}{c}{$P(\text{target})$} \\
\cmidrule(lr){2-4}\cmidrule(lr){5-7}
$d$ & SWAP & HP & $X_n$ & SWAP & HP & $X_n$ \\
\midrule
$ 2$ & $n=2$  & n/a           & n/a           & $0.754$ & n/a     & n/a     \\
$ 3$ & $n=3$  & $\mathrm{HP}_2$  & $X_2$  & $0.836$ & $\mathbf{0.916}$ & $0.738$ \\
$ 4$ & $n=4$  & $\mathrm{HP}_3$  & n/a           & $\mathbf{0.815}$ & $0.805$ & n/a     \\
$ 5$ & $n=5$  & $\mathrm{HP}_4$  & $X_3$  & $\mathbf{0.773}$ & $0.708$ & $0.403$ \\
$ 6$ & $n=6$  & $\mathrm{HP}_5$  & n/a           & $\mathbf{0.768}$ & $0.495$ & n/a     \\
$ 7$ & $n=7$  & $\mathrm{HP}_6$  & $X_4$  & $\mathbf{0.761}$ & $0.414$ & $0.243$ \\
$ 8$ & $n=8$  & $\mathrm{HP}_7$  & n/a           & $\mathbf{0.738}$ & $0.257$ & n/a     \\
$ 9$ & $n=9$  & $\mathrm{HP}_8$  & $X_5$  & $\mathbf{0.719}$ & $0.079$ & $0.065$ \\
$10$ & $n=10$ & $\mathrm{HP}_9$  & n/a           & $\mathbf{0.691}$ & $0.102$ & n/a     \\
$11$ & $n=11$ & $\mathrm{HP}_{10}$ & $X_6$ & $\mathbf{0.705}$ & $0.022$ & $0.050$ \\
$12$ & $n=12$ & $\mathrm{HP}_{11}$ & n/a          & $\mathbf{0.655}$ & $0.003$ & n/a     \\
$13$ & $n=13$ & $\mathrm{HP}_{12}$ & $X_7$ & $\mathbf{0.646}$ & $0.002$ & $0.055$ \\
$14$ & $n=14$ & $\mathrm{HP}_{13}$ & n/a          & $\mathbf{0.604}$ & $0.064$ & n/a     \\
$15$ & $n=15$ & $\mathrm{HP}_{14}$ & $X_8$ & $\mathbf{0.601}$ & $0.000$ & $0.012$ \\
$16$ & $n=16$ & $\mathrm{HP}_{15}$ & n/a          & $\mathbf{0.589}$ & $0.009$ & n/a     \\
$17$ & $n=17$ & $\mathrm{HP}_{16}$ & $X_9$ & $\mathbf{0.552}$ & $0.000$ & $0.001$ \\
$18$ & $n=18$ & $\mathrm{HP}_{17}$ & n/a          & $\mathbf{0.469}$ & $0.002$ & n/a     \\
$19$ & $n=19$ & $\mathrm{HP}_{18}$ & $X_{10}$ & $\mathbf{0.421}$ & $0.000$ & $0.000$ \\
$20$ & $n=20$ & $\mathrm{HP}_{19}$ & n/a          & $\mathbf{0.395}$ & $0.002$ & n/a     \\
$21$ & $n=21$ & $\mathrm{HP}_{20}$ & $X_{11}$ & $\mathbf{0.389}$ & $0.000$ & $0.000$ \\
\bottomrule
\end{tabular}
\caption{Measured hardware target probability $P(\text{target})$ across
the three constructions, indexed by transfer distance $d$. \emph{SWAP}
columns are from \cref{tab:swap-chain-hw}; \emph{HP} columns are from
\cref{tab:hp-rstar}; \emph{X$_n$} columns are from \cref{tab:xn-rstar}.
The bolded entry per row is the largest $P(\text{target})$ among the
three (a tie goes to no bolding). All values are $4096$-shot runs on
\code{ibm\_kingston}.}
\label{tab:transfer-distance}
\end{table}

In the \code{cz} basis, the SWAP chain gives the largest target
probability for every $d \geq 4$, and for $d \geq 12$ it is larger by a
factor of at least $9$. At $d = 3$, $\mathrm{HP}_2$ gives $0.916$ and the
SWAP chain gives $0.836$, although $\mathrm{HP}_2$ uses $14$ two-qubit
gates and the SWAP chain uses $6$. In the fractional basis,
$\mathrm{HP}_3$ also gives a larger target probability than the SWAP chain
at $d = 4$ ($0.849$ against $0.815$); for $d \geq 5$, the SWAP chain is
better in both bases. The SWAP chain uses $3(d - 1)$ two-qubit gates,
fewer than the Trotter circuits at every $d$.

\subsection*{Acknowledgements}

Priscilla Awuah was supported by QDNL for a three-month
research visit to QuSoft at the University of Amsterdam, between March and
May 2026, during which she carried out a master's internship project under
the supervision of Krystal Guo.

Claude Code (Anthropic) was used to implement the Qiskit code for the
computations in this paper, based on code written and inspected by the
authors, and to replicate it for the different graphs we consider. It was
also used to help draft parts of the paper.

\appendix

\section{Per-vertex breakdowns for the Hamming-weighted-path PST experiment}\label{app:hamming-tables}

\Cref{tab:hp-rstar} gives the probabilities of all single-excitation states for the hardware runs on $\mathrm{HP}_n$ in \cref{sec:hamming-hw}, with $r = r^{*}$ from \cref{tab:hp-r-optimal}.

{\small
\begin{longtable}{cccccc}
\caption{Per-vertex single-excitation probabilities for the Hamming-weighted-path PST circuits on $\mathrm{HP}_n$ at the $\hat{P}_{\mathrm{HW}}$-maximizing $r^{*}$ from \cref{tab:hp-r-optimal}, $t = \pi/2$. \emph{Aer noiseless} is the statevector probability after transpilation to \code{ibm\_kingston}'s native \code{cz} basis (no coupling map); the values are identical, to all six decimal places, for the \code{cz}-basis and fractional-basis transpilations of \cref{sec:c4-transpile}, so the column serves both regimes. \emph{Hardware (\code{cz})} and \emph{Hardware (fractional)} are $4096$-shot runs on \code{ibm\_kingston} in the default \code{cz} basis and with fractional gates enabled (\cref{sec:c4-transpile}); the target-vertex row $\ket{e_n}$ is set in bold in each $n$-block. Each $n$-block footer reports the Aer wall-clock time (on a laptop) and the IBM-billed \code{job.usage()} QPU seconds.}
\label{tab:hp-rstar} \\
\toprule
$n$ ($r^{*}$) & qubit & Bit-string & Aer & Hardware (\code{cz}) & Hardware (frac.) \\
\midrule
\endfirsthead

\multicolumn{6}{l}{\textit{(\cref{tab:hp-rstar}, continued)}} \\
\toprule
$n$ ($r^{*}$) & qubit & Bit-string & Aer & Hardware (\code{cz}) & Hardware (frac.) \\
\midrule
\endhead

\midrule
\multicolumn{6}{r}{\textit{continued on next page}} \\
\endfoot

\bottomrule
\endlastfoot

\multirow{6}{*}{$2$ ($r=3$)}
   & $0$ & \texttt{001} & $0.004951$ & $0.010742$ & $0.010254$ \\
   & $1$ & \texttt{010} & $0.000628$ & $0.011230$ & $0.011719$ \\
   & $\mathbf{2}$ & \texttt{100} & $\mathbf{0.994421}$ & $\mathbf{0.915527}$ & $\mathbf{0.896729}$ \\
\cmidrule(lr){2-6}
   & \multicolumn{2}{c}{in-subspace sum} & $1.000000$ & $0.937500$ & $0.918701$ \\
   & \multicolumn{2}{c}{Aer wall time (s)} & $0.17$ & n/a & n/a \\
   & \multicolumn{2}{c}{QPU usage (s)} & n/a & $3$ & $3$ \\
\midrule
\multirow{7}{*}{$3$ ($r=4$)}
   & $0$ & \texttt{0001} & $0.006762$ & $0.014160$ & $0.016113$ \\
   & $1$ & \texttt{0010} & $0.004505$ & $0.010986$ & $0.010010$ \\
   & $2$ & \texttt{0100} & $0.000158$ & $0.012939$ & $0.008057$ \\
   & $\mathbf{3}$ & \texttt{1000} & $\mathbf{0.988575}$ & $\mathbf{0.805420}$ & $\mathbf{0.849365}$ \\
\cmidrule(lr){2-6}
   & \multicolumn{2}{c}{in-subspace sum} & $1.000000$ & $0.843506$ & $0.883545$ \\
   & \multicolumn{2}{c}{Aer wall time (s)} & $0.01$ & n/a & n/a \\
   & \multicolumn{2}{c}{QPU usage (s)} & n/a & $3$ & $3$ \\
\midrule
\multirow{8}{*}{$4$ ($r=4$)}
   & $0$ & \texttt{00001} & $0.000277$ & $0.013184$ & $0.009033$ \\
   & $1$ & \texttt{00010} & $0.021615$ & $0.021973$ & $0.021973$ \\
   & $2$ & \texttt{00100} & $0.010350$ & $0.016602$ & $0.019287$ \\
   & $3$ & \texttt{01000} & $0.000225$ & $0.013428$ & $0.011475$ \\
   & $\mathbf{4}$ & \texttt{10000} & $\mathbf{0.967533}$ & $\mathbf{0.708496}$ & $\mathbf{0.735107}$ \\
\cmidrule(lr){2-6}
   & \multicolumn{2}{c}{in-subspace sum} & $1.000000$ & $0.773682$ & $0.796875$ \\
   & \multicolumn{2}{c}{Aer wall time (s)} & $0.02$ & n/a & n/a \\
   & \multicolumn{2}{c}{QPU usage (s)} & n/a & $3$ & $3$ \\
\midrule
\multirow{9}{*}{$5$ ($r=5$)}
   & $0$ & \texttt{000001} & $0.000931$ & $0.014404$ & $0.010986$ \\
   & $1$ & \texttt{000010} & $0.000028$ & $0.015137$ & $0.009033$ \\
   & $2$ & \texttt{000100} & $0.021464$ & $0.020508$ & $0.019531$ \\
   & $3$ & \texttt{001000} & $0.007514$ & $0.017334$ & $0.010010$ \\
   & $4$ & \texttt{010000} & $0.002368$ & $0.029785$ & $0.010742$ \\
   & $\mathbf{5}$ & \texttt{100000} & $\mathbf{0.967695}$ & $\mathbf{0.495361}$ & $\mathbf{0.553955}$ \\
\cmidrule(lr){2-6}
   & \multicolumn{2}{c}{in-subspace sum} & $1.000000$ & $0.592529$ & $0.614258$ \\
   & \multicolumn{2}{c}{Aer wall time (s)} & $0.02$ & n/a & n/a \\
   & \multicolumn{2}{c}{QPU usage (s)} & n/a & $3$ & $3$ \\
\midrule
\multirow{10}{*}{$6$ ($r=5$)}
   & $0$ & \texttt{0000001} & $0.000827$ & $0.012939$ & $0.006592$ \\
   & $1$ & \texttt{0000010} & $0.004475$ & $0.014648$ & $0.009521$ \\
   & $2$ & \texttt{0000100} & $0.000020$ & $0.009521$ & $0.006348$ \\
   & $3$ & \texttt{0001000} & $0.037302$ & $0.030273$ & $0.022217$ \\
   & $4$ & \texttt{0010000} & $0.011587$ & $0.012939$ & $0.009521$ \\
   & $5$ & \texttt{0100000} & $0.012229$ & $0.023926$ & $0.014893$ \\
   & $\mathbf{6}$ & \texttt{1000000} & $\mathbf{0.933560}$ & $\mathbf{0.414307}$ & $\mathbf{0.502441}$ \\
\cmidrule(lr){2-6}
   & \multicolumn{2}{c}{in-subspace sum} & $1.000000$ & $0.518555$ & $0.571533$ \\
   & \multicolumn{2}{c}{Aer wall time (s)} & $0.03$ & n/a & n/a \\
   & \multicolumn{2}{c}{QPU usage (s)} & n/a & $3$ & $3$ \\
\midrule
\multirow{11}{*}{$7$ ($r=6$)}
   & $0$ & \texttt{00000001} & $0.000123$ & $0.006104$ & $0.005615$ \\
   & $1$ & \texttt{00000010} & $0.000794$ & $0.007080$ & $0.007812$ \\
   & $2$ & \texttt{00000100} & $0.003789$ & $0.010254$ & $0.009521$ \\
   & $3$ & \texttt{00001000} & $0.000008$ & $0.009277$ & $0.006348$ \\
   & $4$ & \texttt{00010000} & $0.033697$ & $0.016602$ & $0.024902$ \\
   & $5$ & \texttt{00100000} & $0.007918$ & $0.013184$ & $0.011719$ \\
   & $6$ & \texttt{01000000} & $0.019714$ & $0.022949$ & $0.035645$ \\
   & $\mathbf{7}$ & \texttt{10000000} & $\mathbf{0.933957}$ & $\mathbf{0.256836}$ & $\mathbf{0.284912}$ \\
\cmidrule(lr){2-6}
   & \multicolumn{2}{c}{in-subspace sum} & $1.000000$ & $0.342285$ & $0.386475$ \\
   & \multicolumn{2}{c}{Aer wall time (s)} & $0.03$ & n/a & n/a \\
   & \multicolumn{2}{c}{QPU usage (s)} & n/a & $3$ & $3$ \\
\midrule
\multirow{12}{*}{$8$ ($r=7$)}
   & $0$ & \texttt{000000001} & $0.000079$ & $0.011963$ & $0.014404$ \\
   & $1$ & \texttt{000000010} & $0.000094$ & $0.011230$ & $0.006592$ \\
   & $2$ & \texttt{000000100} & $0.000768$ & $0.010742$ & $0.007568$ \\
   & $3$ & \texttt{000001000} & $0.003085$ & $0.011963$ & $0.005371$ \\
   & $4$ & \texttt{000010000} & $0.000021$ & $0.007812$ & $0.007812$ \\
   & $5$ & \texttt{000100000} & $0.031014$ & $0.019043$ & $0.010986$ \\
   & $6$ & \texttt{001000000} & $0.005821$ & $0.012695$ & $0.007324$ \\
   & $7$ & \texttt{010000000} & $0.027649$ & $0.015625$ & $0.024902$ \\
   & $\mathbf{8}$ & \texttt{100000000} & $\mathbf{0.931470}$ & $\mathbf{0.078857}$ & $\mathbf{0.223389}$ \\
\cmidrule(lr){2-6}
   & \multicolumn{2}{c}{in-subspace sum} & $1.000000$ & $0.179932$ & $0.308350$ \\
   & \multicolumn{2}{c}{Aer wall time (s)} & $0.04$ & n/a & n/a \\
   & \multicolumn{2}{c}{QPU usage (s)} & n/a & $3$ & $3$ \\
\midrule
\multirow{13}{*}{$9$ ($r=7$)}
   & $0$ & \texttt{0000000001} & $0.000065$ & $0.009033$ & $0.006348$ \\
   & $1$ & \texttt{0000000010} & $0.000473$ & $0.008301$ & $0.005859$ \\
   & $2$ & \texttt{0000000100} & $0.000312$ & $0.005615$ & $0.006836$ \\
   & $3$ & \texttt{0000001000} & $0.001741$ & $0.008789$ & $0.004150$ \\
   & $4$ & \texttt{0000010000} & $0.006394$ & $0.007080$ & $0.007080$ \\
   & $5$ & \texttt{0000100000} & $0.000216$ & $0.005859$ & $0.005615$ \\
   & $6$ & \texttt{0001000000} & $0.044598$ & $0.010742$ & $0.021729$ \\
   & $7$ & \texttt{0010000000} & $0.007655$ & $0.009277$ & $0.011475$ \\
   & $8$ & \texttt{0100000000} & $0.057423$ & $0.026123$ & $0.034180$ \\
   & $\mathbf{9}$ & \texttt{1000000000} & $\mathbf{0.881124}$ & $\mathbf{0.101562}$ & $\mathbf{0.182373}$ \\
\cmidrule(lr){2-6}
   & \multicolumn{2}{c}{in-subspace sum} & $1.000000$ & $0.192383$ & $0.285645$ \\
   & \multicolumn{2}{c}{Aer wall time (s)} & $0.06$ & n/a & n/a \\
   & \multicolumn{2}{c}{QPU usage (s)} & n/a & $3$ & $3$ \\
\midrule
\multirow{14}{*}{$10$ ($r=8$)}
   & $0$ & \texttt{00000000001} & $0.000020$ & $0.004883$ & $0.010010$ \\
   & $1$ & \texttt{00000000010} & $0.000061$ & $0.002197$ & $0.006836$ \\
   & $2$ & \texttt{00000000100} & $0.000448$ & $0.007812$ & $0.005615$ \\
   & $3$ & \texttt{00000001000} & $0.000195$ & $0.003174$ & $0.002930$ \\
   & $4$ & \texttt{00000010000} & $0.001594$ & $0.001709$ & $0.003906$ \\
   & $5$ & \texttt{00000100000} & $0.004647$ & $0.004150$ & $0.007324$ \\
   & $6$ & \texttt{00001000000} & $0.000344$ & $0.006104$ & $0.005127$ \\
   & $7$ & \texttt{00010000000} & $0.040183$ & $0.006592$ & $0.009766$ \\
   & $8$ & \texttt{00100000000} & $0.006124$ & $0.004883$ & $0.004150$ \\
   & $9$ & \texttt{01000000000} & $0.066959$ & $0.006836$ & $0.009521$ \\
   & $\mathbf{10}$ & \texttt{10000000000} & $\mathbf{0.879425}$ & $\mathbf{0.022217}$ & $\mathbf{0.051758}$ \\
\cmidrule(lr){2-6}
   & \multicolumn{2}{c}{in-subspace sum} & $1.000000$ & $0.070557$ & $0.116943$ \\
   & \multicolumn{2}{c}{Aer wall time (s)} & $0.09$ & n/a & n/a \\
   & \multicolumn{2}{c}{QPU usage (s)} & n/a & $3$ & $3$ \\
\midrule
\multirow{15}{*}{$11$ ($r=8$)}
   & $0$ & \texttt{000000000001} & $0.000037$ & $0.007812$ & $0.009521$ \\
   & $1$ & \texttt{000000000010} & $0.000128$ & $0.004395$ & $0.004150$ \\
   & $2$ & \texttt{000000000100} & $0.000203$ & $0.003418$ & $0.006104$ \\
   & $3$ & \texttt{000000001000} & $0.001308$ & $0.001221$ & $0.004395$ \\
   & $4$ & \texttt{000000010000} & $0.000521$ & $0.001709$ & $0.005859$ \\
   & $5$ & \texttt{000000100000} & $0.003120$ & $0.003418$ & $0.004883$ \\
   & $6$ & \texttt{000001000000} & $0.007725$ & $0.002686$ & $0.005859$ \\
   & $7$ & \texttt{000010000000} & $0.001315$ & $0.004883$ & $0.006836$ \\
   & $8$ & \texttt{000100000000} & $0.053787$ & $0.004150$ & $0.008545$ \\
   & $9$ & \texttt{001000000000} & $0.008380$ & $0.003174$ & $0.009277$ \\
   & $10$ & \texttt{010000000000} & $0.112941$ & $0.003418$ & $0.017822$ \\
   & $\mathbf{11}$ & \texttt{100000000000} & $\mathbf{0.810534}$ & $\mathbf{0.003418}$ & $\mathbf{0.057617}$ \\
\cmidrule(lr){2-6}
   & \multicolumn{2}{c}{in-subspace sum} & $1.000000$ & $0.043701$ & $0.140869$ \\
   & \multicolumn{2}{c}{Aer wall time (s)} & $0.10$ & n/a & n/a \\
   & \multicolumn{2}{c}{QPU usage (s)} & n/a & $3$ & $3$ \\
\midrule
\multirow{16}{*}{$12$ ($r=10$)}
   & $0$ & \texttt{0000000000001} & $0.000001$ & $0.001465$ & $0.006104$ \\
   & $1$ & \texttt{0000000000010} & $0.000009$ & $0.000977$ & $0.002686$ \\
   & $2$ & \texttt{0000000000100} & $0.000023$ & $0.001465$ & $0.002441$ \\
   & $3$ & \texttt{0000000001000} & $0.000047$ & $0.001465$ & $0.001465$ \\
   & $4$ & \texttt{0000000010000} & $0.000354$ & $0.000488$ & $0.002930$ \\
   & $5$ & \texttt{0000000100000} & $0.000097$ & $0.000977$ & $0.002197$ \\
   & $6$ & \texttt{0000001000000} & $0.001449$ & $0.001465$ & $0.003906$ \\
   & $7$ & \texttt{0000010000000} & $0.002668$ & $0.000488$ & $0.004150$ \\
   & $8$ & \texttt{0000100000000} & $0.000714$ & $0.001465$ & $0.002686$ \\
   & $9$ & \texttt{0001000000000} & $0.033424$ & $0.000732$ & $0.003906$ \\
   & $10$ & \texttt{0010000000000} & $0.004747$ & $0.001953$ & $0.004395$ \\
   & $11$ & \texttt{0100000000000} & $0.085348$ & $0.001221$ & $0.009766$ \\
   & $\mathbf{12}$ & \texttt{1000000000000} & $\mathbf{0.871119}$ & $\mathbf{0.002197}$ & $\mathbf{0.029297}$ \\
\cmidrule(lr){2-6}
   & \multicolumn{2}{c}{in-subspace sum} & $1.000000$ & $0.016357$ & $0.075928$ \\
   & \multicolumn{2}{c}{Aer wall time (s)} & $0.17$ & n/a & n/a \\
   & \multicolumn{2}{c}{QPU usage (s)} & n/a & $3$ & $3$ \\
\midrule
\multirow{17}{*}{$13$ ($r=3$)}
   & $0$ & \texttt{00000000000001} & $0.303619$ & $0.028809$ & $0.053711$ \\
   & $1$ & \texttt{00000000000010} & $0.060710$ & $0.014404$ & $0.028809$ \\
   & $2$ & \texttt{00000000000100} & $0.002393$ & $0.005859$ & $0.006104$ \\
   & $3$ & \texttt{00000000001000} & $0.001668$ & $0.002686$ & $0.002686$ \\
   & $4$ & \texttt{00000000010000} & $0.000224$ & $0.004150$ & $0.002441$ \\
   & $5$ & \texttt{00000000100000} & $0.013838$ & $0.005859$ & $0.004639$ \\
   & $6$ & \texttt{00000001000000} & $0.002483$ & $0.005371$ & $0.005615$ \\
   & $7$ & \texttt{00000010000000} & $0.040975$ & $0.009033$ & $0.012207$ \\
   & $8$ & \texttt{00000100000000} & $0.003161$ & $0.005859$ & $0.004639$ \\
   & $9$ & \texttt{00001000000000} & $0.026798$ & $0.009033$ & $0.005859$ \\
   & $10$ & \texttt{00010000000000} & $0.000387$ & $0.002930$ & $0.001221$ \\
   & $11$ & \texttt{00100000000000} & $0.003614$ & $0.002441$ & $0.003906$ \\
   & $12$ & \texttt{01000000000000} & $0.004702$ & $0.006592$ & $0.013428$ \\
   & $\mathbf{13}$ & \texttt{10000000000000} & $\mathbf{0.535430}$ & $\mathbf{0.064453}$ & $\mathbf{0.010986}$ \\
\cmidrule(lr){2-6}
   & \multicolumn{2}{c}{in-subspace sum} & $1.000000$ & $0.167480$ & $0.156250$ \\
   & \multicolumn{2}{c}{Aer wall time (s)} & $0.10$ & n/a & n/a \\
   & \multicolumn{2}{c}{QPU usage (s)} & n/a & $3$ & $3$ \\
\midrule
\multirow{18}{*}{$14$ ($r=10$)}
   & $0$ & \texttt{000000000000001} & $0.000003$ & $0.000244$ & $0.000977$ \\
   & $1$ & \texttt{000000000000010} & $0.000018$ & $0.000000$ & $0.000488$ \\
   & $2$ & \texttt{000000000000100} & $0.000036$ & $0.000244$ & $0.000488$ \\
   & $3$ & \texttt{000000000001000} & $0.000138$ & $0.000244$ & $0.000244$ \\
   & $4$ & \texttt{000000000010000} & $0.000234$ & $0.000000$ & $0.000244$ \\
   & $5$ & \texttt{000000000100000} & $0.000284$ & $0.000244$ & $0.000000$ \\
   & $6$ & \texttt{000000001000000} & $0.001650$ & $0.000244$ & $0.000244$ \\
   & $7$ & \texttt{000000010000000} & $0.000552$ & $0.000000$ & $0.000000$ \\
   & $8$ & \texttt{000000100000000} & $0.004629$ & $0.000000$ & $0.000244$ \\
   & $9$ & \texttt{000001000000000} & $0.005987$ & $0.000000$ & $0.000000$ \\
   & $10$ & \texttt{000010000000000} & $0.004475$ & $0.000244$ & $0.000000$ \\
   & $11$ & \texttt{000100000000000} & $0.053471$ & $0.000000$ & $0.000000$ \\
   & $12$ & \texttt{001000000000000} & $0.010552$ & $0.000244$ & $0.000488$ \\
   & $13$ & \texttt{010000000000000} & $0.185497$ & $0.000000$ & $0.000000$ \\
   & $\mathbf{14}$ & \texttt{100000000000000} & $\mathbf{0.732475}$ & $\mathbf{0.000488}$ & $\mathbf{0.000732}$ \\
\cmidrule(lr){2-6}
   & \multicolumn{2}{c}{in-subspace sum} & $1.000000$ & $0.002197$ & $0.004150$ \\
   & \multicolumn{2}{c}{Aer wall time (s)} & $0.36$ & n/a & n/a \\
   & \multicolumn{2}{c}{QPU usage (s)} & n/a & $4$ & $3$ \\
\midrule
\multirow{19}{*}{$15$ ($r=4$)}
   & $0$ & \texttt{0000000000000001} & $0.545576$ & $0.007812$ & $0.003418$ \\
   & $1$ & \texttt{0000000000000010} & $0.055642$ & $0.005127$ & $0.002686$ \\
   & $2$ & \texttt{0000000000000100} & $0.000000$ & $0.002686$ & $0.003174$ \\
   & $3$ & \texttt{0000000000001000} & $0.011260$ & $0.001709$ & $0.002930$ \\
   & $4$ & \texttt{0000000000010000} & $0.000259$ & $0.002930$ & $0.001221$ \\
   & $5$ & \texttt{0000000000100000} & $0.001435$ & $0.003174$ & $0.000977$ \\
   & $6$ & \texttt{0000000001000000} & $0.000017$ & $0.002197$ & $0.001709$ \\
   & $7$ & \texttt{0000000010000000} & $0.000017$ & $0.001709$ & $0.000244$ \\
   & $8$ & \texttt{0000000100000000} & $0.000000$ & $0.002197$ & $0.000732$ \\
   & $9$ & \texttt{0000001000000000} & $0.000017$ & $0.000732$ & $0.000488$ \\
   & $10$ & \texttt{0000010000000000} & $0.000017$ & $0.000732$ & $0.001465$ \\
   & $11$ & \texttt{0000100000000000} & $0.001471$ & $0.001221$ & $0.000977$ \\
   & $12$ & \texttt{0001000000000000} & $0.000364$ & $0.001465$ & $0.001221$ \\
   & $13$ & \texttt{0010000000000000} & $0.016103$ & $0.004395$ & $0.001953$ \\
   & $14$ & \texttt{0100000000000000} & $0.005690$ & $0.002197$ & $0.008545$ \\
   & $\mathbf{15}$ & \texttt{1000000000000000} & $\mathbf{0.362132}$ & $\mathbf{0.008789}$ & $\mathbf{0.001953}$ \\
\cmidrule(lr){2-6}
   & \multicolumn{2}{c}{in-subspace sum} & $1.000000$ & $0.049072$ & $0.033691$ \\
   & \multicolumn{2}{c}{Aer wall time (s)} & $0.25$ & n/a & n/a \\
   & \multicolumn{2}{c}{QPU usage (s)} & n/a & $3$ & $3$ \\
\midrule
\multirow{20}{*}{$16$ ($r=12$)}
   & $0$ & \texttt{00000000000000001} & $0.000000$ & $0.000244$ & $0.000488$ \\
   & $1$ & \texttt{00000000000000010} & $0.000001$ & $0.000000$ & $0.000244$ \\
   & $2$ & \texttt{00000000000000100} & $0.000004$ & $0.000000$ & $0.000000$ \\
   & $3$ & \texttt{00000000000001000} & $0.000014$ & $0.000000$ & $0.000000$ \\
   & $4$ & \texttt{00000000000010000} & $0.000020$ & $0.000000$ & $0.000488$ \\
   & $5$ & \texttt{00000000000100000} & $0.000092$ & $0.000000$ & $0.000000$ \\
   & $6$ & \texttt{00000000001000000} & $0.000117$ & $0.000244$ & $0.000000$ \\
   & $7$ & \texttt{00000000010000000} & $0.000206$ & $0.000000$ & $0.000244$ \\
   & $8$ & \texttt{00000000100000000} & $0.000972$ & $0.000000$ & $0.000244$ \\
   & $9$ & \texttt{00000001000000000} & $0.000350$ & $0.000000$ & $0.000000$ \\
   & $10$ & \texttt{00000010000000000} & $0.003766$ & $0.000000$ & $0.000977$ \\
   & $11$ & \texttt{00000100000000000} & $0.003567$ & $0.000244$ & $0.000732$ \\
   & $12$ & \texttt{00001000000000000} & $0.005161$ & $0.000000$ & $0.000000$ \\
   & $13$ & \texttt{00010000000000000} & $0.043036$ & $0.000000$ & $0.000000$ \\
   & $14$ & \texttt{00100000000000000} & $0.010578$ & $0.000244$ & $0.000000$ \\
   & $15$ & \texttt{01000000000000000} & $0.197042$ & $0.000244$ & $0.000244$ \\
   & $\mathbf{16}$ & \texttt{10000000000000000} & $\mathbf{0.735073}$ & $\mathbf{0.000000}$ & $\mathbf{0.000488}$ \\
\cmidrule(lr){2-6}
   & \multicolumn{2}{c}{in-subspace sum} & $1.000000$ & $0.001221$ & $0.004150$ \\
   & \multicolumn{2}{c}{Aer wall time (s)} & $2.88$ & n/a & n/a \\
   & \multicolumn{2}{c}{QPU usage (s)} & n/a & $4$ & $4$ \\
\midrule
\multirow{21}{*}{$17$ ($r=4$)}
   & $0$ & \texttt{000000000000000001} & $0.204173$ & $0.001465$ & $0.001709$ \\
   & $1$ & \texttt{000000000000000010} & $0.252433$ & $0.001709$ & $0.003418$ \\
   & $2$ & \texttt{000000000000000100} & $0.006530$ & $0.000977$ & $0.001221$ \\
   & $3$ & \texttt{000000000000001000} & $0.023314$ & $0.000244$ & $0.001465$ \\
   & $4$ & \texttt{000000000000010000} & $0.000572$ & $0.000732$ & $0.001221$ \\
   & $5$ & \texttt{000000000000100000} & $0.000055$ & $0.000488$ & $0.000000$ \\
   & $6$ & \texttt{000000000001000000} & $0.000510$ & $0.000488$ & $0.000732$ \\
   & $7$ & \texttt{000000000010000000} & $0.008540$ & $0.001221$ & $0.000977$ \\
   & $8$ & \texttt{000000000100000000} & $0.002260$ & $0.001221$ & $0.000244$ \\
   & $9$ & \texttt{000000001000000000} & $0.014377$ & $0.001709$ & $0.000000$ \\
   & $10$ & \texttt{000000010000000000} & $0.002705$ & $0.001221$ & $0.000732$ \\
   & $11$ & \texttt{000000100000000000} & $0.008828$ & $0.000488$ & $0.001221$ \\
   & $12$ & \texttt{000001000000000000} & $0.000767$ & $0.000488$ & $0.000488$ \\
   & $13$ & \texttt{000010000000000000} & $0.000057$ & $0.000244$ & $0.000488$ \\
   & $14$ & \texttt{000100000000000000} & $0.000879$ & $0.000488$ & $0.000488$ \\
   & $15$ & \texttt{001000000000000000} & $0.025847$ & $0.000488$ & $0.001953$ \\
   & $16$ & \texttt{010000000000000000} & $0.023378$ & $0.001465$ & $0.001465$ \\
   & $\mathbf{17}$ & \texttt{100000000000000000} & $\mathbf{0.424774}$ & $\mathbf{0.002197}$ & $\mathbf{0.001221}$ \\
\cmidrule(lr){2-6}
   & \multicolumn{2}{c}{in-subspace sum} & $1.000000$ & $0.017334$ & $0.019043$ \\
   & \multicolumn{2}{c}{Aer wall time (s)} & $1.94$ & n/a & n/a \\
   & \multicolumn{2}{c}{QPU usage (s)} & n/a & $3$ & $3$ \\
\midrule
\multirow{22}{*}{$18$ ($r=12$)}
   & $0$ & \texttt{0000000000000000001} & $0.000001$ & $0.000000$ & $0.000000$ \\
   & $1$ & \texttt{0000000000000000010} & $0.000003$ & $0.000000$ & $0.000000$ \\
   & $2$ & \texttt{0000000000000000100} & $0.000011$ & $0.000000$ & $0.000244$ \\
   & $3$ & \texttt{0000000000000001000} & $0.000024$ & $0.000000$ & $0.000000$ \\
   & $4$ & \texttt{0000000000000010000} & $0.000052$ & $0.000000$ & $0.000000$ \\
   & $5$ & \texttt{0000000000000100000} & $0.000132$ & $0.000000$ & $0.000000$ \\
   & $6$ & \texttt{0000000000001000000} & $0.000140$ & $0.000000$ & $0.000244$ \\
   & $7$ & \texttt{0000000000010000000} & $0.000472$ & $0.000000$ & $0.000244$ \\
   & $8$ & \texttt{0000000000100000000} & $0.000552$ & $0.000000$ & $0.000244$ \\
   & $9$ & \texttt{0000000001000000000} & $0.000878$ & $0.000000$ & $0.000000$ \\
   & $10$ & \texttt{0000000010000000000} & $0.002787$ & $0.000000$ & $0.000000$ \\
   & $11$ & \texttt{0000000100000000000} & $0.001470$ & $0.000000$ & $0.000000$ \\
   & $12$ & \texttt{0000001000000000000} & $0.009209$ & $0.000000$ & $0.000000$ \\
   & $13$ & \texttt{0000010000000000000} & $0.007078$ & $0.000000$ & $0.000000$ \\
   & $14$ & \texttt{0000100000000000000} & $0.016265$ & $0.000000$ & $0.000000$ \\
   & $15$ & \texttt{0001000000000000000} & $0.059629$ & $0.000000$ & $0.000000$ \\
   & $16$ & \texttt{0010000000000000000} & $0.025755$ & $0.000000$ & $0.000000$ \\
   & $17$ & \texttt{0100000000000000000} & $0.316053$ & $0.000000$ & $0.000000$ \\
   & $\mathbf{18}$ & \texttt{1000000000000000000} & $\mathbf{0.559489}$ & $\mathbf{0.000000}$ & $\mathbf{0.000000}$ \\
\cmidrule(lr){2-6}
   & \multicolumn{2}{c}{in-subspace sum} & $1.000000$ & $0.000000$ & $0.000977$ \\
   & \multicolumn{2}{c}{Aer wall time (s)} & $12.71$ & n/a & n/a \\
   & \multicolumn{2}{c}{QPU usage (s)} & n/a & $4$ & $4$ \\
\midrule
\multirow{23}{*}{$19$ ($r=4$)}
   & $0$ & \texttt{00000000000000000001} & $0.386538$ & $0.000977$ & $0.000244$ \\
   & $1$ & \texttt{00000000000000000010} & $0.083055$ & $0.000977$ & $0.000732$ \\
   & $2$ & \texttt{00000000000000000100} & $0.000156$ & $0.000488$ & $0.000244$ \\
   & $3$ & \texttt{00000000000000001000} & $0.002777$ & $0.000000$ & $0.000244$ \\
   & $4$ & \texttt{00000000000000010000} & $0.000000$ & $0.000000$ & $0.000000$ \\
   & $5$ & \texttt{00000000000000100000} & $0.001677$ & $0.000000$ & $0.000000$ \\
   & $6$ & \texttt{00000000000001000000} & $0.000055$ & $0.000488$ & $0.000488$ \\
   & $7$ & \texttt{00000000000010000000} & $0.008270$ & $0.000244$ & $0.000244$ \\
   & $8$ & \texttt{00000000000100000000} & $0.002230$ & $0.000732$ & $0.000000$ \\
   & $9$ & \texttt{00000000001000000000} & $0.016206$ & $0.000000$ & $0.000244$ \\
   & $10$ & \texttt{00000000010000000000} & $0.006287$ & $0.000977$ & $0.000244$ \\
   & $11$ & \texttt{00000000100000000000} & $0.030122$ & $0.000488$ & $0.000000$ \\
   & $12$ & \texttt{00000001000000000000} & $0.005242$ & $0.001221$ & $0.000732$ \\
   & $13$ & \texttt{00000010000000000000} & $0.029824$ & $0.000732$ & $0.000000$ \\
   & $14$ & \texttt{00000100000000000000} & $0.001912$ & $0.000732$ & $0.000000$ \\
   & $15$ & \texttt{00001000000000000000} & $0.006434$ & $0.000000$ & $0.000244$ \\
   & $16$ & \texttt{00010000000000000000} & $0.000000$ & $0.000977$ & $0.000000$ \\
   & $17$ & \texttt{00100000000000000000} & $0.011152$ & $0.000732$ & $0.000244$ \\
   & $18$ & \texttt{01000000000000000000} & $0.005591$ & $0.000732$ & $0.000244$ \\
   & $\mathbf{19}$ & \texttt{10000000000000000000} & $\mathbf{0.402472}$ & $\mathbf{0.001953}$ & $\mathbf{0.000244}$ \\
\cmidrule(lr){2-6}
   & \multicolumn{2}{c}{in-subspace sum} & $1.000000$ & $0.012451$ & $0.004395$ \\
   & \multicolumn{2}{c}{Aer wall time (s)} & $38.77$ & n/a & n/a \\
   & \multicolumn{2}{c}{QPU usage (s)} & n/a & $3$ & $3$ \\
\midrule
\multirow{24}{*}{$20$ ($r=5$)}
   & $0$ & \texttt{000000000000000000001} & $0.365942$ & $0.000244$ & $0.000244$ \\
   & $1$ & \texttt{000000000000000000010} & $0.199443$ & $0.000000$ & $0.000488$ \\
   & $2$ & \texttt{000000000000000000100} & $0.008785$ & $0.000000$ & $0.000244$ \\
   & $3$ & \texttt{000000000000000001000} & $0.051057$ & $0.000000$ & $0.000000$ \\
   & $4$ & \texttt{000000000000000010000} & $0.002438$ & $0.000244$ & $0.000000$ \\
   & $5$ & \texttt{000000000000000100000} & $0.008701$ & $0.000000$ & $0.000000$ \\
   & $6$ & \texttt{000000000000001000000} & $0.000226$ & $0.000000$ & $0.000000$ \\
   & $7$ & \texttt{000000000000010000000} & $0.000080$ & $0.000000$ & $0.000244$ \\
   & $8$ & \texttt{000000000000100000000} & $0.000034$ & $0.000000$ & $0.000000$ \\
   & $9$ & \texttt{000000000001000000000} & $0.001215$ & $0.000244$ & $0.000000$ \\
   & $10$ & \texttt{000000000010000000000} & $0.000208$ & $0.000000$ & $0.000000$ \\
   & $11$ & \texttt{000000000100000000000} & $0.001890$ & $0.000244$ & $0.000000$ \\
   & $12$ & \texttt{000000001000000000000} & $0.000136$ & $0.000244$ & $0.000000$ \\
   & $13$ & \texttt{000000010000000000000} & $0.000312$ & $0.000488$ & $0.000000$ \\
   & $14$ & \texttt{000000100000000000000} & $0.000009$ & $0.000000$ & $0.000000$ \\
   & $15$ & \texttt{000001000000000000000} & $0.002104$ & $0.000000$ & $0.000000$ \\
   & $16$ & \texttt{000010000000000000000} & $0.001023$ & $0.000244$ & $0.000000$ \\
   & $17$ & \texttt{000100000000000000000} & $0.025887$ & $0.000244$ & $0.000000$ \\
   & $18$ & \texttt{001000000000000000000} & $0.008687$ & $0.000000$ & $0.000244$ \\
   & $19$ & \texttt{010000000000000000000} & $0.159510$ & $0.000244$ & $0.000000$ \\
   & $\mathbf{20}$ & \texttt{100000000000000000000} & $\mathbf{0.162311}$ & $\mathbf{0.000488}$ & $\mathbf{0.000000}$ \\
\cmidrule(lr){2-6}
   & \multicolumn{2}{c}{in-subspace sum} & $1.000000$ & $0.002930$ & $0.001465$ \\
   & \multicolumn{2}{c}{Aer wall time (s)} & $111.05$ & n/a & n/a \\
   & \multicolumn{2}{c}{QPU usage (s)} & n/a & $3$ & $3$ \\

\end{longtable}
}

\section{Per-vertex breakdowns for the $X_n$ peak-state-transfer experiment}\label{app:xn-tables}

\Cref{tab:xn-rstar} gives the probabilities of all single-excitation states for the hardware runs on $X_n$ in \cref{sec:xn-hw}, with $r = r^{*}$ from \cref{tab:xn-r-optimal}.

{\small
\begin{longtable}{cccccc}
\caption{Per-vertex single-excitation probabilities for the $X_n$ peak-ST circuits at the $\hat{P}_{\mathrm{HW}}$-maximizing $r^{*}$ from \cref{tab:xn-r-optimal}, $t = \pi$. \emph{Aer noiseless} is the statevector probability after transpilation to \code{ibm\_kingston}'s native \code{cz} basis (no coupling map); the values are identical, to all six decimal places, for the \code{cz}-basis and fractional-basis transpilations of \cref{sec:c4-transpile}, so the column serves both regimes. \emph{Hardware (\code{cz})} and \emph{Hardware (fractional)} are $4096$-shot runs on \code{ibm\_kingston} in the default \code{cz} basis and with fractional gates enabled (\cref{sec:c4-transpile}); the target-vertex row $\ket{e_{w_n}}$ is set in bold in each $n$-block. Each $n$-block footer reports the Aer wall-clock time (on a laptop) and the IBM-billed \code{job.usage()} QPU seconds.}
\label{tab:xn-rstar} \\
\toprule
$n$ ($r^{*}$) & qubit & Bit-string & Aer & Hardware (\code{cz}) & Hardware (frac.) \\
\midrule
\endfirsthead

\multicolumn{6}{l}{\textit{(\cref{tab:xn-rstar}, continued)}} \\
\toprule
$n$ ($r^{*}$) & qubit & Bit-string & Aer & Hardware (\code{cz}) & Hardware (frac.) \\
\midrule
\endhead

\midrule
\multicolumn{6}{r}{\textit{continued on next page}} \\
\endfoot

\bottomrule
\endlastfoot

\multirow{7}{*}{$2$ ($r=5$)}
   & $0$ & \texttt{0001} & $0.020263$ & $0.024658$ & $0.020752$ \\
   & $1$ & \texttt{0010} & $0.075738$ & $0.066406$ & $0.076660$ \\
   & $2$ & \texttt{0100} & $0.004922$ & $0.013916$ & $0.007080$ \\
   & $\mathbf{3}$ & \texttt{1000} & $\mathbf{0.899077}$ & $\mathbf{0.738037}$ & $\mathbf{0.766113}$ \\
\cmidrule(lr){2-6}
   & \multicolumn{2}{c}{in-subspace sum} & $1.000000$ & $0.843018$ & $0.870605$ \\
   & \multicolumn{2}{c}{Aer wall time (s)} & $0.27$ & n/a & n/a \\
   & \multicolumn{2}{c}{QPU usage (s)} & n/a & $3$ & $3$ \\
\midrule
\multirow{9}{*}{$3$ ($r=6$)}
   & $0$ & \texttt{000001} & $0.001517$ & $0.012207$ & $0.011230$ \\
   & $1$ & \texttt{000010} & $0.017223$ & $0.027832$ & $0.021484$ \\
   & $2$ & \texttt{000100} & $0.073960$ & $0.053955$ & $0.058350$ \\
   & $3$ & \texttt{001000} & $0.070348$ & $0.053955$ & $0.055664$ \\
   & $4$ & \texttt{010000} & $0.000934$ & $0.021729$ & $0.012695$ \\
   & $\mathbf{5}$ & \texttt{100000} & $\mathbf{0.836018}$ & $\mathbf{0.402588}$ & $\mathbf{0.490234}$ \\
\cmidrule(lr){2-6}
   & \multicolumn{2}{c}{in-subspace sum} & $1.000000$ & $0.572266$ & $0.649658$ \\
   & \multicolumn{2}{c}{Aer wall time (s)} & $0.02$ & n/a & n/a \\
   & \multicolumn{2}{c}{QPU usage (s)} & n/a & $3$ & $3$ \\
\midrule
\multirow{11}{*}{$4$ ($r=7$)}
   & $0$ & \texttt{00000001} & $0.002144$ & $0.008789$ & $0.010254$ \\
   & $1$ & \texttt{00000010} & $0.000006$ & $0.008789$ & $0.007568$ \\
   & $2$ & \texttt{00000100} & $0.000536$ & $0.013428$ & $0.010742$ \\
   & $3$ & \texttt{00001000} & $0.035734$ & $0.032715$ & $0.028076$ \\
   & $4$ & \texttt{00010000} & $0.106107$ & $0.040039$ & $0.048828$ \\
   & $5$ & \texttt{00100000} & $0.068488$ & $0.028320$ & $0.040283$ \\
   & $6$ & \texttt{01000000} & $0.006711$ & $0.015869$ & $0.013428$ \\
   & $\mathbf{7}$ & \texttt{10000000} & $\mathbf{0.780275}$ & $\mathbf{0.243408}$ & $\mathbf{0.321533}$ \\
\cmidrule(lr){2-6}
   & \multicolumn{2}{c}{in-subspace sum} & $1.000000$ & $0.391357$ & $0.480713$ \\
   & \multicolumn{2}{c}{Aer wall time (s)} & $0.03$ & n/a & n/a \\
   & \multicolumn{2}{c}{QPU usage (s)} & n/a & $3$ & $3$ \\
\midrule
\multirow{13}{*}{$5$ ($r=8$)}
   & $0$ & \texttt{0000000001} & $0.000509$ & $0.007568$ & $0.006592$ \\
   & $1$ & \texttt{0000000010} & $0.000133$ & $0.010010$ & $0.005127$ \\
   & $2$ & \texttt{0000000100} & $0.006211$ & $0.008789$ & $0.004150$ \\
   & $3$ & \texttt{0000001000} & $0.000459$ & $0.006836$ & $0.005127$ \\
   & $4$ & \texttt{0000010000} & $0.000086$ & $0.008545$ & $0.007568$ \\
   & $5$ & \texttt{0000100000} & $0.038899$ & $0.011230$ & $0.016846$ \\
   & $6$ & \texttt{0001000000} & $0.118611$ & $0.025879$ & $0.037354$ \\
   & $7$ & \texttt{0010000000} & $0.075448$ & $0.016846$ & $0.031006$ \\
   & $8$ & \texttt{0100000000} & $0.050861$ & $0.014404$ & $0.032959$ \\
   & $\mathbf{9}$ & \texttt{1000000000} & $\mathbf{0.708782}$ & $\mathbf{0.065186}$ & $\mathbf{0.173340}$ \\
\cmidrule(lr){2-6}
   & \multicolumn{2}{c}{in-subspace sum} & $1.000000$ & $0.175293$ & $0.320068$ \\
   & \multicolumn{2}{c}{Aer wall time (s)} & $0.06$ & n/a & n/a \\
   & \multicolumn{2}{c}{QPU usage (s)} & n/a & $3$ & $3$ \\
\midrule
\multirow{15}{*}{$6$ ($r=4$)}
   & $0$ & \texttt{000000000001} & $0.000023$ & $0.008057$ & $0.006592$ \\
   & $1$ & \texttt{000000000010} & $0.039734$ & $0.016357$ & $0.022461$ \\
   & $2$ & \texttt{000000000100} & $0.025368$ & $0.010986$ & $0.017334$ \\
   & $3$ & \texttt{000000001000} & $0.002529$ & $0.005127$ & $0.005371$ \\
   & $4$ & \texttt{000000010000} & $0.019471$ & $0.006836$ & $0.007080$ \\
   & $5$ & \texttt{000000100000} & $0.014651$ & $0.004639$ & $0.004639$ \\
   & $6$ & \texttt{000001000000} & $0.031875$ & $0.008301$ & $0.005371$ \\
   & $7$ & \texttt{000010000000} & $0.026985$ & $0.010010$ & $0.006104$ \\
   & $8$ & \texttt{000100000000} & $0.068129$ & $0.015381$ & $0.019531$ \\
   & $9$ & \texttt{001000000000} & $0.102768$ & $0.014893$ & $0.020508$ \\
   & $10$ & \texttt{010000000000} & $0.233259$ & $0.033691$ & $0.044434$ \\
   & $\mathbf{11}$ & \texttt{100000000000} & $\mathbf{0.435207}$ & $\mathbf{0.050049}$ & $\mathbf{0.062744}$ \\
\cmidrule(lr){2-6}
   & \multicolumn{2}{c}{in-subspace sum} & $1.000000$ & $0.184326$ & $0.222168$ \\
   & \multicolumn{2}{c}{Aer wall time (s)} & $0.06$ & n/a & n/a \\
   & \multicolumn{2}{c}{QPU usage (s)} & n/a & $3$ & $3$ \\
\midrule
\multirow{17}{*}{$7$ ($r=4$)}
   & $0$ & \texttt{00000000000001} & $0.046173$ & $0.019531$ & $0.008545$ \\
   & $1$ & \texttt{00000000000010} & $0.025188$ & $0.010498$ & $0.023193$ \\
   & $2$ & \texttt{00000000000100} & $0.159755$ & $0.021240$ & $0.038086$ \\
   & $3$ & \texttt{00000000001000} & $0.009834$ & $0.006348$ & $0.018799$ \\
   & $4$ & \texttt{00000000010000} & $0.012914$ & $0.004150$ & $0.010498$ \\
   & $5$ & \texttt{00000000100000} & $0.000035$ & $0.002197$ & $0.002197$ \\
   & $6$ & \texttt{00000001000000} & $0.005114$ & $0.004639$ & $0.004395$ \\
   & $7$ & \texttt{00000010000000} & $0.000000$ & $0.003174$ & $0.002930$ \\
   & $8$ & \texttt{00000100000000} & $0.005347$ & $0.002441$ & $0.005615$ \\
   & $9$ & \texttt{00001000000000} & $0.000068$ & $0.001465$ & $0.000732$ \\
   & $10$ & \texttt{00010000000000} & $0.013139$ & $0.007080$ & $0.003418$ \\
   & $11$ & \texttt{00100000000000} & $0.011819$ & $0.004639$ & $0.002686$ \\
   & $12$ & \texttt{01000000000000} & $0.012060$ & $0.006592$ & $0.010254$ \\
   & $\mathbf{13}$ & \texttt{10000000000000} & $\mathbf{0.698554}$ & $\mathbf{0.055420}$ & $\mathbf{0.072021}$ \\
\cmidrule(lr){2-6}
   & \multicolumn{2}{c}{in-subspace sum} & $1.000000$ & $0.149414$ & $0.203369$ \\
   & \multicolumn{2}{c}{Aer wall time (s)} & $0.10$ & n/a & n/a \\
   & \multicolumn{2}{c}{QPU usage (s)} & n/a & $3$ & $3$ \\
\midrule
\multirow{19}{*}{$8$ ($r=5$)}
   & $0$ & \texttt{0000000000000001} & $0.017323$ & $0.001953$ & $0.000977$ \\
   & $1$ & \texttt{0000000000000010} & $0.003800$ & $0.001221$ & $0.003174$ \\
   & $2$ & \texttt{0000000000000100} & $0.000385$ & $0.002441$ & $0.000977$ \\
   & $3$ & \texttt{0000000000001000} & $0.038512$ & $0.004150$ & $0.003418$ \\
   & $4$ & \texttt{0000000000010000} & $0.039760$ & $0.002686$ & $0.002930$ \\
   & $5$ & \texttt{0000000000100000} & $0.014048$ & $0.003174$ & $0.003418$ \\
   & $6$ & \texttt{0000000001000000} & $0.026079$ & $0.002441$ & $0.002686$ \\
   & $7$ & \texttt{0000000010000000} & $0.004486$ & $0.001709$ & $0.000977$ \\
   & $8$ & \texttt{0000000100000000} & $0.021495$ & $0.000977$ & $0.001709$ \\
   & $9$ & \texttt{0000001000000000} & $0.002543$ & $0.001465$ & $0.001221$ \\
   & $10$ & \texttt{0000010000000000} & $0.027595$ & $0.001465$ & $0.004395$ \\
   & $11$ & \texttt{0000100000000000} & $0.001330$ & $0.000732$ & $0.001465$ \\
   & $12$ & \texttt{0001000000000000} & $0.045576$ & $0.002930$ & $0.003418$ \\
   & $13$ & \texttt{0010000000000000} & $0.006117$ & $0.001953$ & $0.000977$ \\
   & $14$ & \texttt{0100000000000000} & $0.040285$ & $0.005615$ & $0.006104$ \\
   & $\mathbf{15}$ & \texttt{1000000000000000} & $\mathbf{0.710666}$ & $\mathbf{0.012451}$ & $\mathbf{0.022949}$ \\
\cmidrule(lr){2-6}
   & \multicolumn{2}{c}{in-subspace sum} & $1.000000$ & $0.047363$ & $0.060791$ \\
   & \multicolumn{2}{c}{Aer wall time (s)} & $0.29$ & n/a & n/a \\
   & \multicolumn{2}{c}{QPU usage (s)} & n/a & $3$ & $3$ \\
\midrule
\multirow{21}{*}{$9$ ($r=6$)}
   & $0$ & \texttt{000000000000000001} & $0.038358$ & $0.000244$ & $0.001221$ \\
   & $1$ & \texttt{000000000000000010} & $0.000400$ & $0.000000$ & $0.001221$ \\
   & $2$ & \texttt{000000000000000100} & $0.077755$ & $0.000488$ & $0.001709$ \\
   & $3$ & \texttt{000000000000001000} & $0.001862$ & $0.000000$ & $0.001953$ \\
   & $4$ & \texttt{000000000000010000} & $0.004057$ & $0.000000$ & $0.000977$ \\
   & $5$ & \texttt{000000000000100000} & $0.004197$ & $0.000000$ & $0.001465$ \\
   & $6$ & \texttt{000000000001000000} & $0.002686$ & $0.000000$ & $0.001221$ \\
   & $7$ & \texttt{000000000010000000} & $0.006983$ & $0.000000$ & $0.000244$ \\
   & $8$ & \texttt{000000000100000000} & $0.008307$ & $0.000244$ & $0.001709$ \\
   & $9$ & \texttt{000000001000000000} & $0.007553$ & $0.000488$ & $0.000732$ \\
   & $10$ & \texttt{000000010000000000} & $0.014809$ & $0.000488$ & $0.001221$ \\
   & $11$ & \texttt{000000100000000000} & $0.009213$ & $0.000244$ & $0.002197$ \\
   & $12$ & \texttt{000001000000000000} & $0.028900$ & $0.000488$ & $0.002686$ \\
   & $13$ & \texttt{000010000000000000} & $0.009955$ & $0.000244$ & $0.001709$ \\
   & $14$ & \texttt{000100000000000000} & $0.059620$ & $0.000000$ & $0.002441$ \\
   & $15$ & \texttt{001000000000000000} & $0.000013$ & $0.000244$ & $0.000000$ \\
   & $16$ & \texttt{010000000000000000} & $0.069954$ & $0.000000$ & $0.003418$ \\
   & $\mathbf{17}$ & \texttt{100000000000000000} & $\mathbf{0.655381}$ & $\mathbf{0.000977}$ & $\mathbf{0.009277}$ \\
\cmidrule(lr){2-6}
   & \multicolumn{2}{c}{in-subspace sum} & $1.000000$ & $0.004150$ & $0.035400$ \\
   & \multicolumn{2}{c}{Aer wall time (s)} & $2.18$ & n/a & n/a \\
   & \multicolumn{2}{c}{QPU usage (s)} & n/a & $3$ & $3$ \\
\midrule
\multirow{23}{*}{$10$ ($r=7$)}
   & $0$ & \texttt{00000000000000000001} & $0.029116$ & $0.000000$ & $0.000732$ \\
   & $1$ & \texttt{00000000000000000010} & $0.020235$ & $0.000000$ & $0.000244$ \\
   & $2$ & \texttt{00000000000000000100} & $0.051152$ & $0.000000$ & $0.000244$ \\
   & $3$ & \texttt{00000000000000001000} & $0.000439$ & $0.000000$ & $0.000244$ \\
   & $4$ & \texttt{00000000000000010000} & $0.061109$ & $0.000000$ & $0.000000$ \\
   & $5$ & \texttt{00000000000000100000} & $0.002573$ & $0.000000$ & $0.000244$ \\
   & $6$ & \texttt{00000000000001000000} & $0.023194$ & $0.000000$ & $0.000488$ \\
   & $7$ & \texttt{00000000000010000000} & $0.000036$ & $0.000000$ & $0.000488$ \\
   & $8$ & \texttt{00000000000100000000} & $0.005308$ & $0.000000$ & $0.000000$ \\
   & $9$ & \texttt{00000000001000000000} & $0.000867$ & $0.000000$ & $0.000244$ \\
   & $10$ & \texttt{00000000010000000000} & $0.000255$ & $0.000488$ & $0.000244$ \\
   & $11$ & \texttt{00000000100000000000} & $0.003615$ & $0.000000$ & $0.000000$ \\
   & $12$ & \texttt{00000001000000000000} & $0.001596$ & $0.000000$ & $0.000000$ \\
   & $13$ & \texttt{00000010000000000000} & $0.009255$ & $0.000000$ & $0.000244$ \\
   & $14$ & \texttt{00000100000000000000} & $0.014295$ & $0.000000$ & $0.000488$ \\
   & $15$ & \texttt{00001000000000000000} & $0.017477$ & $0.000000$ & $0.000000$ \\
   & $16$ & \texttt{00010000000000000000} & $0.059022$ & $0.000000$ & $0.000732$ \\
   & $17$ & \texttt{00100000000000000000} & $0.004303$ & $0.000000$ & $0.000244$ \\
   & $18$ & \texttt{01000000000000000000} & $0.115035$ & $0.000000$ & $0.000488$ \\
   & $\mathbf{19}$ & \texttt{10000000000000000000} & $\mathbf{0.581118}$ & $\mathbf{0.000000}$ & $\mathbf{0.000732}$ \\
\cmidrule(lr){2-6}
   & \multicolumn{2}{c}{in-subspace sum} & $1.000000$ & $0.000488$ & $0.006104$ \\
   & \multicolumn{2}{c}{Aer wall time (s)} & $63.53$ & n/a & n/a \\
   & \multicolumn{2}{c}{QPU usage (s)} & n/a & $4$ & $3$ \\
\midrule
\multirow{25}{*}{$11$ ($r=8$)}
   & $0$ & \texttt{0000000000000000000001} & $0.016821$ & $0.000000$ & $0.000000$ \\
   & $1$ & \texttt{0000000000000000000010} & $0.026210$ & $0.000000$ & $0.000000$ \\
   & $2$ & \texttt{0000000000000000000100} & $0.023938$ & $0.000000$ & $0.000244$ \\
   & $3$ & \texttt{0000000000000000001000} & $0.012843$ & $0.000000$ & $0.000000$ \\
   & $4$ & \texttt{0000000000000000010000} & $0.034219$ & $0.000000$ & $0.000488$ \\
   & $5$ & \texttt{0000000000000000100000} & $0.000006$ & $0.000000$ & $0.000244$ \\
   & $6$ & \texttt{0000000000000001000000} & $0.045213$ & $0.000000$ & $0.000244$ \\
   & $7$ & \texttt{0000000000000010000000} & $0.001447$ & $0.000000$ & $0.000244$ \\
   & $8$ & \texttt{0000000000000100000000} & $0.036117$ & $0.000000$ & $0.000000$ \\
   & $9$ & \texttt{0000000000001000000000} & $0.001111$ & $0.000000$ & $0.000000$ \\
   & $10$ & \texttt{0000000000010000000000} & $0.023521$ & $0.000000$ & $0.000000$ \\
   & $11$ & \texttt{0000000000100000000000} & $0.000177$ & $0.000000$ & $0.000000$ \\
   & $12$ & \texttt{0000000001000000000000} & $0.012549$ & $0.000000$ & $0.000244$ \\
   & $13$ & \texttt{0000000010000000000000} & $0.000301$ & $0.000000$ & $0.000000$ \\
   & $14$ & \texttt{0000000100000000000000} & $0.003073$ & $0.000000$ & $0.000000$ \\
   & $15$ & \texttt{0000001000000000000000} & $0.004189$ & $0.000000$ & $0.000000$ \\
   & $16$ & \texttt{0000010000000000000000} & $0.001842$ & $0.000000$ & $0.000000$ \\
   & $17$ & \texttt{0000100000000000000000} & $0.016407$ & $0.000000$ & $0.000000$ \\
   & $18$ & \texttt{0001000000000000000000} & $0.047778$ & $0.000000$ & $0.000000$ \\
   & $19$ & \texttt{0010000000000000000000} & $0.011884$ & $0.000000$ & $0.000000$ \\
   & $20$ & \texttt{0100000000000000000000} & $0.175447$ & $0.000000$ & $0.000000$ \\
   & $\mathbf{21}$ & \texttt{1000000000000000000000} & $\mathbf{0.504908}$ & $\mathbf{0.000000}$ & $\mathbf{0.000000}$ \\
\cmidrule(lr){2-6}
   & \multicolumn{2}{c}{in-subspace sum} & $1.000000$ & $0.000000$ & $0.001709$ \\
   & \multicolumn{2}{c}{Aer wall time (s)} & $354.37$ & n/a & n/a \\
   & \multicolumn{2}{c}{QPU usage (s)} & n/a & $4$ & $4$ \\

\end{longtable}
}

\printbibliography

\end{document}